%% file: draft_prx_accepted.tex
\documentclass[pra,aps,11pt,nofootinbib,groupedaddress]{revtex4-2} 
\usepackage{graphicx}
\usepackage[nointegrals]{wasysym} %
\usepackage[export]{adjustbox}
\usepackage{amsmath,amsfonts,amssymb,latexsym}
\usepackage{hhline}
\usepackage{bm}
\usepackage{verbatim}
\usepackage{enumitem}
\usepackage{mathrsfs}
\usepackage{slashed}
\usepackage{empheq}
\usepackage{physics}

\input{ACPreamble}

\newcommand{\ZS}[1]{\textcolor{blue}{\it [ZS: #1]}}

\newcommand{\bs}{\boldsymbol}

\allowdisplaybreaks

\begin{document}


\title{Doping a fractional quantum anomalous Hall insulator}

\author{Zhengyan Darius Shi}%
\email{zdshi@mit.edu}
\affiliation{
Department of Physics, Massachusetts Institute of Technology,
Cambridge, Massachusetts 02139, USA
}

\author{T. Senthil}
\email{senthil@mit.edu}
\affiliation{
Department of Physics, Massachusetts Institute of Technology,
Cambridge, Massachusetts 02139, USA
}%

\date{September 30, 2024}

\begin{abstract}
       We study novel itinerant phases that can be accessed by doping a fractional quantum anomalous Hall (FQAH) insulator, with a focus on the experimentally observed Jain states at lattice filling $\nu = p/(2p+1)$. Unlike in the lowest Landau level, where charge motion is confined into cyclotron orbits, the charged excitations in the FQAH occupy Bloch states with well-defined crystal momenta. At a nonzero doping density, this enables the formation of itinerant states of the doped anyons just beyond the FQAH plateau region.  Specializing to the vicinity of $\nu = 2/3$, we describe a few possible such itinerant states. These include a topological superconductor with chiral neutral fermion edge modes as well as a more exotic Pair Density Wave (PDW) superconductor with non-abelian topological order. A Fermi liquid metal with a doping-induced period-3 charge density wave also occurs naturally in our analysis. This Fermi liquid (as well as the PDW) arises from pairing instabilities of a composite Fermi liquid metal that can emerge near filling $2/3$. Though inspired by the theory of anyon superconductivity, we explain how our construction is qualitatively different. At a general Jain filling $\nu = p/(2p+1)$, the same analytical framework leads to a wider variety of phases including higher-charge superconductors and generalized composite Fermi liquids. We predict unusual physical signatures associated with each phase and analyze the crossover between different temperature regimes. These results provide a proof-of-principle that exotic itinerant phases can be stabilized by correlations intrinsic to the FQAH setup. 
\end{abstract}

\maketitle
\newpage 
\tableofcontents

\section{Introduction}

Two-dimensional electronic systems in a strong magnetic field exhibit an amazing variety of correlated phenomena at fractional filling of the lowest Landau level, broadly referred to as the fractional quantum Hall (FQH) effect. Four decades after their initial discovery, these phenomena remain the most robust demonstrations of many important concepts in modern condensed matter physics. Salient examples include fractionalized charge and statistics in topologically ordered phases~\cite{Laughlin1983_FQHtheory,Jain1989_CFframework,Stormer1999_FQHreview} as well as emergent Fermi surfaces and non-Fermi liquid physics in composite Fermi liquids (CFL)~\cite{Halperin1993_HLRtheory}. 

One key ingredient for these exotic phases is the topology associated with non-interacting electronic states. While the origin of topology in the continuum is the formation of isolated Landau levels induced by strong magnetic fields, the same topological features persist in lattice systems \textit{without a magnetic field} whenever the active energy bands have a nonzero Chern number~\cite{Thouless1982_TKNN,Haldane1988_QAH}. The robust topological states of matter realized at zero field in a fractionally filled Chern band are collectively known as the fractional quantum anomalous Hall (FQAH) phases.\footnote{We use the terminology FQAH to refer to situations where quantum Hall phenomena occur in systems that are microscopically time reversal invariant. A more general situation is where quantum Hall effects occur in lattice systems, with or without microscopic time reversal symmetry, and we will use the terminology ``Fractional Chern Insulators" for this general case. } Theoretically, FQAH was first conceptualized and established for idealized model systems in a large body of numerical work about a decade ago~\cite{Neupert2010_FQAH,Sun2011_FCI,Sheng2011_FCI,Tang2011_FQAH,Wang2011_FQAH,Regnault2011_FCI,Bergholtz2013_FCIreview,Parameswaran2013_FCIreview}. However, realistic experimental proposals only appeared much later, after the advent of two-dimensional Van der Waals materials with tunable topological bands~\cite{Zhang2018_nearly_flat,Ledwith2019_FCITBG,Repellin2019_TBGChern,Wilhelm2020_TBGFCI,Abouelkomsan2019_MoireFCI,Wu2018_TMD_topo,Yu2019_TMD_topo,Devakul2021_TMD_topo} and robust ferromagnetism 
(see Refs.~\cite{Zhang2018_nearly_flat, Bultinck2020_QAHFM_TBG,zhang2019twisted, Repellin2020_moireFM_theory, liu2021orbital,Devakul2021_TMD_topo} for early theoretical work and Refs.~\cite{Sharpe2019_TBG_FM,Anderson2023_TMD_FM} for experimental realizations in twisted bilayer graphene and twisted Transition Metal Dichalcogenides). The concerted experimental effort in this direction culminated in the recent landmark discovery of FQAH in twisted MoTe$_2$~\cite{Cai2023_FQAHTMD,Park2023_FQAHTMD,Xu2023_FQAHTMD,Zeng2023_FQAHTMD} and rhombohedral pentalayer graphene~\cite{Lu2023_FQAHPenta}.



In this work, we explore itinerant phases that can be accessed by doping an FQAH insulator, with a focus on the experimentally observed fermionic Jain states at lattice filling $\nu = p/(2p+1)$. The charged excitations of the FQAH state are anyons with fractional electric charge. At low doping density, the anyons localize due to a combination of disorder and long-range Coulomb interactions. This localization, of course, is what leads to the quantum Hall plateau as a function of density. We are interested in what happens once the anyon density becomes large enough that we exit the plateau. In the conventional realization of quantum Hall physics in large magnetic fields, the kinetic energy of the anyons is suppressed by the magnetic field; thus the localization effects are very strong, and the plateau persists till the basic quantum Hall physics responsible for the formation of the anyons is itself destroyed. In contrast, as we explain below, the anyonic excitations in an FQAH insulator have a nontrivial dispersion. The resulting kinetic energy competes against the interaction and disorder effects that favor localization. 
It is then natural to ask about the possibility of a delocalized many-body state of the doped anyons. Clearly, if such a delocalized state occurs, it will destroy the quantum Hall plateau.   

The problem of a doped FQAH state has interesting similarities (and differences) with the problem of 
a doped Mott insulator, which shows a number of striking phenomena including high-temperature superconductivity~\cite{Lee2004_dopedMott}. In both cases, the key questions involve the competition between the interaction effects that lead to the fractionally filled insulator, and the electron kinetic energy. At small doping, the density of charge carriers in both cases is determined by the dopant density rather than the total electron density. 
In contrast to a generic Mott insulating state, the FQAH state breaks time-reversal symmetry and has fractionally charged anyons as excitations. Thus, it is interesting to explore the fate of these doped anyons in the hopes of finding novel many-body phenomena.  

A key feature of the FQAH setting -- in contrast to a Landau level -- is the presence of discrete lattice translation symmetries. Many-body energy eigenstates can thus be labeled by the Bloch momenta of the corresponding Brillouin zone. A state with a definite Bloch momentum is a simultaneous eigenstate of the two elementary translations $T_1, T_2$ along two lattice vectors $\bf{e_1}, \bf{e_2}$. Excitations created by local operators must also transform linearly under all physical symmetries including lattice translations, and hence have definite Bloch momenta in the Brillouin zone. However, a single anyonic excitation cannot be created by local operators. This enables it to exhibit the phenomenon of symmetry fractionalization\footnote{A more exotic general possibility - discussed in the literature on Symmetry Enriched Topological Order - in that some symmetries act by permuting distinct anyons. This is not directly pertinent to the phases we discuss in this paper.}, namely fractional quantum numbers; indeed famously the anyons carry fractional electric charge. Less famous, though certainly known, is the statement that the anyons also fractionalize lattice translation symmetry~\cite{Cheng2016_SET_translation,Bultinck2018_LSM_zerofield}. This manifests itself as a projective action of lattice translations on the anyons. For an anyon $a$, we have 
\begin{equation} 
T_1 T_2 T^{-1}_1T^{-1}_2 [a] = e^{i\phi_a}
\end{equation} 
where $\phi_a$ is a phase that depends on the anyon $a$ and is fixed for any given FQAH phase. For concreteness, consider 
the $\frac{2}{3}$-filling FQAH state with quantized Hall conductivity $\sigma_{xy} = \frac{2}{3} \frac{e^2}{h}$. This phase has the same topological order as the particle-hole conjugate of the familiar $1/3$ Laughlin state, and has two distinct nontrivial anyons with fractional charge. For these anyons,  the phase $\phi_a = \frac{2\pi}{3}, \frac{4\pi}{3}$ mod $2\pi$.  Thus, though the state preserves lattice translation symmetry, the anyons move in an enlarged unit cell that is 3 times the size of the microscopic unit cell. In momentum space, this means that the anyon states are labeled by Bloch momenta in a reduced Brillouin zone that is $1/3$ the size of the microscopic one. The projective realization of the translation symmetry implies that there are at least 3 distinct species of anyons that are related by the action of $T_{1,2}$ with identical dispersions in the reduced Brillouin zone. 

Our work is inspired by old ideas on novel many body states of anyons at finite density, the most dramatic example being  an anyon superconductor~\cite{Laughlin1988_anyonSC,Fetter1989_anyonSC_RPA,Lee1989_anyonSC,Chen1989_anyonSC,
halperin1989consequences,Wen1990_anyonSC}. This old literature mostly considered the problem of an interacting anyon fluid without reference to the emergence of these anyons from a particular microscopic system. 
In our work, we are interested in an anyon fluid where the anyons emerge through the formation of specific FQAH states in a lattice electronic system. The parent FQAH phase enforces a specific pattern of $U(1)$ and lattice translation symmetry enrichment in the anyon fluid, which leads to crucial differences with the older analyses of ``anyonic quantum matter".
\begin{figure}[h!]
    \centering
    \includegraphics[width=0.45\textwidth]{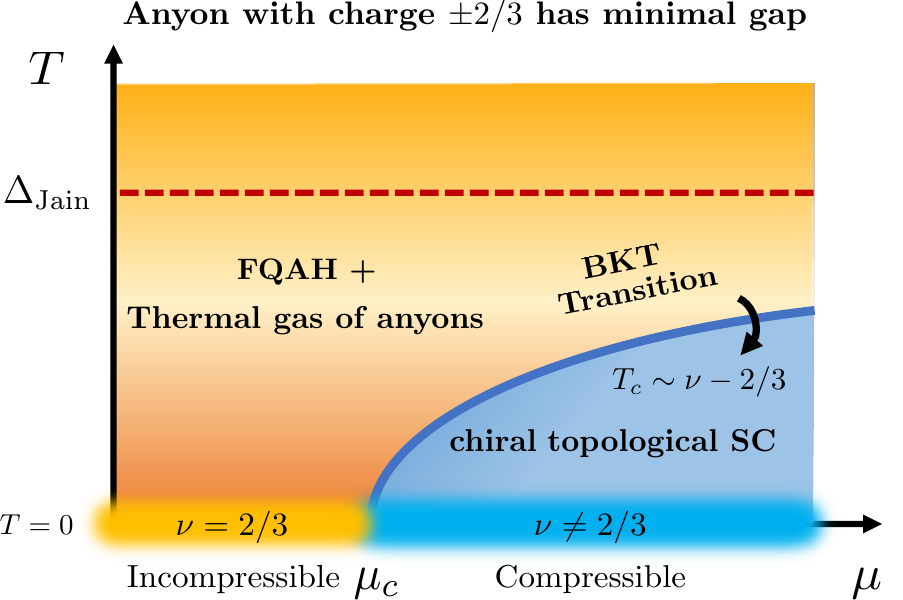}
    \includegraphics[width=0.45\textwidth]{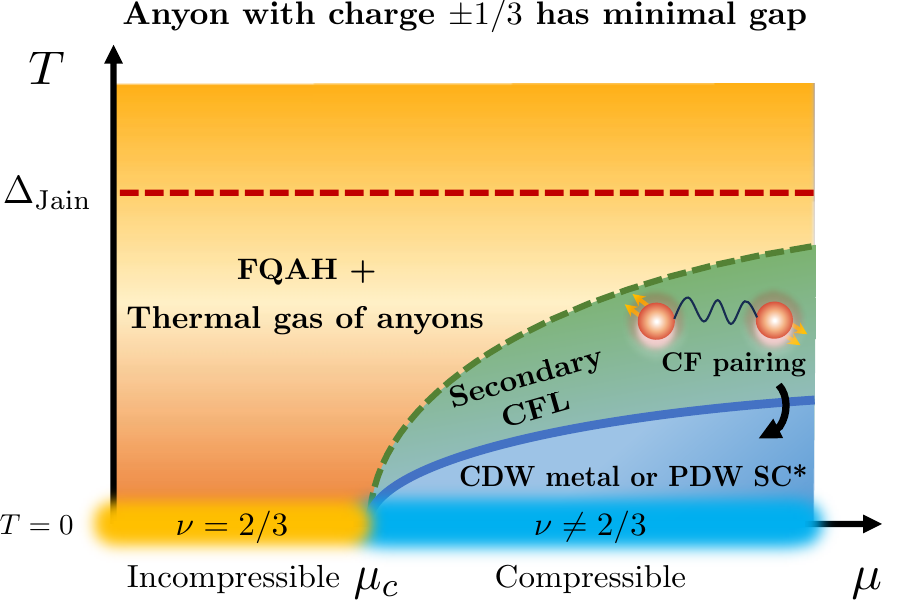}
    \caption{A summary of the itinerant phases that arise in the vicinity of the $\nu = 2/3$ Jain state, upon varying the chemical potential $\mu$ and the temperature $T$. Below the energy gap $\Delta_{\rm Jain}$ of the Jain state, the system behaves like an FQAH insulator + a thermal gas of anyons whose density is determined by $\mu$. At sufficiently low $T$, depending on the energy gaps of different anyons, the thermal gas undergoes a phase transition into one of the compressible phases depicted in the blue regions.}
    \label{fig:finiteT_phase_diagram_intro}
\end{figure} 

\begin{figure}[h!]
    \centering
    \includegraphics[width=0.9\textwidth]{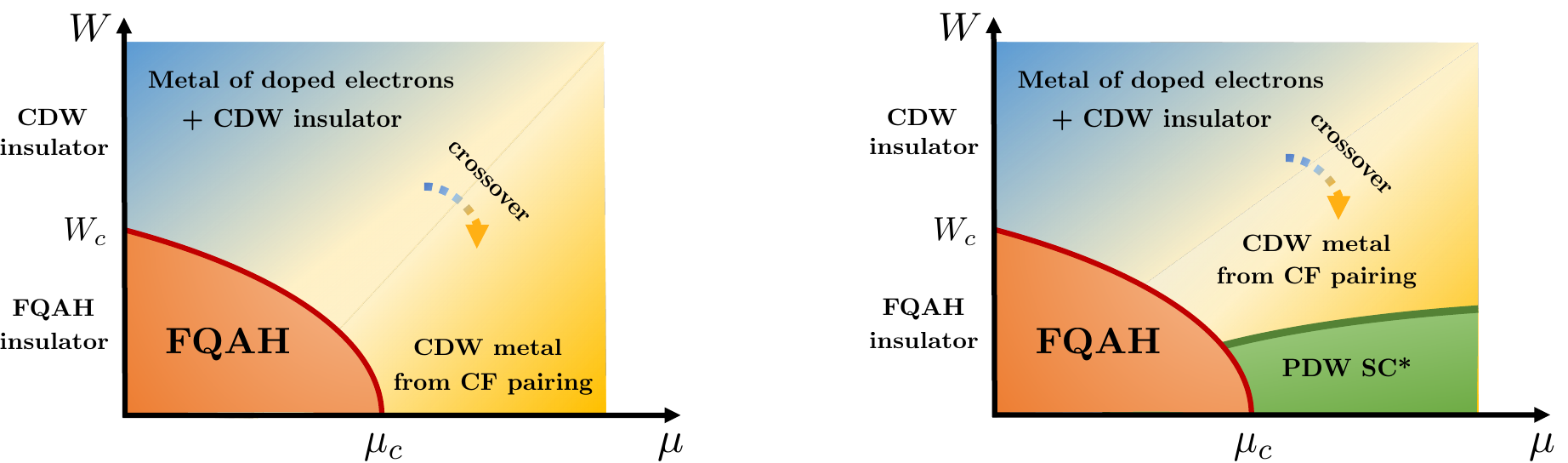}
    \caption{Two possible zero-temperature phase diagrams when the charge $q = \frac{1}{3}$ anyon has the lowest energy gap. $\mu$ and $W$ are dimensionless measures of the chemical potential and bandwidth respectively. The FQAH state at the origin is gapped and incompressible. As a result, it always survives in a finite region in the $\mu$-$W$ plane. On the horizontal axis, increasing $\mu$ drives a transition into a CDW metal or a PDW SC* with non-abelian topological order. Our analysis shows that the conditions for the formation of the PDW SC* are much more stringent than for the CDW metal. Depending on whether these conditions can be satisfied in the accessible region of $\mu$ and $W$, two different phase diagrams are possible, as illustrated above. On the vertical axis, increasing the bandwidth $W$ drives the FQAH state into a CDW insulator. Doping then gives a Fermi liquid of doped electrons on top of the CDW insulator. Despite their distinct microscopic origins, the metallic states near the horizontal and vertical axes turn out to be adiabatically connected.}
    \label{fig:ZeroT_dopef1_phase_diagram}
\end{figure}

In common with the older literature, though through mechanisms that are in general different, we will find a variety of superconducting states, depending on the parent FQAH insulator and the specific anyon that enters the doped system. These include topological superconductors with neutral chiral edge modes, as well as more exotic superconductors (denoted SC$^*$~\cite{Senthil2000_SC*}) that themselves have topological order with nontrivial anyons that descend from those of the ``parent'' FQAH state. Near the $2/3$ state, the SC$^*$ that we find is also a ``Pair Density Wave" (PDW) that spontaneously breaks lattice translation symmetry.  These superconducting states all feature the condensation of charge-$2e$ Cooper pairs, and hence have flux quantization in units of $\frac{h}{2e}$. In addition, we find a Fermi liquid state coexisting with a period-3 charge order that enlarges the unit cell. This Fermi liquid (as well as the exotic PDW SC$^*$) emerges at low temperature out of a finite-temperature non-Fermi liquid state that may be understood as a composite Fermi liquid formed by the doped anyons near $\nu = 2/3$. A summary of these phases is provided in Fig.~\ref{fig:finiteT_phase_diagram_intro} and Fig.~\ref{fig:ZeroT_dopef1_phase_diagram}. 

The same analysis can be applied to a more general Jain filling $\nu = p/(2p+1)$, where a richer variety of delocalized phases emerge.\footnote{A discussion of superconductivity near the $p = 1$ case using flux-attached composite fermions is in Ref. \cite{Tang2013_anyon_hop}.} In addition to the PDW SC* with non-abelian topological order and the CDW metal, we find charge-$|p| e$ superconductors with abelian topological order at even $p$, and generalized composite Fermi liquids (which may have further pairing instabilities) at odd $p$. These more exotic phases become increasingly fragile at large $p$ as their transition temperatures scale as $1/p^2$. 

The superconducting states we find always occur in the presence of (spontaneous) time-reversal (TRS) symmetry breaking. In the context of moire materials, this happens through polarization of the valley (and spin, if it is independent) degree of freedom. We thus have the unusual situation in which a superconductor coexists with a spontaneous orbital magnetization. Within the standard BCS theory, the absence of TRS suppresses Cooper pairing. The route to superconductivity described here is strongly non-BCS, and hence can easily allow for coexistence with TRS breaking. We remark that very recent experiments on four-layer rhombohedral graphene find strong evidence for a valley-polarized superconductor in a range of parameters~\cite{Han2024_chiralSC_penta}. While our analysis is likely not directly applicable to this system (which has no moire potential), these experiments show that such superconductors with broken TRS can exist not just in principle, but in practice as well. 


In what theoretical framework can we access the physics of a doped FQAH state? 
In this work, our interest is in understanding the landscape of possible itinerant phases of a doped FQAH state without committing to a specific setup. Therefore, we will rely on an analytic effective field theoretic description and defer detailed numerical studies on microscopic models to future work. The minimal effective field theory for the FQAH state is a Topological Quantum Field Theory (TQFT) described by a Chern-Simons (CS) Lagrangian that captures the topological order of the FQAH state. 
By coupling the TQFT to a background electromagnetic probe field, the global $U(1)$ charge conservation symmetry can also be readily incorporated. However, including the implementation of the microscopic lattice translation symmetry is more complicated and involves introducing extra background gauge fields associated with lattice translation. As this information is important for us to keep track of, we will not work with the minimal CS theory and instead obtain the FQAH state through a (generalization of) composite fermion construction. 

Historically, composite fermions were introduced~\cite{Jain1989_CFframework} into the study of fractional quantum Hall phenomena through a flux attachment procedure where an electron is bound to an even number of flux quanta. While this procedure has great success for treating electrons in Landau levels, in more general situations (especially the lattice systems of interest to us here), it is more useful to obtain composite fermions through a parton construction where we formally write the electron as a product of three fermionic partons ${c = f_1 f_2 f_3}$ (a more detailed review is provided in Section~\ref{sec:parton}). The parton representation is redundant and must be accompanied by fluctuating gauge fields $a_i$ to which the partons couple.  Within this framework, strongly correlated phases for the electronic system can be described by mean-field phases for the partons, with long-range entanglement encoded in the dynamical gauge fluctuations. This representation becomes equivalent to flux attachment in Landau-level problems, but easily generalizes to other situations. We will see that it provides a flexible framework for accessing the itinerant phases obtained by doping an FQAH insulator. 

For the fermionic Jain state at filling ${\nu = p/(2p+1)}$, the mean-field description is where all three partons form Chern insulators with Chern numbers ${C_1 = p, C_2 = C_3 = 1}$.\footnote{One can prove that for every $p$, this is the unique choice of Chern numbers that gives rise to the Jain topological orders up to permutations.} Upon doping, the nature of the many-body ground state is determined by the active parton Chern band closest to the chemical potential. Physically, the distinct partons 
evolve into distinct anyons of the parent FQAH phase. Thus the formal discussion of which parton band is doped corresponds to the very physical question of which anyon excitation is doped into the system when the electron filling is changed. This physical picture leads us to two families of itinerant phases as illustrated in Fig.~\ref{fig:finiteT_phase_diagram_intro}.

While the stability of these itinerant phases can be established within our effective field theory approach, which of them is realized near an experimentally observed Jain state depends on the intricate interplay between disorder, interactions, and microscopic band structures. The main message of our work is that novel itinerant states specific to the FQAH setup, including superconductors, exist in the vicinity of Jain fractions and exhibit smoking-gun phenomenological signatures. These results also carry over to doped fractional Chern insulators (FCI) stabilized by a nonzero magnetic field, which have been realized in several different experimental platforms~\cite{Spanton2017_FCI,Xie2021_FCI,Aronson2024_FCI}. The various phases we have discussed, such as topological superconductors and charge-ordered Fermi liquids, should be pertinent to these systems as well at small doping.

The rest of this paper will be organized as follows. In Section~\ref{sec:anyon_dispersion}, we make the basic point that anyons in the FQAH insulator are endowed with a nontrivial dispersion due to the projective action of lattice translations. In Section~\ref{sec:parton}, we briefly review the parton-based construction of FQAH states at Jain fillings, which automatically implements the projective action of lattice translation. Next, we dope the parton theory in Section~\ref{sec:two_fate} and motivate the two families of states that naturally arise, depending on which anyon has the lowest excitation gap. To elucidate their conceptual origin without excessive formalism, we first restrict to the simplest case ${p = - 2}$ and derive the phenomenological features associated with each family in Section~\ref{sec:scenario1_SC} and Section~\ref{sec:scenario2_FL}. These constructions are then generalized to arbitrary $p$ in Appendix~\ref{app:generalization}. Future directions are discussed in Section~\ref{sec:discussion}.

\section{The origin of anyon dispersion: lattice translation symmetry and its projective action}\label{sec:anyon_dispersion}

We begin by explaining the origin of anyon dispersion in the FQAH insulator, which plays a fundamental role in the construction of itinerant phases throughout this paper. In a general topological order enriched with a $U(1)$ charge conservation symmetry and a $\mathbb{Z}_x \times \mathbb{Z}_y$ lattice translation symmetry, the translation operators $T_x, T_y$ act projectively on the anyonic excitations
~\cite{jalabert1991spontaneous,senthil2000z,Cheng2016_SET_translation,Bultinck2018_LSM_zerofield}. To illustrate this general phenomenon, we consider the simplest case of a $\nu = 1/3$ FQAH insulator on the square lattice, which exhibits the same topological order as the Laughlin state. In the translation-invariant ground state wavefunction, there is a background anyon $a$ with charge $q_a = 1/3$ in each unit cell (see Ref.~\cite{Lu2017_filling_Hall} for a transparent physical discussion). Now, let us add an elementary anyon excitation $b$ on top of the ground state. Upon transporting the additional anyon $b$ around a single unit cell, the wavefunction accumulates a braiding phase $\theta_{ab} = 2\pi n_{ab}/3$ where $n_{ab}$ is an integer that depends on the choice of $b$. This implies that translation operators act on anyon $b$ as $T_x T_y T_x^{-1} T_y^{-1} = e^{2\pi i n_{ab}/3}$. For example, we get $n_{ab} = 1, 2$ when $b = a, a^2$ respectively. One can check that $n_{ab}$ is not a multiple of 3 for any choice of nontrivial anyon $b$ in this topological order. As a result, the original lattice translation symmetry acts projectively on every nontrivial anyon.

\begin{figure}[!h]
    \centering
    \includegraphics[width=0.9\linewidth]{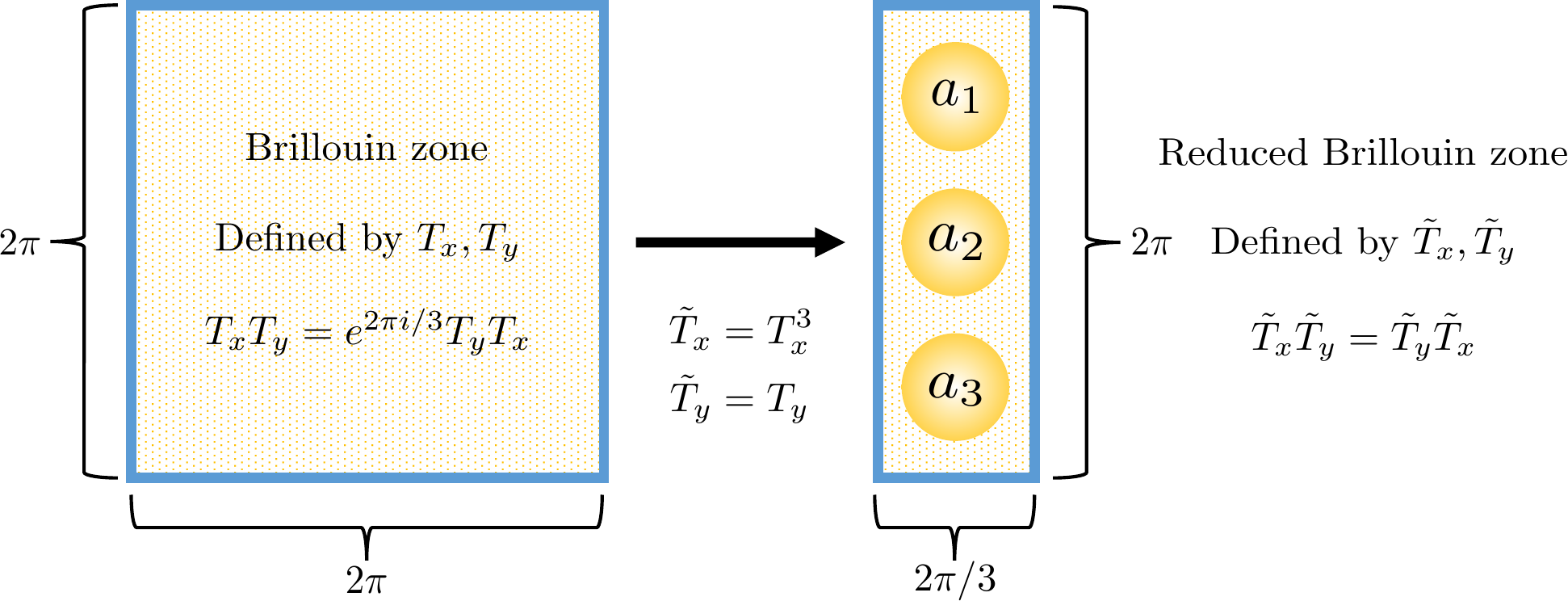}
    \caption{In a $U(1) \times \mathbb{Z}^2$ symmetry-enriched topological order, lattice translation operators $T_x, T_y$ (generating the $\mathbb{Z}^2$) act projectively on the anyons that carry fractional $U(1)$ charge. However, at rational filling, there is always a modified set of commuting translation operators $\tilde T_x, \tilde T_y$ under which the anyons carry a well-defined momentum. In the reduced Brillouin zone defined by $\tilde T_x, \tilde T_y$, the anyon splits into three degenerate species that transform under $T_x, T_y$ as in \eqref{eq:translation_action_anyon}. This ``fractionalized momentun" of the anyons in the FQAH states distinguishes them from conventional FQH states realized in Landau levels in a large magnetic field. }
    \label{fig:projective_translation}
\end{figure}

The presence of these projective phases means that anyon excitations do not carry a well-defined crystal momentum in the original Brillouin zone $k_x, k_y \in [0,2\pi] \times [0, 2\pi]$. However, since the projective phase is commensurate, we can define modified translation operators $\tilde T_x = T_x^3, \tilde T_y = T_y$ that commute with each other. These modified translation operators define a tripled unit cell and a reduced Brillouin zone $\tilde k_x, \tilde k_y \in [0,2\pi/3] \times [0, 2\pi]$. Inside the reduced Brillouin zone, the anyon must furnish a faithful representation of the projective translation symmetry generated by $T_x, T_y$. The minimum dimension of an irreducible representation of this projective algebra is three. Therefore, we will generically have three degenerate species of anyons $a_{n=1,2,3}(\tilde k_x, \tilde k_y)$ as illustrated in Fig.~\ref{fig:projective_translation}.\footnote{In the presence of additional lattice symmetries, it is possible to have a higher number of degenerate species that transform in a larger reducible representation of the projective algebra. See Ref.~\onlinecite{Tang2013_anyon_hop} for an example on the triangular lattice, where a six-fold degeneracy is enforced by a $C_6$ lattice rotational symmetry. By modifying the hopping parameters, we can reduce the rotational symmetry from $C_6$ to $C_3$ and lift the six-fold degeneracy. However, the projective lattice translation algebra requires that a three-fold degeneracy always survives. In current experimental realizations (twisted MoTe$_2$ and pentalayer graphene), we only have $C_3$ symmetry for valley-polarized electrons. Thus we limit our analysis to this case, and consider only the minimal 3-fold degeneracy of anyon species.} The original lattice translation operators act on these species as
\begin{equation}\label{eq:translation_action_anyon}
    \begin{aligned}
    T_x&: a_n(\tilde k_x, \tilde k_y) \rightarrow a_{n+1}(\tilde k_x, \tilde k_y) \, e^{i \tilde k_x/3} \,, \\ \quad T_y&: a_n(\tilde k_x, \tilde k_y) \rightarrow a_n(\tilde k_x, \tilde k_y) \, e^{2\pi n i/3} \, e^{i \tilde k_y} \,, 
    \end{aligned}
\end{equation}
where $a_{n+3}$ is identified with $a_n$. Since $T_x, T_y$ commute with the microscopic Hamiltonian, these three species of anyons have degenerate energy dispersion in the reduced Brillouin zone. The above discussion generalizes easily to an arbitrary Jain state, where the projective phase is now an integer multiple of $2\pi/(2p+1)$. The Brillouin zone gets reduced by a factor of $|2p+1|$ and each anyon splits into at least $|2p+1|$ species $a_n$ that transform under $T_x, T_y$ as
\begin{equation}
    \begin{aligned}
    T_x&: a_n(\tilde k_x, \tilde k_y) \rightarrow a_{n+1}(\tilde k_x, \tilde k_y) \, e^{i \tilde k_x/|2p+1|} \,, \\ \quad T_y&: a_n(\tilde k_x, \tilde k_y) \rightarrow a_n(\tilde k_x, \tilde k_y) \, e^{2\pi n i/|2p+1|} \, e^{i \tilde k_y} \,, 
    \end{aligned}
\end{equation}
where $a_n$ is identified with $a_{n+|2p+1|}$ and $\tilde k_x, \tilde k_y \in [0, 2\pi/|2p+1|, 0, 2\pi]$. 

Except in fine-tuned limits, the energy of these anyon species $\epsilon_n(\tilde k_x, \tilde k_y)$ is generally momentum-dependent. As a simple toy model, one can start with a Landau level and broaden it into a Chern band by introducing a periodic potential that encloses one flux quantum per unit cell. When the strength of the periodic potential $V$ is much weaker than the repulsive electron-electron interaction energy $U$, the topological order of the Laughlin state survives and we can ask about the dispersion of anyons. A quasi-hole wavefunction of the Laughlin state is Gaussian-localized at some spatial coordinate $\eta = \eta_x + i \eta_y$, with a decay length set by the magnetic length $l_B$. Since the unit cell encloses one flux quantum, the set of quasi-hole wavefunctions centered at the minima of the periodic potential $V(\bs{r})$ form a complete non-orthogonal basis of the single-quasihole sector of the Hilbert space. From here, one can set up a tight-binding model, treating the almost-orthogonal basis states as ``atomic orbitals". This tight-binding model generates a dispersion for the anyons $\epsilon(\tilde k_x, \tilde k_y) \sim V$, with a prefactor set by the overlap between neighboring orbitals. When a finite density of anyons are excited, the momentum dependence of their energy dispersion suppresses the tendency towards spatial localization and sets the stage for a variety of itinerant phases that we explore in the rest of this paper. 

\section{Parton construction of fractional quantum anomalous Hall insulators at the Jain fractions}\label{sec:parton}

A systematic theoretical framework that allows us to access the dynamics of a finite density of itinerant anyons is the parton construction, which is familiar from the study of other strong correlation problems (see e.g. the review Ref.~\onlinecite{Lee2004_dopedMott}). The basic philosophy of the parton construction is to rewrite the electron operator $c$ as a product of partons $f_i$. This decomposition introduces an unphysical gauge redundancy that must be removed by the inclusion of emergent fluctuating gauge fields $a_i$. Fractionalized phases can then be realized by considering different mean-field ansatzes for $f_i$ and $a_i$. 
An important merit of the parton construction is that constraints arising from the lattice translation symmetry, such as those derived in Section~\ref{sec:anyon_dispersion}, are automatically incorporated. Moreover, in contrast to the more familiar flux attachment approach, the parton construction always generates topological terms in the low-energy effective Lagrangian that respect flux quantization and correctly capture the global properties of the topological phase. Both of these features will play an important role in the detailed analysis in Sections~\ref{sec:scenario1_SC} through~\ref{sec:scenario2_FL}. With this basic strategy in mind, let us review a concrete construction of the experimentally observed principal Jain sequence at fractional filling $\nu = p/(2p+1)$. A discussion of more general applications of the parton construction to fractionalized phases in the FQAH setting can be found in~\cite{Song2023_QPT_FQAH}. 

For the fermionic Jain states, we decompose the electron as $c(\bs{r}) = f_1(\bs{r}) f_2(\bs{r}) f_3(\bs{r})$. This decomposition introduces a local $SU(3)$ gauge redundancy under which the partons transform as
\begin{equation}
    \begin{aligned}
    f_i(\bs{r}) &\rightarrow U_{ij}(\bs{r}) f_j(\bs{r}) \,, \\
    c(\bs{r}) &\rightarrow c(\bs{r}) \det U(\bs{r}) \,,
    \end{aligned}
\end{equation}
where $U(\bs{r}) \in SU(3)$. To describe the Jain sequence, we consider a mean-field ansatz in which the full SU(3) gauge symmetry is broken to a $U(1) \times U(1)$ subgroup generated by two parameters, $\theta_1$ and $\theta_2$, which act on the partons as
\begin{equation}
    \begin{aligned}
    f_1(\bs{r}) &\rightarrow e^{i\theta_1(\bs{r})} f_1(\bs{r}) \,,\\ 
    f_2(\bs{r}) &\rightarrow e^{i \theta_2(\bs{r}) - i \theta_1(\bs{r})} f_2(\bs{r}) \,,\\
    f_3(\bs{r}) &\rightarrow e^{-i\theta_2(\bs{r})} f_3(\bs{r}) \,.
    \end{aligned}
\end{equation}
The Fock space associated with $f_1, f_2, f_3$ is much larger than the microscopic electronic Hilbert space. To remove the unphysical degrees of freedom, we must introduce a pair of dynamical $U(1)$ gauge fields $a$ and $b$ that transform as $a \rightarrow a - d \theta_1, b \rightarrow b - d \theta_2$.\footnote{\label{footnote_metric} In our subsequent discussion we will pass to a continuum field theory of Lagrangians such as these. In addition, it will be extremely useful to place this continuum theory on a general oriented space-time 3-manifold of the form $M_3 = M_2 \times M_1$ (where $M_n$ denotes an $n$-manifold) with a metric $g$. This will enable us to keep track of important topological data such as the quantized thermal Hall conductivity. The restriction to manifolds of this form is appropriate as our theory is non-relativistic. 
We will also need to specify whether any abelian gauge field is an ordinary $U(1)$ gauge field or a $\text{spin}_{\mathbb{C}}$ connection. The latter has a modified flux quantization condition around cycles. For a review, see Ref.~\cite{senthil2019duality} and references therein. As emphasized in Refs.~\cite{metlitski2015s,Seiberg2016_TQFTgappedbdry}, this distinction cleanly enables keeping track of gauge fields whose elementary source fields are fermions from those whose source fields are bosons. Thus the external probe field $A$ is a $\text{spin}_{\mathbb{C}}$ connection as its elementary sources are electrons. Similarly since $f_i$'s are fermions, $a$ must be a $\text{spin}_{\mathbb{C}}$ connection and $b$ must be an ordinary $U(1)$ gauge field.} In terms of these gauge fields, the low energy effective Lagrangian takes the general form
\begin{equation}
    L_{\rm eff} = L[f_1, a] + L[f_2, b - a] + L[f_3, A - b] \,,
\end{equation}
where $f_3$, the only parton that couples to the external gauge field $A$, carries the physical charge of the microscopic electron. At fractional filling $\nu = p/(2p+1)$, the equations of motion for the gauge fields enforce the constraint $\nu_1 = \nu_2 = \nu_3 = \nu$. 

At the mean-field level, the ground state of this theory is determined by the emergent fluxes carried by $a$ and $b$. In order to construct an incompressible phase at fractional filling, each of the fermionic partons must see a mean-field flux per unit cell proportional to $2\pi/(2p+1)$.  For concreteness, let us choose the mean-field flux assignment
\begin{equation}\label{eq:mean_field_flux_Jain}
    \frac{1}{2\pi} \ev{\nabla \times \bs{a}} = \frac{1}{2p+1} \,, \quad \frac{1}{2\pi} \ev{\nabla \times \bs{b}} = \frac{p+1}{2p+1} \,.
\end{equation}
In the presence of this average flux, the microscopic lattice translation operators $T_x, T_y$ act projectively on the low-energy theory. A well-defined lattice translation symmetry can only be defined using modified lattice translation operators $\tilde T_x = T_x^n, \tilde T_y = T_y^m$ such that $n, m \in \mathbb{Z}$ and $nm = |2p+1|$. These modified translation operators define a new unit cell that is enlarged by a factor of $|2p+1|$ relative to the microscopic unit cell. In this enlarged unit cell, the partons fill an integer number of Chern bands and can form ordinary Chern insulators. The principal Jain sequence is realized by a mean-field state in which the Chern insulators for $f_1, f_2, f_3$ have Chern numbers ${C_1 = p, C_2 = C_3 = 1}$. After integrating out the gapped partons, the remaining gauge fluctuations are captured by a continuum topological quantum field theory (TQFT), which we couple to a background $\textrm{spin}_{\mathbb{C}}$ connection $A$ and a metric $g$ (see footnote \ref{footnote_metric})
\begin{equation}\label{eq:Jain_TQFT}
    \begin{aligned}
    L_{\rm eff} &= \frac{p}{4\pi} a\, da + \frac{1}{4\pi} (b-a)\, d(b-a) + \frac{1}{4\pi} (A - b)\, d (A - b) + 2(2+p)\, \textrm{CS}_g \,,
    \end{aligned}
\end{equation}
where $\textrm{CS}_g$ is a gravitational Chern-Simons term that must accompany an ordinary Chern-Simons term for a $\textrm{spin}_{\mathbb{C}}$ connection (see e.g. the discussion in Ref. \onlinecite{seiberg2016duality}). The composite Fermi liquid state at $\nu = 1/2$ can be accessed by taking the $p \rightarrow \infty$ limit of \eqref{eq:mean_field_flux_Jain} such that $f_2$ and $f_3$ see $\pm \pi$-flux while $f_1$ sees zero flux. In this flux background, $f_1$ forms a Fermi liquid, while $f_2$ and $f_3$ form $C = 1$ Chern insulators. The combined effect of $f_2$ and $f_3$ is to generate a $U(1)_{2}$ bosonic Laughlin state that couples to the $f_1$ parton and implements a lattice version of the flux attachment procedure~\cite{Jain1989_CFframework}. This allows us to identify $f_1$ as the zero-field analogue of the familiar composite fermion.

Now let us derive several physical properties associated with the TQFT in \eqref{eq:Jain_TQFT}. On a manifold without curvature, the gravitational Chern-Simons term can be dropped and the remaining TQFT simplifies to
\begin{equation}
    L_{\rm eff} = \frac{1}{4\pi} \sum_{I,J} \alpha_I K_{IJ} d \alpha_J - \frac{1}{2\pi} \alpha_2 d A + \frac{1}{4\pi} A d A \,,
\end{equation}
where $\alpha_1, \alpha_2$ are identified with $a, b$ and the K-matrix takes the form
\begin{equation}
    K = \begin{pmatrix}
        p+1 & - 1 \\ -1 & 2 
    \end{pmatrix} \,, \quad K^{-1} = \begin{pmatrix}
        \frac{2}{2p+1} & \frac{1}{2p+1} \\ \frac{1}{2p+1} & \frac{p+1}{2p+1} 
    \end{pmatrix} \,.
\end{equation}
From this description, we identify the physical charge vector as $t_I = (0, -1)^T$. A simple calculation shows that this charge vector reproduces the correct Hall conductivity
\begin{equation}
    \sigma_H = 1 - \sum_{I,J} t_I (K^{-1})_{IJ} t_J = 1 - \frac{p+1}{2p+1} = \frac{p}{2p+1} \,.
\end{equation}
The anyons in this theory can be labeled by the integer-quantized charges $l_1, l_2$ that they carry under $a, b$. From the K-matrix, we can determine the physical charge and self-statistics associated with each anyon (since $a$ is a $\text{spin}_{\mathbb{C}}$ connection, sources of $a$ additionally induce a $\pi$-shift in the self-statistics):
\begin{equation}
    \begin{aligned}
        q_{(l_1,l_2)} &= - \sum_{I,J} t_I (K^{-1})_{IJ} l_J = \frac{1}{2p+1} \left[l_1 + (p+1) l_2\right] \,, \\
        \theta_{(l_1,l_2)} &= - \sum_{I,J} l_I (K^{-1})_{IJ} l_J + l_1\pi = -\frac{\pi}{2p+1} \left[2 l_1^2 + (p+1) l_2^2 + 2 l_1 l_2\right] + l_1 \pi \,.
    \end{aligned}
\end{equation}
Returning to the parton Lagrangian, we see that every parton can bind with emergent fluxes of $a$ and $b$ to form deconfined anyon excitations of the Jain state:
\begin{enumerate}
    \item $f_1$ carries $+1$ charge under $a$ and maps to ${(l_1, l_2) = (1, 0)}$. The physical charge of this anyon is ${q = 1/(2p+1)}$ and the self-statistics is ${\theta = \pi (2p-1)/(2p+1)}$. 
    \item $f_2$ carries charge $\mp 1$ under $a, b$ respectively. Therefore, it maps to $(l_1, l_2) = (-1, 1)$ with charge $q=p/(2p+1)$ and statistics $\theta = \pi \left[-(p+1)/(2p+1) - 1\right] = \pi p/(2p+1) \mod 2 \pi$. 
    \item $f_3$ carries charge $\pm 1$ under $A, b$ respectively and maps to a bound state of the electron $c$ and an anyon with $(l_1, l_2) = (0, -1)$. The total physical charge is $q = -(p+1)/(2p+1) + 1 = p/(2p+1)$ and the self-statistics is $\theta = \pi\left[-(p+1)/(2p+1) + 1 \right] = \pi p/(2p+1)$. 
\end{enumerate}
The anti-particles of $f_i$ carry the opposite charge but the same fractional statistics. For the $\nu = 2/3$ state, all the nontrivial anyons can be obtained from either $f_i$ or their anti-particles. As we will see, these basic topological properties will provide some simple intuition for the technical results in the next few sections.

\section{Fate of the Jain state under doping: preliminaries}\label{sec:two_fate}

Having reviewed the parton construction for the Jain states, let us now hole-dope the system to ${\nu = p/(2p+1) - \delta}$ where $\delta$ is a small positive number. The story for electron doping is similar and will not be discussed separately in this paper.  

In real systems, the presence of  spatial disorder will localize the doped holes at low doping so that the quantized topological properties of the Jain state persist to finite $\delta$.  Further, even without disorder, long-range Coulomb interactions will always favor a Wigner crystal state of the doped charges, 
and this will be pinned once weak disorder is included. Both of these possibilities lead to extended plateaus along the doping axis. Here we are interested in what happens beyond the plateau, assuming that the basic physics of the ``parent" FQAH state is still operative.  We thus consider the situation in which the doped charges enter in the form of anyons forming an itinerant fluid (as a result of quantum melting of any putative anyon Wigner crystal). Clearly this is more likely in the FQAH (or FCI) context where the kinetic energy of the anyons is not suppressed by an orbital magnetic field.  We expect that such an itinerant anyon fluid can form in some range of doping near the FQAH plateau before the basic physics of the parent FQAH is destroyed in favor of, say, a different (F)QAH state at some other filling.   

With spatial localization excluded, we return to the clean parton effective theory and analyze the effects of doping. In this paper we explore the simplest possibility, namely that there is a single parton band that is partially filled in the doped system. This will enable us to describe some (but certainly not all) possible fates of the doped FQAH state. In principle, the band structure of $f_i$ 
is non-universal and can take arbitrarily complicated forms. When different bands overlap, doping affects the mean-field states of all three partons, thereby complicating  the analysis. 

Thus we proceed under a ``single active parton" assumption: all the parton bands near the chemical potential are well-separated in energy, so that hole-doping only affects the top valence band associated with one of the three partons, leaving the incompressible states formed by the remaining two partons unmodified. Since the mean-field Chern insulators formed by $f_2$ and $f_3$ have the same Chern number, activating $f_2$ and $f_3$ give identical predictions for the universal low energy physics. We are thus left with two distinct possibilities -- activating $f_1$ or activating $f_2/f_3$. In Sections~\ref{sec:scenario1_SC} and \ref{sec:scenario2_FL}, we consider these two possibilities for the simplest case where $\nu = 2/3$ ($p = -2$). The generalization to arbitrary $p$ is straightforward and will be treated in Appendix~\ref{app:generalization}. 

Physically, the question of which parton band gets activated is directly related to the energetics of distinct anyon species in the system. As demonstrated before, at $\nu = 2/3$, $f_1^\dagger$ sources an anyon with physical charge $q = \frac{1}{3}$ and self-statistics $-\frac{\pi}{3}$, henceforth denoted $a_{\frac{1}{3}}$. The anti-particle of this anyon (corresponding to $f_1$) with charge $q = - \frac{1}{3}$ and self-statistics $\theta = -\frac{\pi}{3}$ will be denoted $\bar{a}_{\frac{1}{3}}$. The $f_2/f_3$ field sources an anyon with physical charge $q = \frac{2}{3}$ and self-statistics $\frac{2\pi}{3}$. We will denote this anyon $a_{\frac{2}{3}}$, and its antiparticle $\bar{a}_{\frac{2}{3}}$. If the energy gap of $a_{2/3}$ is larger than or equal to twice the energy gap of $a_{1/3}$, an effective theory that only activates the $f_1$ band is preferred. On the other hand, if the energy gap of $a_{2/3}$ is smaller than twice the energy gap of $a_{1/3}$, a pair of isolated $a_{1/3}$ anyons would always bind into an $a_{2/3}$ anyon and an effective theory that only activates the $f_2/f_3$ band becomes appropriate. Following this physical picture, we say that activating the $f_1$ band corresponds to ``doping in the $a_{1/3}$ anyon" while activating the $f_2/f_3$ band corresponds to ``doping in the $a_{2/3}$ anyon". 

We emphasize that ``doping in an anyon $a$" does not imply that all other anyon species are projected out. Instead, all finite-energy excitations that can be obtained from the elementary anyon $a$ through fusion are allowed in the low energy effective theory. If we are doping in the $a_{1/3}$ anyon, then the fusion products $a_{1/3}^n$ can be created by acting with $n$ powers of $f_1$ in a localized spatial region. These fusion products generate the entire anyon spectrum of the Jain state. If the $a_{2/3}$ anyon is doped into the system, then we have a ``pair-binding'' situation in which it is cheaper to create a single $a_{2/3}$ anyon than a pair of $a_{1/3}$ anyons with the same total charge. In this case, the $a_{1/3}$ anyon is projected out and the finite-energy anyon excitations are fusion products of $a_{2/3}$, which do not generate the entire anyon spectrum. This physical distinction will be reflected in the effective Lagrangians we write down in Section~\ref{sec:scenario1_SC} and Section~\ref{sec:scenario2_FL}.

\section{Doping in the \texorpdfstring{$a_{\frac{2}{3}}$}{} anyon near \texorpdfstring{$\nu = 2/3$}{}: chiral \texorpdfstring{$p-ip$}{} superconductor}\label{sec:scenario1_SC}
We begin by studying the possibility that the doped state has a finite density of the charge $2/3$ anyon that we denoted as $a_{\frac{2}{3}}$. We note that, strictly speaking, for electron doping, {\em i.e}, at fillings $\nu = \frac{2}{3} -\delta$ with $\delta < 0$, the dopant charges will be $a_{\frac{2}{3}}$ while for hole-doping ($\delta > 0$), the dopant charges will be the $\bar{a}_{\frac{2}{3}}$ anyon. Nevertheless we will refer to both as doping the $a_{\frac{2}{3}}$ anyon, and focus on the case of hole doping.

\subsection{Formation of the superconductor}\label{subsec:scenario1_SCformation}

Within the parton description, doping in the $a_{\frac{2}{3}}$ anyon corresponds to doping into the $f_3$ parton band,  while $f_1$ and $f_2$ continue to form $C_1 = -2, C_2 = 1$ Chern insulators at the new filling $\nu_1 = \nu_2 = 2/3 - \delta$. By the Streda formula, in order for the Chern insulators to persist, the mean-field flux seen by $f_1$ must be increased by $\delta/2$ while the mean-field flux seen by $f_2$ must be reduced by $\delta$ relative to the flux seen at $\nu_1 = \nu_2 = 2/3$. This condition translates to the mean-field flux assignment
\begin{equation}
    \frac{1}{2\pi} \ev{\nabla \times \bs{a}} = -\frac{1}{3} + \frac{\delta}{2} \,, \quad \frac{1}{2\pi} \ev{\nabla \times \bs{b}} = \frac{1}{3} - \frac{\delta}{2} \,. 
\end{equation}
Since the Chern insulators are gapped, we can integrate out the partons $f_1, f_2$ and obtain a simpler effective Lagrangian
\begin{equation}\label{eq:dopef3_prelude}
    L_{\rm eff} = -\frac{2}{4\pi} a da + \frac{1}{4\pi} (b - a) d (b - a) + L[f_3, A - b] - 2 \textrm{CS}_g  \,.
\end{equation}
After integrating out $a$, we obtain a level-2 Chern-Simons term for $b$, which can be interpreted as a $\nu = 2$ bosonic IQH state formed by $\Phi = f_1 f_2$. Note that integrating out $a$ also generates an additional gravitational Chern-Simons term $2 \mathrm{CS}_g$ which cancels the background Chern-Simons term in \eqref{eq:dopef3_prelude}. Therefore, the resulting effective Lagrangian takes the simpler form
\begin{equation}
    L_{\rm eff} = \frac{2}{4\pi} b db + L[f_3, A - b]  \,.
\end{equation}
Now let us think about the fate of $f_3$ with zero external magnetic field (i.e. $\nabla \times \bs{A} = 0$). After tripling the unit cell, the flux seen by $f_3$ is $2\pi (-1 + \frac{3}{2} \delta)$ while the lattice filling is $2 - 3 \delta$. By the Streda formula, part of the $f_3$ partons with total lattice filling $2 + \frac{3}{2} \delta$ (in the tripled unit cell) will fill the $C_3 = 1$ Chern band and continue to form a Chern insulator. The remaining $f_3$ partons have lattice filling $- \frac{9}{2} \delta$ and see an emergent magnetic field with strength $\frac{3}{2} \delta$. Since $\delta > 0$ is small, it is useful to perform a particle-hole transformation such that $\psi$ creates holes relative to the filled $C_3 = 1$ Chern band for $f_3$. After integrating out the filled band, we obtain the Lagrangian for $\psi$:
\begin{equation}
\label{eq:dopedf3}
    L_{\rm eff} = \frac{2}{4\pi} bdb + \frac{1}{4\pi} (b -A) d (b -A) + 2 \textrm{CS}_g + L[\psi, b - A] \,. 
\end{equation}
In the effective theory above, single electron excitations are gapped out. In the gapless sector,  the elementary gauge-invariant excitation that carries physical charge is the Cooper pair, which can be represented as $\psi^3 \mathcal{M}^{\dagger}_b$ where $\mathcal{M}^{\dagger}_b$ is a monopole operator that inserts $2\pi$-flux of $b$. From the Lagrangian, we see that $\psi^3$ carries charge $3$ under $A$ and charge $-3$ under $b$. On the other hand, the Chern-Simons term for $b$ implies that the bare monopole operator $M^\dagger_b$ carries charge $3$ under $b$ and charge $-1$ under $A$. Therefore the composite operator $\psi^3 \mathcal{M}^{\dagger}_b$ carries charge 2 under $A$ but no charge under $b$, allowing us to identify it with the gauge-invariant Cooper pair operator. Since the minimal-charge gauge-invariant excitation is a Cooper pair, it is natural for them to condense at finite density and form a superconductor. This expectation is confirmed by a careful analysis of the $\psi$ sector. Since $\nu_{\psi} = \frac{9}{2} \delta$ and the flux seen by $\psi$ is $-\frac{3}{2} \delta$, the $\psi$ fermions fill $-3$ Landau levels at the mean-field level and form the $\nu = -3$ integer quantum Hall state, with a gap proportional to $\delta$. Going beyond mean-field, we can integrate out the filled Landau levels and generate the low energy effective Lagrangian 
\begin{equation}\label{eq:SC_final_lagrangian}
    \begin{aligned}
    L_{\rm eff} &= \frac{2}{4\pi} bdb + \frac{1}{4\pi} (b -A) d (b -A) + 2 \textrm{CS}_g - \frac{3}{4\pi} (b-A) d (b-A) - 6 \textrm{CS}_g  \\
    &= \frac{2}{2\pi} b d A - \frac{2}{4\pi} A d A - 4 \textrm{CS}_g \,.
    \end{aligned}
\end{equation}
The integral over $b$ Higgses the $U(1)$ gauge group associated with the external gauge field $A$ down to $\mathbb{Z}_2$. Therefore, this state is a charge-2 superconductor. The presence of the gravitational Chern-Simons term signals a quantized thermal Hall conductivity $\kappa_{xy}$. Note that the thermal Hall response $\kappa_Q = \frac{1}{2} \frac{\pi^2}{3} k_B^2 T$ of a single $p+ip$ topological superconductor corresponds to a term $\textrm{CS}_g$. Thus the superconductor we get has a thermal Hall response $\kappa_{xy} = - 4\kappa_Q$. The sign of the thermal Hall conductivity is negative: the meaning of this  is that the sign is negative relative to the sign of the electrical Hall conductivity of the parent FQAH state at $2/3$ filling. Note that this FQAH state has $\kappa_{xy} = 0$ due to the presence of counter-propagating edge states. 

Thus we find a charge-2 chiral topological superconductor with a quantized thermal Hall conductivity in the doped state. The nonzero $\kappa_{xy}$ implies the presence of four species of neutral chiral edge Majorana fermions that counter-propagate (relative to the direction of the charged edge mode of the FQAH state). 

When a nonzero external magnetic field is turned on, we expect a vortex lattice state with vortex density proportional to the field strength. This is most easily understood by working with the bosonic vortices $\tilde \phi$ which are dual to the charge-2 Cooper pair $\phi$. Reintroducing gapped vortices into the effective Lagrangian, we have 
\begin{equation}
    L_{\rm eff} = L[\tilde \phi, b] + \frac{2}{2\pi} b d A - \frac{2}{4\pi} A d A - 4 \textrm{CS}_g \,,
\end{equation}
where $L[\tilde \phi, b]$ describes a gapped insulator with respect to $b$. In a vortex lattice state, $\rho_{\tilde \phi}(\bs{Q})$ condenses at a wavevector $\bs{Q}$ determined by the external magnetic field. From the perspectives of fermionic partons, this means that $\rho_{f_1}(\bs{Q}) = \rho_{f_2}(\bs{Q}) = \rho_{f_3}(\bs{Q})$ must condense at the same wavevector $\bs{Q}$. This equivalence between a vortex lattice state of the superconductor and a charge density wave insulator of the $f_i$ partons also makes sense if we directly think about the parton dynamics. In the presence of an external magnetic flux, the doped $f_i$ partons no longer fill complete Landau levels. The small density of partons that spill over feel a big effective magnetic field and prefer to crystallize in the presence of repulsive interactions. 

\subsection{Pairing symmetry of the superconductor}\label{subsec:scenario1_pairingsym}

The low energy effective field theory for the superconductor in \eqref{eq:SC_final_lagrangian} strongly constrains its pairing symmetry. In a general 2+1D system with internal degrees of freedom labeled by an index $\alpha$, a charge-2 superconducting order parameter can be represented by a matrix
\begin{equation}
    [\Delta_{\bs{k}}]_{\alpha, \beta} \equiv \ev{\psi^{\dagger}_{\alpha, \bs{k}} \psi^{\dagger}_{\beta, -\bs{k}}} \,, \quad \Delta_{\bs{k}}^T = - \Delta_{-\bs{k}} \,.
\end{equation}
Assuming the lattice point group symmetry is given by $G$ and that the normal state doesn't break this symmetry, the orbital part of the order parameter can be classified by irreps of $G$. With an eye towards applications to pentalayer graphene, we take the point group to be $G = C_3$ for which there are only three irreducible representations: (1) the trivial representation (minimal angular momentum $l = 0$), (2) the two dimensional representation $R_{2\pi i g/3}$ (minimal angular momentum $l = 1$), (3) the two dimensional representation $R_{2\pi i (2g)/3}$ (minimal angular momentum $l = 2$). We can refer to these cases as $s, p, d$-wave respectively. 

In current experimental FQAH systems~\cite{Xu2023_FQAHTMD,Lu2023_FQAHPenta},  the parent state of the superconductor is fully flavor polarized  (valley in TMD  and spin + valley in graphene-based systems) in the vicinity of $\nu = \frac{2}{3}$. Then  the flavor indices can be neglected and Fermi statistics demands the pairing wavefunction to be odd under parity. This limits us to $p$-wave pairing. To  match the TQFT and its associated gapless edge states,  the doped superconductor must be  smoothly connected to a stack of 4 chiral $p-ip$ topological superconductors. 

We note that a somewhat similar analysis appeared~\cite{Chen1989_anyonSC} in the old theory of anyon superconductivity (ASC), where the microscopic Hamiltonian describes a gas of anyons with fractional statistics $\theta = \pi (1- \frac{1}{n})$ where $n$ is an integer. In our case we have anyons with statistics $\frac{2\pi}{3}$ so that, at first sight, it might seem to be the $n = 3$ anyon superconductor. However, this identification is incorrect. As we explain in Appendix~\ref{app:anyonSC}, the model of Ref.~\cite{Chen1989_anyonSC} does not correspond to {\em any} microscopic electron system for odd $n$.   We also show that modifications of the model in Ref.~\cite{Chen1989_anyonSC} 
with other couplings to the background $A,g$ are allowed in electronic systems, and will have distinct physical properties. Our results can be incorporated into this modified model. If we regard our system just as anyons at finite density without reference to the underlying FQAH state, then we need to invoke a peculiar (but legal) coupling to the background $A,g$. We also highlight further differences with other old results on anyon superconductivity in Appendix~\ref{app:anyonSC}. Thus their results are not directly relevant to our system. Further, as we show in Appendix~\ref{app:generalization}, the TQFT for a charge-$|p|$ superconductor obtained from doping the $\nu = p/(2p+1)$ Jain state is completely different from the TQFT for a charge-$|p|$ ASC. For these reasons, we will refrain from using the terminology of ``anyon superconductivity" to avoid unnecessary confusion. 

\subsection{Finite temperature phenomenology}\label{subsec:scenario1_finiteT}

At $T = 0$ the Cooper pair field (which is represented in the low energy theory as a monopole operator that creates $2\pi$ flux of $b$) has long range order.  At a nonzero temperature, as usual, this will be replaced by power law order till a BKT transition at a temperature $T_{\rm BKT}$. The transition temperature $T_{\rm BKT}$ will be on the order of the phase stiffness $K$. Since the topological superconductor is formed by a small density $\sim \delta$ of doped anyons,  we expect $K \sim \delta$, and hence $T_{\rm BKT} \sim \delta$.   

\begin{figure}[!h]
    \centering
    \includegraphics[width=0.8\linewidth]{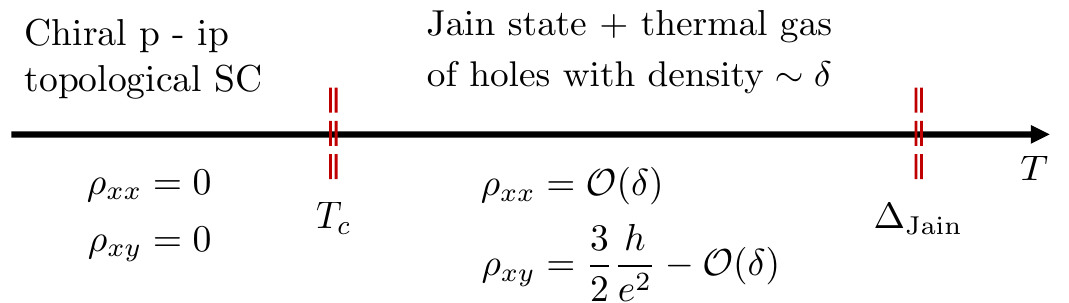}
    \caption{The finite temperature phase diagram at $\nu = 2/3 - \delta$ in scenario 1. For $T < \Delta_{\rm Jain}$, the system behaves like the Jain state at $\nu = 2/3$ coexisting with a thermal gas of holes with density $\delta$. The phase transition into the chiral SC state belongs to the KT universality class and occurs at $T_c \sim \delta$. A sharp signature of the transition is a rapid drop in the Hall resistivity of $\mathcal{O}(h/e^2)$.}
    \label{fig:finiteT_phase_diagram_scenario1}
\end{figure}

Besides $T_c$, there is an additional energy scale $\Delta_{\rm Jain}$, which corresponds to the many-body gap of the Jain state at $\nu = 2/3$. Since $\delta$ is small, there is a parametrically large temperature regime $T_c < T < \Delta_{\rm Jain}$ where the system behaves like a $\nu = 2/3$ Jain state with a thermal gas of doped anyons.  As $T$ increases past $\Delta_{\rm Jain}$, thermal fluctuations destroy quantization and the entire system crosses over to a thermal gas at filling $\nu = 2/3 - \delta$. The full finite temperature phase diagram is shown in Fig.~\ref{fig:finiteT_phase_diagram_scenario1}. 

The fragility of our topological superconductor at asymptotically small $\delta$ implies  that it may lose to other competing states in a realistic experimental setting. On the other hand, at sufficiently large $\delta$, the filling may approach other rational fractions where other incompressible quantum Hall phases can be stabilized. Therefore, the topological superconductor is expected to only emerge in some intermediate range of $\delta$. This is analogous to the situation in cuprate high temperature superconductors, where a host of ordered phases intervene between the Mott insulator at half-filling ($p = 0$), and the superconducting dome that onsets around hole-doping $p = 0.1$~\cite{Lee2004_dopedMott}.

\section{Doping in the \texorpdfstring{$a_{\frac{1}{3}}$}{} anyon near \texorpdfstring{$\nu = 2/3$}{}
}\label{sec:scenario2_FL}
Having understood the chiral topological superconductor, we now turn to the second possibility where the charge enters the system as $a_{\frac{1}{3}}$ anyons (or their antiparticles) with charge $\frac{1}{3}$.  In the parton description, this corresponds to the situation where the active band that is doped into is associated with the $f_1$ parton, while $f_2, f_3$ continue to form $C=1$ Chern insulators at the new filling $\nu_2 = \nu_3 = \frac{2}{3} - \delta$. By the Streda formula, the new mean-field flux seen by $f_2$ and $f_3$ must be reduced by $\delta$ relative to the flux at $\nu_2 = \nu_3 = 2/3$. This condition translates to the mean-field flux assignment
\begin{equation}
    \frac{1}{2\pi} \ev{\nabla \times \bs{a}} = -\frac{1}{3} + 2 \delta \,, \quad \frac{1}{2\pi} \ev{\nabla \times \bs{b}} = \frac{1}{3} + \delta \,. 
\end{equation}
Since $f_2, f_3$ are gapped, we can integrate them out to generate topological terms for $b-a$ and $A-b$. The resulting effective Lagrangian takes the form
\begin{equation}
    \begin{aligned}
    L_{\rm eff} &= L[f_1, a] + \frac{1}{4\pi} (b - a) d(b-a) + \frac{1}{4\pi} (A - b) d(A - b) + 4 \textrm{CS}_g \,,
    \end{aligned}
\end{equation}
where the gravitational Chern-Simons term arises from properly treating $a, A$ as $\textrm{spin}_{\mathbb{C}}$ connections. 

In the tripled unit cell where all anyons carry well-defined crystal momentum, $f_1$ is at filling $\nu_1 = 2 - 3\delta$ while the flux of the gauge field $a$ is $(- 1 + 6 \delta) 2\pi$. By the Streda formula, a fraction of the $f_1$ partons at filling $2 - 12 \delta$ continue to form a $C_1 = -2$ Chern insulator. This filled band can be integrated out and we get a $\nu = -2$ IQH effect for the emergent gauge field $a$. What about the remaining fraction of $f_1$ partons at filling $\nu_1 - (2-12 \delta) = 9 \delta$? Due to the projective action of lattice translation, the $f_1$ band structure at $\delta = 0$ enjoys a three-fold degeneracy in the Brillouin zone. Therefore, the $9 \delta$ partons split into three mini-pockets with creation operators $d^{\dagger}_{i = 1, 2, 3}$, each at half-filling relative to the emergent $2\pi (6 \delta)$ flux that it sees. In terms of the doped particles, the Lagrangian takes the form
\begin{equation}\label{eq:sCFL_start}
    L_{\rm eff} = \sum_{i=1}^3 L[d_i, a] - \frac{2}{4\pi} a da + \frac{1}{4\pi} (b-a) d (b-a) + \frac{1}{4\pi} (A-b) d (A-b) \,. 
\end{equation}

At intermediate temperatures, the natural state formed by each $d_i$ pocket at half-filling relative to an effective magnetic field is the famous composite Fermi liquid (CFL). Since $d_i$ itself originates from the composite fermion $f_1$, we will refer to the resulting state as a secondary CFL. We construct the secondary CFL explicitly in Sections~\ref{subsec:scenario2_CFLformation} through~\ref{subsec:transport_2ndCFL} and show that it is a non-Fermi liquid metal with dynamical critical exponent $z = 3$. In Section~\ref{subsec:pairing_2ndCFL}, we go to asymptotically low temperatures and study the pairing instabilities of the secondary CFL within a renormalization group framework. Unlike for the usual CFL, gauge fluctuations in the secondary CFL always enhance the pairing susceptibility between composite fermions in different pockets. This interpocket pairing Higgses all the emergent gauge fields and results in a CDW metal with no charge fractionalization. On the other hand, if the microscopic couplings are arranged so that the effective BCS attraction within the same pocket far exceeds the gauge-field-mediated attraction between different pockets, pairing occurs for each pocket separately and each $d_i$ forms a Moore-Read state~\cite{Moore1991_nonabelian,Read1999_pair}. The resulting phase describes a superconductor with broken lattice translation and non-abelian topological order, henceforth referred to as the PDW SC*. The finite temperature phenomenology for both types of pairing instabilities is discussed in Section~\ref{subsec:scenario2_FL_finiteT}.


\subsection{Formation of the secondary CFL}\label{subsec:scenario2_CFLformation}


Following the standard strategy for describing a zero-field CFL state, we start with \eqref{eq:sCFL_start}, fractionalize $d_i$ as $d_i = z_i \psi_i$, and introduce three emergent gauge fields $\alpha_{i=1,2,3}$ such that the bosonic parton $z_i$ couples to $a - \alpha_i$ and the fermionic parton $\psi_i$ couples to $\alpha_i$. The effective Lagrangian for each $d_i$ then takes the form
\begin{equation}
    L[d_i, a] = L[\psi_i, \alpha_i] + L[z_i, a - \alpha_i] \,.
\end{equation}
We know from the previous section that $\nabla \times \bs{a}/(2\pi) = 6 \delta$, while the density of $z_i, \psi_i$ is locked to $3\delta$. To obtain the CFL state for $d_i$, we consider a mean-field in which $\nabla \times \bs{\alpha_i} = 0$, such that $\psi_i$ sees zero flux on average and $z_i$ is at filling $1/2$. At this filling, it is natural to put $z_i$ in the $U(1)_{2}$ bosonic Laughlin state and $\psi_i$ in the Fermi liquid state. This choice gives the CFL Lagrangian:
\begin{equation}
    \begin{aligned} 
    L_{\rm CFL}[d_i, a]
    = L_{\rm FL}[\psi_i, \alpha_i] + \frac{1}{2\pi} (a - \alpha_i) d \beta_i - \frac{2}{4\pi} \beta_i d \beta_i \,.
    \end{aligned}
\end{equation}
At zero external magnetic field (i.e. $\nabla \times \bs{A} = 0$), the physical charge density is $\rho = -\frac{1}{2\pi} \nabla \times \bs{b}$. Therefore, the microscopic Coulomb interactions enter the Lagrangian as a kinetic term for $b$
\begin{equation}
    L_{\rm Coulomb}[b] = \frac{1}{2} \frac{1}{(2\pi)^2} \int_{t, \bs{r}, \bs{r}'} \nabla \times \bs{b}(\bs{r}, t) \, V(\bs{r} - \bs{r}') \, \nabla \times \bs{b}(\bs{r}', t) \,. 
\end{equation}
After including the Coulomb interaction, the full Lagrangian takes the form
\begin{equation}\label{eq:2ndCFL_Lag}
    \begin{aligned}
    L_{\rm eff} = L_{\rm Coulomb}[b] + \sum_{i=1}^3 \, L_{\rm CFL}[d_i, a] - \frac{2}{4\pi} ada + \frac{1}{4\pi} (b-a) d (b-a) + \frac{1}{4\pi} (A-b) d (A-b) \,.
    \end{aligned}
\end{equation}
As the secondary CFs labeled by $\psi_i$ see zero magnetic field on average, the mean-field state of $\psi_i$ is simply a Fermi liquid. Going beyond mean field, we expect the fluctuations of $\alpha_i, \beta_i, a, b$ to destroy the fermionic quasiparticles associated with $\psi_i$ and create a secondary CFL with non-Fermi liquid features. These dynamical features will be derived in the next section.

\subsection{Fluctuation effects in the secondary CFL}

For the purpose of understanding Fermi surface dynamics, we can work on a flat spacetime and integrate out $\beta_i$ and $b$ in favor of $\alpha_i, a, A$ without worrying about flux quantization. The equations of motion for $b$ and $\beta_i$ impose
\begin{equation}
    \frac{1}{2\pi} d (b-a) + \frac{1}{2\pi} d (b-A) = 0 \quad \rightarrow \quad b = \frac{a + A}{2} \,, \quad 2 d \beta_i = d a - d \alpha_i \,. 
\end{equation}
Working at $\nabla \times \bs{A} = 0$, we can plug the solutions of these equations back into the Lagrangian and obtain
\begin{equation}
    \begin{aligned}
    L =& \sum_{i=1}^3 \left\{L[\psi_i,\alpha_i] + \frac{1}{8\pi} (\alpha_i - a) d (\alpha_i - a)\right\} - \frac{2}{4\pi} a da + \frac{1}{8\pi} (A - a) d (A - a) + L_{\rm Coulomb}[a] \,. 
    \end{aligned}
\end{equation}
Inspecting the Lagrangian above, we see that the self Chern-Simons terms for $a$ cancel out, leaving us with 
\begin{equation}
    \begin{aligned}
    L = \sum_{i=1}^3 &\left\{L[\psi_i, \alpha_i] + \frac{1}{8\pi} \alpha_i d \alpha_i \right\} \,+\, L_{\rm Coulomb}[a] - \frac{1}{4\pi} a \left(dA + \sum_{i=1}^3 d \alpha_i\right) + \frac{1}{8\pi} A dA \,.
    \end{aligned}
\end{equation}
We are now in a position to study the effects of Coulomb interactions. It is most convenient to work in the transverse gauge and decompose the gauge field $a$ into $a_0$ and the transverse component $a_T$. Since the action for $a$ is quadratic, we can integrate it out to generate an effective action for $\alpha$:
\begin{equation}
    \begin{aligned}
    &L_{\rm Coulomb}[a] \, -\,  \frac{1}{4\pi} a \,\sum_{i=1}^3 d \alpha_i \rightarrow \, - \frac{1}{4\pi} a_0 \sum_i (-iq) \alpha_{i, T} + \frac{1}{2} \int \sum_{i,j} \alpha_{i,0} \frac{|\bs{q}|}{g^2} \alpha_{j,0} \,,
    \end{aligned}
\end{equation}
where $g$ is some real constant determined by parameters in the Coulomb interaction. Combining the above result with the rest of the Lagrangian, we get
\begin{equation}
    \begin{aligned}
    L &= \sum_{i=1}^3 \left\{L[\psi_i, \alpha_i] + \frac{1}{8\pi} \alpha_i d \alpha_i\right\} + \frac{1}{8\pi} A dA - \frac{1}{4\pi} a_0 \nabla \times \left(\sum_{i=1}^3 \bs{\alpha_i} + \bs{A}\right) + \frac{1}{2} \sum_{i,j} \int \alpha_{i,0} \frac{|\bs{q}|}{g^2} \alpha_{j,0} \,.
    \end{aligned}
\end{equation}
The above Lagrangian does not contain a Maxwell term for $\alpha_i$. However, as in any theory where emergent gauge fields are coupled to Fermi surfaces, quantum fluctuations of $\psi_i$ generate an effective kinetic term for $\alpha_i$ (see~\cite{Lohneysen2006_HertzMillis_review} and references therein)
\begin{equation}
    \delta L = \sum_{i=1}^3 \alpha_{i,T}^* \left(\chi q^2 + \gamma \frac{|\omega|}{q}\right) \alpha_{i,T} + \alpha_{i,0}^* \, m^2 \, \alpha_{i,0} \,.
\end{equation}
This effective kinetic term has several important consequences:
\begin{enumerate}
    \item The dynamically generated kinetic term for $\alpha_{i,0}$ dominates over the term arising from the long-range Coulomb interactions. Physically, this is because interactions in the density channel are screened by the Fermi surface. 
    \item The dynamically generated kinetic term for $\alpha_{i,T}$ has $z = 3$ scaling. Therefore, the dimensionless gauge couplings associated with $\alpha_{i,T}$ flow to finite fixed-point values that can be computed within a controlled expansion in powers of $\epsilon = z - 2$~\cite{Mross2010_NFL,Metlitski2014_PairingNFL}.
    \item The equations of motion for ${a_0}$ enforces a constraint ${\sum_{i=1}^3 \alpha_{i,T} = - A_T}$. Therefore, the three gauge fields $\alpha_{i,T}$ are not independent. 
\end{enumerate}
From these observations, we conclude that the combination of electronic interactions and gauge fluctuations creates a non-Fermi liquid state in which three species of composite fermions are strongly coupled to three emergent gauge fields with dynamical critical exponent $z = 3$. An important subtlety is that the \textit{three emergent gauge fields are not independent in the low energy theory}. This subtlety has dramatic effects on transport phenomena and instabilities of the secondary CFL state which we clarify in Section~\ref{subsec:transport_2ndCFL} and Section~\ref{subsec:pairing_2ndCFL}.

\subsection{Transport in the secondary CFL}\label{subsec:transport_2ndCFL}

To understand transport, it is convenient to reintroduce the gauge fields $\beta_i$ which describe the $U(1)_{2}$ Laughlin sectors in \eqref{eq:2ndCFL_Lag}. The effective Lagrangian that includes the most IR-singular gauge fluctuations takes the form
\begin{equation}\label{eq:2ndCFL_Lag_simpler}
    \begin{aligned}
    L&_{\rm eff} = \sum_{i=1}^3 L[\psi_i, \alpha_i] - \frac{1}{2\pi} \alpha_i d \beta_i - \frac{2}{4\pi} \beta_i d \beta_i + \frac{1}{8\pi} A dA \\
    &+\alpha_{i,T}^* \left(\chi q^2 + \gamma \frac{|\omega|}{q}\right) \alpha_{i,T} - \frac{1}{4\pi} a_0 \nabla \times \left(\bs{A} + \sum_{i=1}^3 \bs{\alpha}_i\right)  \,. 
    \end{aligned}
\end{equation}
We define parton response functions $\Pi_{\psi_i}, \Pi_{\beta_i}$ via
\begin{equation}
    \bs{J}_{\psi_i} = \Pi_{\psi_i} \bs{\alpha}_i \,, \quad \bs{J}_{\beta_i} = \Pi_{\beta_i} \bs{\alpha}_i \,, 
\end{equation}
where $\bs{J}_{\beta_i}$ is the spatial-part of the 1-form current $J_{\beta_i} = - \frac{1}{2\pi} \star d \beta_i$. The integration over $a_0$ provides a constraint $\sum_i \bs{\alpha}_i = - \bs{A}$. Using this constraint, we can integrate out $\bs{\alpha}_3$ in favor of $\bs{\alpha_1}, \bs{\alpha_2}$ in the Lagrangian. This means that the physical charge current can be identified with 
\begin{equation}
    J_{\rm phy} = - J_{\psi_3} - J_{\beta_3} + \Pi_{U(1)_2} \bs{A} \,,
\end{equation}
where $\Pi_{U(1)_2}$ is the response function of the $U(1)_2$ state that generates the background Chern-Simons term $\frac{1}{8\pi} A dA$. 
To fix the physical conductivity, we observe that the original effective Lagrangian enjoys an $S_3$ symmetry that permutes the three internal indices. Using the $S_3$ symmetry, we can set $\Pi_{\psi_i} = \Pi_{\psi}, \Pi_{\beta_i} = \Pi_{\beta}$ for all $i$. Moreover, at the level of expectation values, the $S_3$ symmetry also implies that $\bs{\alpha}_i = \bs{\alpha}$ for all $i$ and $3\bs{\alpha} = -\bs{A}$. Plugging this result into the physical current, we immediately obtain the total physical conductivity
\begin{equation}
    \bs{J} = -\bs{J}_{\psi_3} - \bs{J}_{\beta_3} + \Pi_{U(1)_{2}} \bs{A} = \frac{\Pi_{\psi} + \Pi_{\beta}}{3} \bs{A} + \Pi_{U(1)_{2}} \bs{A} \,.
\end{equation}
Since $\beta$ is also in the $U(1)_{2}$ state, the conductivity of the secondary CFL simplifies to
\begin{equation}
    \sigma_{\rm sCFL} \approx \frac{2}{3} \frac{e^2}{h} \begin{pmatrix}
        0 & 1 \\ -1 & 0
    \end{pmatrix} + \frac{1}{3} \sigma_{\psi} \,. 
\end{equation}
This formula has a simple interpretation: the first term is the quantized Hall conductivity associated with the background $\nu = 2/3$ Jain state; the second term arises from the non-Fermi liquid sector and captures the response of the $\psi$ fermions to the \textit{internal gauge fields} they see.

The response function $\sigma_{\psi}$ is generally difficult to compute in a strongly interacting metal. However, recent theoretical developments have shown that in a large class of non-Fermi liquids (including Fermi surfaces coupled to fluctuating gauge fields), $\sigma_{\psi}$ does not receive singular corrections from infrared gauge fluctuations and essentially retains the Fermi liquid form~\cite{Shi2022_gifts,Guo2022_clean_NFLtransport,Shi2024_transport_exp}. In the absence of disorder, an immediate consequence of this result is that $\sigma_{\mathrm{sCFL}, xx} = \sigma_{\psi, xx} = \infty$, in contrast to the vanishing longitudinal conductivity in the conventional CFL state. In the presence of weak disorder, the diagonal conductivity becomes a finite function of $\delta$
\begin{equation}
    \sigma_{\psi, xx} \approx \sigma_{\psi, yy} \approx \Sigma_0 \frac{e^2 \delta}{2\pi \hbar} \,, \quad \Sigma_0 = \frac{\hbar \tau_{\rm tr} \rho_0}{m_{\rm tr}} \,, 
\end{equation}
where $\tau_{\rm tr}$ is the transport scattering time, $m_{\rm tr}$ is the transport effective mass of the active $f_1$ band, and $\rho_0$ is the microscopic electron density at unit lattice filling. The constant $\Sigma_0^{-1}$ is a dimensionless measure of the disorder strength, which vanishes as the scattering time diverges. 

Unlike $\sigma_{\psi, xx}$, the Hall component $\sigma_{\psi, xy}$ is independent of $\Sigma_0$ and instead sensitive to the band topology. The leading contribution to $\sigma_{\psi, xy}$ is given by Haldane's formula 
\begin{equation}
    \sigma_{\psi, xy} = \sum_{\alpha} \frac{\nu_{\alpha} e^2}{h} \,, \quad \nu_{\alpha} = \frac{1}{2\pi} \int_{\rm BZ} d^2 \bs{k} \, F_{\alpha}(\bs{k}) \cdot N_n(\bs{k}, \mu, T) \,,
\end{equation}
where $F_{\alpha}$ is the Berry curvature in the $\alpha$-th band and $N_n$ is the Fermi-Dirac distribution function~\cite{Haldane2004_FSBerry}. When $\delta$ is small, the $\psi$ fermions only occupy a single band and the summation over band indices can be dropped. If we crudely approximate the Berry curvature as a constant $\bar F = \mathcal{O}(1)$ near the bottom of the band, the integral over $\bs{k}$ simply gives the total density, which is fixed to be $ \sim \delta$ by Luttinger's theorem. 

The qualitative behavior of the physical resistivity tensor depends on the relative sizes of the two dimensionless parameters $\delta$ and $\Sigma_0$. At very low doping, $\sigma_{\psi, xx}$ and $\sigma_{\psi, xy}$ are both proportional to $\delta$ and it is useful to expand the physical resistivity matrix to leading order in $\delta$
\begin{equation}
    \rho \approx \frac{h}{e^2} \, \begin{pmatrix}
        \frac{3 h}{4 e^2} \sigma_{\psi, yy} & - \frac{3}{2} + \frac{3 h}{4 e^2} \sigma_{\psi, xy} \\  \frac{3}{2} - \frac{3 h}{4 e^2} \sigma_{\psi, xy} &\frac{3 h}{4 e^2} \sigma_{\psi, xx}
    \end{pmatrix} + \mathcal{O}(\delta^2) \,, \quad \delta \ll \Sigma_0^{-1} \ll 1 \,. 
\end{equation}
At moderate doping and relatively weak disorder, it is possible to enter a ``double-scaling" regime where $\Sigma_0$ is large, $\delta$ is small, while the product $\Sigma_0\, \delta$ is $\mathcal{O}(1)$. Within such a regime, the Hall component $\sigma_{\psi,xy}$ is suppressed relative to the diagonal components $\sigma_{\psi, xx}, \sigma_{\psi, yy}$ and the physical resistivity matrix can be approximated as
\begin{equation}
    \rho \approx \frac{3h}{e^2 \left(4 + \frac{h^4}{e^4} \sigma_{xx} \sigma_{yy}\right)} \, \begin{pmatrix}
        \frac{h}{e^2} \sigma_{yy} & - 2 \\ 2 & \frac{h}{e^2} \sigma_{xx}
    \end{pmatrix} \,, \quad \delta \sim \left(\frac{\hbar \tau_{\rm tr} \rho_0}{m_{\rm tr}}\right)^{-1} \ll 1 \,. 
\end{equation}
A notable feature of this result is that the magnitude of $\rho_{xy}$ always decreases as we hole-dope away from the Jain state at $\nu = 2/3$. This trend is a peculiar feature of the double-scaling regime. 

\subsection{Pairing instability of the secondary CFL}\label{subsec:pairing_2ndCFL}

In order for the transport phenomena in Section~\ref{subsec:transport_2ndCFL} to be observed, the secondary CFL state must survive down to asymptotically low temperatures. At first sight, this seems reasonable, as time-reversal and inversion symmetries are spontaneously broken in the doped FQAH regime and there is no BCS instability. However, at low dopant density, the composite fermions live near the band minimum and their dispersion relations enjoy an approximate rotational symmetry. This means that the superconducting susceptibilities can be quite strong even without a symmetry-protected matching between $\epsilon(\bs{k})$ and $\epsilon(-\bs{k})$. In what follows, we will consider the idealized scenario in which the low energy fermion dispersion is exactly rotationally invariant. Deviations from this idealization will not be visible except at the lowest temperature scales.

In the standard CFL with rotationally invariant fermion dispersion, it is well-known that gauge fluctuations suppress the na\"{i}ve BCS instability between composite fermions~\cite{Metlitski2014_PairingNFL}. However, an important feature of the secondary CFL Lagrangian in \eqref{eq:2ndCFL_Lag} is that the three gauge fields $\alpha_i$ are \textit{not independent}. Therefore, we must perform a more careful renormalization group analysis on the BCS interactions $V_{\rm BCS, ij}$ between secondary composite fermions $\psi_i, \psi_j$ to determine the correct ground state. Surprisingly, we find that gauge fluctuations tend to suppress $V_{\rm BCS, ii}$ but enhance $V_{\rm BCS, i \neq j}$. Therefore, the dominant instability is an ``interpocket" pairing between composite fermions $\psi_i, \psi_j$ with $i \neq j$, while ``intrapocket" pairing can only be realized if the microscopic parameters are arranged so that the UV couplings satisfy $V_{\rm BCS,ii} \gg V_{\rm BCS, i \neq j}$. 

The ground states induced by these two pairing instabilities are drastically different. While interpocket pairing leads to a conventional CDW metal with no fractionalization, intrapocket pairing gives rise to a PDW superconductor with non-abelian topological order. In what follows, we will first present the RG analysis and then demonstrate how these two distinct phases arise from the pairing between secondary composite fermions. 

\subsubsection{RG analysis: identifying the dominant instability}

We start by recalling the simplified Lagrangian \eqref{eq:2ndCFL_Lag_simpler} which is valid in the transverse gauge
\begin{equation}
    \begin{aligned}
    L&_{\rm eff} = \sum_{i=1}^3 L[\psi_i, \alpha_i] - \frac{1}{2\pi} \alpha_i d \beta_i - \frac{2}{4\pi} \beta_i d \beta_i + \frac{1}{8\pi} A dA \\
    &+\alpha_{i,T}^* \left(\chi q^2 + \gamma \frac{|\omega|}{q}\right) \alpha_{i,T} - \frac{1}{4\pi} a_0 \nabla \times \left(\sum_{i=1}^3 \bs{\alpha}_i + \bs{A}\right)  \,. 
    \end{aligned}
\end{equation}
For the RG analysis, it is useful to turn off the external gauge field $A$ and integrate over $a_0$ to enforce the constraint $\alpha_{1,T} + \alpha_{2,T} + \alpha_{3,T} = 0$. Without loss of generality, let us use this constraint to replace $\alpha_{3,T}$ with $-\alpha_{1,T} - \alpha_{2,T}$ and then perform a rotation of the gauge field $\alpha_{\pm,T} = \frac{1}{2} (\alpha_{1, T} \pm \alpha_{2,T})$. To analyze the pairing instability in a controlled fashion, we adopt the large $N$, small $\epsilon$ expansion introduced in~\cite{Mross2010_NFL}, where each $\psi_i$ gets upgraded to $N$ flavors of fermions $\psi_{i,n}$ with $n =1, \ldots, N$ and the gauge field dispersion $q^2$ is replaced by $q^{1+\epsilon}$. A controlled expansion is obtained by taking $N \rightarrow \infty, \epsilon \rightarrow 0$ with the product $\epsilon N$ held fixed. The deformed Lagrangian takes the form
\begin{equation}\label{eq:secondaryCFL_Lag_Mrossexp}
    \begin{aligned}
    L &= L[\psi_{1,n}, \alpha_{1,0}, \alpha_{+,T} + \alpha_{-,T}] - \frac{iq}{8\pi} \alpha_{1,0} (\alpha_{+,T} + \alpha_{-,T}) \\
    &+ L[\psi_{2,n}, \alpha_{2,0}, \alpha_{+,T} - \alpha_{-,T}] -\frac{iq}{8\pi} \alpha_{2,0} (\alpha_{+,T} - \alpha_{-,T}) \\
    &+ L[\psi_{3,n}, \alpha_{3,0}, - 2\alpha_{+,T}]  + \frac{iq}{4\pi} \alpha_{3,0} \alpha_{+,T} \\
    &+ 6 N \alpha_{+,T}^* \left(\chi_+ q^{1+\epsilon} + \gamma_+ \frac{|\omega|}{q}\right) \alpha_{+, T} \\
    &+ 2 N \alpha_{-,T}^* \left(\chi_- q^{1+\epsilon} + \gamma_- \frac{|\omega|}{q}\right) \alpha_{-, T} \,.
    \end{aligned}
\end{equation}
Following conventions in the literature, we define the 
dimensionless gauge couplings
\begin{equation}
    g_{\pm} = \frac{\Lambda^{-\epsilon} v_F}{2 \chi_{\pm} (2\pi)^2} \,,
\end{equation}
where $\Lambda$ is a UV momentum cutoff. To analyze the pairing instability, we also add explicit BCS interactions involving Cooper pairs formed by electrons from $i$ and $j$ pockets
\begin{equation}
    S_{\rm BCS, ij} = - \frac{1}{4} \int \prod_{a=1}^4 \frac{d^3 k_a}{(2\pi)^3} \psi_i^{\dagger}(k_1) \psi_j^{\dagger}(k_2) \psi_i(k_3) \psi_j(k_4) (2\pi)^3 \delta(k_1 + k_2 - k_3 - k_4) V_{\rm BCS, ij}(\bs{k}_1, \bs{k}_2; \bs{k}_3, \bs{k}_4) \,.
\end{equation}
In this expression, $k_i$ are spacetime momenta while $\bs{k}_i$ are the spatial components. Since pairing occurs near the Fermi surface, we can take $\bs{k}_i$ to lie on the Fermi surface and restrict the spatial momenta to satisfy the BCS condition $\bs{k}_1 = -\bs{k}_2, \bs{k}_3 = - \bs{k}_4$. Therefore, the relevant BCS couplings are $V_{\rm BCS,ij}(\theta_1, \theta_2) \equiv V_{\rm BCS,ij}(\bs{k}_1, - \bs{k}_1; \bs{k}_2, - \bs{k}_2)$ where $\theta_1, \theta_2$ are the angles associated with $\bs{k}_1, \bs{k}_2$. To simplify the RG analysis, we will further assume rotational invariance of the fermion dispersion so that the coupling is only a function of $\theta_1 - \theta_2$. This allows us to resolve the BCS interactions into different angular momentum channels
\begin{equation}
    V_{\rm BCS, ij}(\theta_1, \theta_2) = \sum_{m=-\infty}^{\infty} V_{\rm BCS, ij}^{(m)} e^{im(\theta_1 - \theta_2)} \,. 
\end{equation}
From here, one can construct the RG flow of $V_{\rm BCS, ij}^{(m)}$ and $g_{\pm}$. The basic idea is to divide the Fermi surface into a large number of patches, each with width $\Lambda_{\rm patch}$. In a single RG step, we split each patch into two identical patches with width $\Lambda_{\rm patch}/2$ and integrate out gauge fluctuations that connect nearby patches. At leading order in large $N$, the effect of this integration is to renormalize both $g_{\pm}$ within each smaller patch and induce BCS interactions between nearby patches. Since the normalizations of gauge and BCS interactions in our problem are identical to those appearing in~\cite{Metlitski2014_PairingNFL}, we can simply follow the RG procedure in~\cite{Metlitski2014_PairingNFL} and read off the $\beta$-functions for $V_{\rm BCS, ij}^{(m)}$:
\begin{equation}
    \begin{aligned}
    \frac{d V^{(m)}_{\rm BCS, 11}}{d l} &= - \left(V^{(m)}_{\rm BCS, 11}\right)^2 + g_+ + g_- \,,\\
    \frac{d V^{(m)}_{\rm BCS, 22}}{d l} &= - \left(V^{(m)}_{\rm BCS, 22}\right)^2 + g_+ + g_-\,,\\
    \frac{d V^{(m)}_{\rm BCS, 33}}{d l} &= - \left(V^{(m)}_{\rm BCS, 33}\right)^2 + 4g_+ \,,\\
    \frac{d V^{(m)}_{\rm BCS, 12}}{d l} &= - \left(V^{(m)}_{\rm BCS, 12}\right)^2 + g_+ - g_-\,, \\
    \frac{d V^{(m)}_{\rm BCS, 23}}{d l} &= - \left(V^{(m)}_{\rm BCS, 23}\right)^2 - 2 g_+ \,,\\
    \frac{d V^{(m)}_{\rm BCS, 13}}{d l} &= - \left(V^{(m)}_{\rm BCS, 13}\right)^2 - 2 g_+ \,.
    \end{aligned}
\end{equation}
These RG equations admit a simple physical interpretation. The universal negative quadratic term represents the usual BCS instability in a Fermi liquid. The additional term due to $g_{\pm}$ mediates a BCS repulsion/attraction when the sign is positive/negative. When $i = j$, the gauge contribution is always repulsive. This repulsion can be understood using Amp\`{e}re's law: $U(1)$ gauge fields mediate an attraction/repulsion between parallel currents running in the same/opposite directions. For a single pocket $i$, the gauge field mediates a current-current interaction between fermions near momentum $\bs{k}$ and $-\bs{k}$. Since these currents are anti-parallel and carry the same charge, they must repel. On the other hand, if $i \neq j$, different fermion pockets could carry different gauge charges under the gauge fields $\alpha_{\pm}$. For example, for $i=1$ and $j=2$, we see from \eqref{eq:secondaryCFL_Lag_Mrossexp} that $\psi_1, \psi_2$ carry the same charge under $\alpha_+$ but opposite charges under $\alpha_-$. This means that $\alpha_+$ induces virtual currents of $\psi_1, \psi_2$ running in opposite directions, while $\alpha_-$ induces virtual currents of $\psi_1, \psi_2$ running in the same direction. This basic intuition explains the sign structure of various terms on the RHS of the RG equations.

Since the physical model has $\epsilon = 1$, both $g_+$ and $g_-$ must flow to $\mathcal{O}(1)$ positive fixed point values. Moreover, the manifest $S_3$ symmetry in the Lagrangian \eqref{eq:2ndCFL_Lag} implies that the renormalization of the three interpocket BCS channels must be identical (the same goes for the three intrapocket BCS channels), implying a nontrivial relation between the fixed point values $g_{\pm}^*$
\begin{equation}
    \{g^*_+ + g^*_- = 4 g^*_+ \,, \quad g^*_+ - g^*_- = - 2 g^*_+\} \quad \rightarrow \quad g^*_+ = \frac{1}{3} g^*_- \,.
\end{equation}
Plugging this relation back into the RG equations, we find that the intrapocket BCS couplings $V^{(m)}_{\rm BCS, ii}$ all flow to finite fixed point values, while the interpocket BCS couplings $V^{(m)}_{\rm BCS, i \neq j}$ have runaway flows that induce a pairing instability. Therefore, with generic microscopic couplings, interpocket pairing is the most likely low-temperature phase, while intrapocket pairing can only be induced when the bare $V^{(m)}_{\rm BCS, ii}$ is much stronger than $V^{(m)}_{\rm BCS, i\neq j}$. This situation bears some resemblance to the half-filled Landau level, where paired states of the standard composite fermions can be realized with an appropriate choice of pseudopotentials, despite the fact that gauge fluctuations in the standard CFL state suppress all pairing channels~\cite{Metlitski2014_PairingNFL}. Motivated by this analogy, we will consider both types of paired states in the following sections and analyze their dynamical properties. 

\subsubsection{Interpocket pairing: CDW metal + level-1 IQH}\label{subsubsec:pairing_inter}

We first determine the low-temperature phase obtained from the interpocket pairing of secondary composite fermions. Without loss of generality, let us pair $\psi_2$ and $\psi_3$, leaving $\psi_1$ untouched (the analysis is identical for pairing between $\psi_1, \psi_3$ or $\psi_1, \psi_2$). From the existing literature on $1/2 + 1/2$ quantum Hall bilayers, we know that two CFLs described by fermionic partons $\psi_2, \psi_3$ favor the $p + i p$ pairing channel in the limit of strong interlayer repulsion~\cite{Sodemann2017_CFLbilayer}. Here, we make the simplifying assumption that the same interlayer pairing channel persists away from the strong interaction limit. Under this assumption, the pairing of $\psi_2, \psi_3$ generates an exciton condensate for $d_2, d_3$, which produces a level-1 Chern-Simons term for the common gauge field $a$ to which they couple (see Appendix~\ref{app:CFpairing_review} for a derivation). After adding the level-1 Chern-Simons term, the original Lagrangian in terms of $d_i$ simplifies to
\begin{equation}\label{eq:CFpairing_2/3_intermediate}
    \begin{aligned}
    L &= L_{\rm CFL}[d_1, a] + \frac{1}{4\pi} a da + 2 \mathrm{CS}_g - \frac{2}{4\pi} ada + \frac{1}{4\pi} (b-a) d (b-a) + \frac{1}{4\pi} (A - b) d (A-b) \\
    &= L_{\rm CFL}[d_1, a] - \frac{1}{2\pi} a db + \frac{2}{4\pi} b db - \frac{1}{2\pi} b d A + \frac{1}{4\pi} A dA + 2 \mathrm{CS}_g  \,.
    \end{aligned}
\end{equation}
We now make a parton decomposition $d_1 = z_1 \psi_1$ and put $z_1$ in a $U(1)_{2}$ state, which can be described by introducing additional gauge fields $\alpha_1, \beta_1$
\begin{equation}
    \begin{aligned}
    L = L[\psi_1, \alpha_1] + \frac{1}{2\pi} (\alpha_1 - a) d \beta_1 - \frac{2}{4\pi} \beta_1 d \beta_1 + \frac{2}{4\pi} b db - \frac{1}{2\pi} (a + A) db + \frac{1}{4\pi} A dA + 2 \mathrm{CS}_g \,. 
    \end{aligned}
\end{equation}
Remarkably, the gauge field $a$ now acts as a Lagrange multiplier. Integrating over $a$ Higgses $b + \beta_1$ and further integrating over $b$ allows us to replace every instance of $b$ with $-\beta_1$: 
\begin{equation}
    L = L[\psi_1, \alpha_1] + \frac{1}{2\pi} (\alpha_1 + A) d \beta_1 + \frac{1}{4\pi} A d A + 2 \mathrm{CS}_g\,. 
\end{equation}
Finally, the integral over $\beta_1$ Higgses $\alpha_1 + A$. As a result, the final Lagrangian takes the form
\begin{equation}\label{eq:2ndCFL_finalFL}
    L = L[\psi_1, -A] + \frac{1}{4\pi} A d A + 2 \mathrm{CS}_g \,.
\end{equation}
Since lattice translation acts projectively on the three partons $d_1, d_2, d_3$, an exciton condensate between any two of them spontaneously breaks the lattice translation symmetry. Therefore, the final state described by \eqref{eq:2ndCFL_finalFL} is a CDW metal stacked on top of a decoupled $\nu = 1$ IQH state.

With all fluctuating gauge fields Higgsed, the total conductivity takes the simple form $\sigma = \sigma_{\psi} + \sigma_{1}$ where $\sigma_1$ is the conductivity of a level-1 IQH state. The conductivity of $\psi$ is proportional to the filling $\delta$. At small doping, the total Hall conductivity is therefore close to $\sigma_{xy} = 1$, deviating strongly from the proximate $\nu = 2/3$ Jain state. However, there is no jump in the conductivity as a function of doping because the transition temperature from the secondary CFL to this Fermi liquid state also vanishes as $\delta \rightarrow 0$, as we will see in Section~\ref{subsec:scenario2_FL_finiteT}.

The CDW metal we have obtained has a simple interpretation. At filling $\nu = \frac{2}{3}$, a natural competing state to the FQAH is a period-3 CDW insulator coexisting with an IQAH component. This insulating state can be regarded as a commensurate crystal of holes doped into the $\nu = 1$ state. Upon tuning parameters (say, a displacement field), the FQAH can transition to the IQAH-CDW state~\cite{Song2023_QPT_FQAH,patri2024extended}. Now consider doping this IQAH-CDW state away from filling $2/3$. One possibility is that the CDW component evolves into an incommensurate hole crystal~\cite{patri2024extended}. However an alternate possibility - if the commensuration effect is strong
- is that the CDW stays commensurate and the extra holes with density $\sim \delta$ are accomodated by forming a Fermi liquid metal with a Fermi surface. This metallic state is smoothly connected to the one we found above through the interpocket pairing of the secondary CFL. Thus we expect that the phase boundary between the FQAH and the IQAH-CDW that emanates from $\nu = \frac{2}{3}$ curves toward the FQAH regime upon doping. This is depicted in Fig.~\ref{fig:ZeroT_dopef1_phase_diagram}. 

\subsubsection{Intrapocket pairing: PDW superconductor with non-abelian topological order}\label{subsubsec:pairing_intra}

We now consider the alternative scenario in which each $\psi_i$ pocket undergoes BCS pairing. Since $d_i = z_i \psi_i$ is at half-filling relative to the emergent flux it sees, we will draw an analogy with the half-filled Landau level and assume that the preferred pairing channel is again $p+ip$. This pairing channel puts each of the three $d_i$ pockets into a Moore-Read state, which can be described by a non-abelian Chern-Simons theory~\cite{Moore1991_nonabelian,Read1999_pair}. The total Lagrangian involving $d_i$ can therefore be rewritten as 
\begin{equation}
    \begin{aligned}
    &L_{\rm MR} = \sum_{i=1}^3 \bigg[- \frac{2}{4\pi} \Tr (\tilde c_i d \tilde c_i + \frac{2}{3} \tilde c_i^3) + \frac{2}{4\pi} (\Tr \tilde c_i) d (\Tr \tilde c_i) + \frac{1}{2\pi} (\Tr \tilde c_i) d \tilde b_i - \frac{1}{4\pi} \tilde b_i d \tilde b_i + \frac{1}{2\pi} a d \tilde b_i\bigg] \,,
    \end{aligned}
\end{equation}
where each $\tilde b_i/\tilde c_i$ is a $U(1)/U(2)$ gauge field (this precise formulation is reviewed in~\cite{Seiberg2016_TQFTgappedbdry}). Gluing this TQFT back to the remaining Lagrangian for $f_2$ and $f_3$, we find 
\begin{equation}
    \begin{aligned}
    L_{\rm eff} &= \sum_{i=1}^3 \bigg[- \frac{2}{4\pi} \Tr (\tilde c_i d \tilde c_i + \frac{2}{3} \tilde c_i^3) + \frac{2}{4\pi} (\Tr \tilde c_i) d (\Tr \tilde c_i) + \frac{1}{2\pi} (\Tr \tilde c_i) d \tilde b_i - \frac{1}{4\pi} \tilde b_i d \tilde b_i + \frac{1}{2\pi} a d \tilde b_i \bigg] \\
    &\hspace{1cm}- \frac{2}{4\pi} a da + \frac{1}{4\pi} (b-a) d (b-a) + \frac{1}{4\pi} (A-b) d (A-b) \,.
    \end{aligned}
\end{equation}
To simplify the TQFT above, we observe that $\tilde b_i$ forms an invertible $U(1)_{1}$ state which can be integrated out. The remaining Lagrangian takes the form
\begin{equation}
    \begin{aligned}
    L_{\rm eff} &= \sum_{i=1}^3 \bigg[- \frac{2}{4\pi} \Tr (\tilde c_i d \tilde c_i + \frac{2}{3} \tilde c_i^3) + \frac{3}{4\pi} (\Tr \tilde c_i) d (\Tr \tilde c_i) + \frac{1}{2\pi} a d \Tr \tilde c_i \bigg] \\
    &\hspace{1cm} + \frac{1}{4\pi} a da + \frac{1}{4\pi} (b-a) d (b-a) + \frac{1}{4\pi} (A-b) d (A-b) \,.
    \end{aligned}
\end{equation}
Next, we rewrite the level-1 Chern-Simons terms for the $\text{spin}_{\mathbb{C}}$ connections $a, b-a, A-b$ as the topological response of $U(1)_{1}$ states formed by three $U(1)$ gauge fields $\alpha, \beta, \gamma$. For example, 
\begin{equation}
    \frac{1}{4\pi} a da \quad \rightarrow \quad - \frac{1}{4\pi} \alpha d \alpha + \frac{1}{2\pi} a d \alpha \,. 
\end{equation}
Substituting this representation into the Lagrangian, we see that $a, b$ become Lagrange multipliers that impose the constraints
\begin{equation}
    \alpha - \beta + \sum_i \Tr \tilde c_i = 0 \,, \quad \beta = \gamma \,. 
\end{equation}
These constraints allow us to integrate out $a, b, \beta, \gamma$ and obtain a new effective Lagrangian
\begin{equation}\label{eq:MooreRead_SCfinal}
    \begin{aligned}
        L_{\rm eff} = \sum_{i=1}^3 &\left[- \frac{2}{4\pi} \Tr (\tilde c_i d \tilde c_i + \frac{2}{3} \tilde c_i^3) + \frac{3}{4\pi} (\Tr \tilde c_i) d (\Tr \tilde c_i) \right] \\
        &- \frac{2}{4\pi}\left(\alpha + \sum_i \Tr \tilde c_i\right) d \left(\alpha + \sum_i \Tr \tilde c_i\right) - \frac{1}{4\pi} \alpha d \alpha + \frac{1}{2\pi} \left(\alpha + \sum_i \Tr \tilde c_i\right) d A \,. 
    \end{aligned}
\end{equation}
We now show that this complicated gauge theory secretly describes a charge-2 superconductor. To see that, it suffices to examine the equations of motion for the abelian factors $\Tr \tilde c_i$ and $\alpha$
\begin{equation}
    \begin{aligned}
        4 d \Tr \tilde c_i - 4 \sum_j d \Tr \tilde c_j - 4 d \alpha + 2 d A &= 0 \,, \\
        -3 d \alpha - 2 \sum_j d \Tr \tilde c_j + d A &= 0 \,. 
    \end{aligned}
\end{equation}
After summing over $i$ in the first equation and multiplying the second equation by 4, we obtain
\begin{equation}
    \begin{aligned}
        - 8 \sum_j d \Tr \tilde c_j - 12 d \alpha + 6 d A &= 0 \,, \\
        -12 d \alpha - 8 \sum_j d \Tr \tilde c_j + 4 d A &= 0 \,. 
    \end{aligned}
\end{equation}
Subtracting these equations immediately gives $2dA = 0$, which implies the formation of a charge-2 superconducting condensate. 

The nature of this superconductor is drastically different from the chiral topological superconductor that we found in Section~\ref{sec:scenario1_SC}. First, following the discussion in Section~\ref{sec:anyon_dispersion}, we observe that the operators $d_i$ which create partons in three degenerate pockets furnish an irreducible representation of the projective lattice translation symmetry $T_x T_y T_x^{-1} T_y^{-1} = e^{-2\pi i/3}$:
\begin{equation}
    \begin{aligned}
    T_x&: d_n(\tilde k_x, \tilde k_y) \rightarrow d_{n+1}(\tilde k_x, \tilde k_y) \, e^{i \tilde k_x/3} \,, \\ 
    T_y&: d_n(\tilde k_x, \tilde k_y) \rightarrow d_n(\tilde k_x, \tilde k_y)\, e^{-2\pi n i/3} \, e^{i\tilde k_y}  \,,
    \end{aligned}
\end{equation}
where $d_{n+3}$ is identified with $d_n$ and $\tilde k_x, \tilde k_y$ take values in the reduced Brillouin zone $[0,2\pi/3] \times [0,2\pi]$. Now we recall that the Moore-Read state for $d_i$ is constructed by binding two units of flux to each $d_i$ to form composite fermions $\psi_i$ and then condensing Cooper pairs of $\psi_i$ in the $p+ip$ channel. Since $\psi_i$ inherits the projective action of lattice translation from $d_i$, the Cooper pairing of $\psi_i$ necessarily breaks the lattice translation symmetry spontaneously, leading to an exotic pair density wave (PDW) state. Moreover, a careful analysis of \eqref{eq:MooreRead_SCfinal} reveals that the superconductor coexists with a non-abelian topological order. An interesting effect is that gluing the abelian Jain state to the Moore-Read states formed by $d_i$ creates new non-abelian anyons with enlarged quantum dimensions. The complete anyon content of this theory is described in Appendix~\ref{app:nonabelian_TO}.

\subsection{Finite temperature phenomenology}\label{subsec:scenario2_FL_finiteT}


Let us now construct a complete finite-temperature phase diagram at small doping $\delta$. As the system cools down from high temperature, the first energy scale it encounters is the energy gap $\Delta_{\rm Jain}$ of the Jain state at filling $\nu = 2/3$. When $T$ decreases below $\Delta_{\rm Jain}$, the doped system behaves as a Jain state coexisting with a thermal gas of doped holes at lattice filling $\delta$. For $\delta \ll 1$, the transport properties of the system in this regime are identical to the Jain state up to $\mathcal{O}(\delta)$ corrections. The classical dynamics of the doped holes persist until $T$ reaches a lower energy scale $E_F(\delta) \sim \delta \ll \Delta_{\rm Jain}$, which is the quantum degeneracy temperature of a Fermi gas at lattice filling $\delta$. The evolution from $T > E_F(\delta)$ to $T< E_F(\delta)$ is not a sharp phase transition, but a smooth crossover during which the doped holes gradually develop quantum coherence and start to behave as three pockets of interacting composite Fermi liquids. Finally, at an even lower temperature $T = T_c$, there is a sharp phase transition into the ultimate low-temperature phase, which can either be the CDW metal in Section~\ref{subsubsec:pairing_inter} or the PDW superconductor in Section~\ref{subsubsec:pairing_intra}. 
\begin{figure}[!h]
    \centering
    \includegraphics[width=0.9\linewidth]{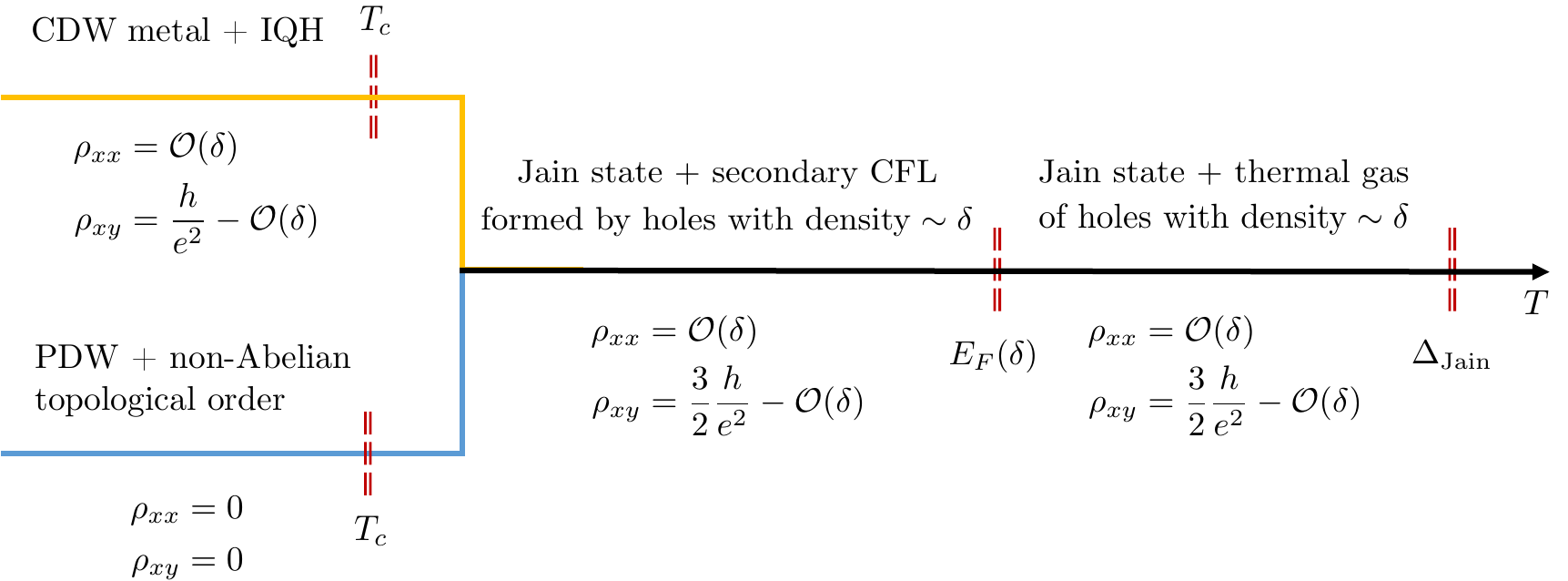}
    \caption{The finite temperature phase diagram at $\nu = 2/3 - \delta$ when the $a_{\frac{1}{3}}$ anyon is doped into the FQAH state at $2/3$-filling. For $T < \Delta_{\rm Jain}$, the system behaves like the Jain state at $\nu = 2/3$ coexisting with a thermal gas of anyonic holes with density $\delta$. As $T$ decreases below a ``degeneracy'' temperature $E_F(\delta)$ at density $\delta$, the anyons fractionalize into composite fermions and form a secondary CFL with non-Fermi liquid characteristics. Depending on microscopic interactions, there is a further phase transition at an even lower $T_c$ into a CDW metal with a coexisting IQH state or a PDW superconductor with coexisting non-abelian topological order. The $\mathcal{O}(h/e^2)$ difference in Hall resistivity between these states provides a sharp experimental diagnostic.}
    \label{fig:finiteT_phase_diagram_scenario2}
\end{figure}

The qualitative picture in the previous paragraph misses one additional subtlety. As is the case for any CFL state, although the CF Fermi surface becomes sharply defined at $T \lesssim E_F(\delta)$, the non-Fermi liquid signatures of the CFL state do not manifest themselves until we reach a lower energy scale $T_{\rm NFL}$. The phenomenology of the secondary CFL regime depends crucially on the relative size of $T_{\rm NFL}$ and $T_c$. 

If the low-temperature phase is a CDW metal coexisting with an IQH state, the transition from the secondary CFL to the low-temperature phase is driven by the pairing of composite fermions from \textit{different pockets}. Since the interpocket attraction between composite fermions is mediated by critical gauge fluctuations, the effective BCS coupling diverges much faster than the logarithmic RG flow associated with the conventional mechanism for superconductivity in a Fermi liquid. As a result, the transition temperature is generally a power-law function $T_c(V_{\rm BCS, inter}) \sim E_F(\delta) \left[V_{\rm BCS, inter}/E_F(\delta)\right]^{r}$ of the bare attractive BCS interaction $V_{\rm BCS, inter}$, rather than an inverse exponential $e^{-C/V_{\rm BCS, inter}}$. Within a controlled expansion for non-Fermi liquids, it has been previously shown that $T_c(V_{\rm BCS, inter})$ is comparable to $T_{\rm NFL}$~\cite{Metlitski2014_PairingNFL}. 
This is not surprising, as superconductivity and non-Fermi liquid phenomena are induced by the same gauge fluctuations in this particular system. 

On the other hand, if the ultimate low-temperature phase is a PDW superconductor + non-abelian topological order, then the transition from the secondary CFL into the low-temperature phase is driven by the pairing of composite fermions in the \textit{same pocket}. Since intrapocket pairing is suppressed by gauge fluctuations, stabilizing this low-temperature phase requires a large bare intrapocket attractive interaction between composite fermions. As a result, the physics that determines the values of $T_c$ in this scenario has no simple connection to the gauge fluctuations that set $T_{\rm NFL}$ and no quantitative relation between $T_c$ and $T_{\rm NFL}$ can be deduced. 

Putting these two cases together, we see that there is no parametric separation between $T_{\rm NFL}$ and $T_c$ in general and the non-Fermi liquid features of the secondary CFL need not be observable. Therefore, the most robust feature in the finite-temperature phase diagram, as illustrated in Fig.~\ref{fig:finiteT_phase_diagram_scenario2}, is the $\mathcal{O}(h/e^2)$ jump in Hall resistivity across $T_c$, both for the CDW metal and the PDW superconductor.

\section{Discussion}\label{sec:discussion}
In this paper, we explored several novel itinerant phases that can be stabilized in a doped FQAH insulator beyond the quantum Hall plateau. 
Near the prominent $\nu = 2/3$ state in current experiments, the low-temperature phase that emerges depends on which anyon has the lowest excitation gap and gets doped into the system as the charge filling is varied. If charge-$2/3$ anyons are doped in, a chiral topological superconductor (with four chiral Majorana edge modes) arises. 
If charge-$1/3$ anyons are doped in instead, we find a generalized composite Fermi liquid at intermediate temperatures which undergoes composite fermion pairing at low temperature. The dominant pairing instability naturally lead to a Fermi liquid metal coexisting with a doping-induced period-3 charge order, although a more exotic PDW superconductor with intrinsic non-abelian topological order can also be stabilized with appropriate interactions in the BCS channel. We discussed the phenomenology of these distinct possibilities, both at zero and at nonzero temperatures. 

In light of these results, it may be interesting to probe the low temperature limit at fillings close to $2/3$ in the current experimental realizations of FQAH phenomena. The energy scale for the onset of the itinerant ground states is set by the doping $\delta$ which is a small fraction of the total filling. The novel states we found will also compete with localized states induced by a combination of Coulomb interactions and disorder. Nevertheless, with improvements in sample quality, it may be possible to reach these itinerant states. Existing experiments find a decrease of the Hall resistivity on either side of $2/3$ filling. While the decrease on the electron-doped side of $2/3$ is natural, the same decrease on the hole-doped side is more surprising.  Our analysis in terms of parton bands gives a number of possible explanations for this behavior. 

The broader message 
in this paper is that the doped FQAH state provides a wonderful platform for the realization of fractionalized itinerant phases. Beyond the cases that we considered, other interesting possibilities certainly also exist. 
For instance, the doped charges may just go in as electrons (or holes) and not as anyons. Then we will get a ``fractionalized Fermi liquid"~\cite{senthil2003fractionalized,senthil2004weak} with a small Fermi pocket of electrons or holes while preserving translation and charge $U(1)$ symmetries. The Fermi surface coexists with the topological order of the FQAH state. This metallic phase violates the conventional Luttinger theorem but satisfies a modified one that relates the Fermi surface area to the density of doped charges. It would also be fruitful to explore the possible doping-induced continuous phase transitions from the gapped FQAH states to these itinerant phases. These transitions would be necessarily ``beyond Landau" as they describe the destruction of a topological order and the simultaneous emergence of itinerant degrees of freedom, sometimes with concomitant translation symmetry-breaking. 

The physics of the doped FQAH state - like the much explored doped Mott insulator - is thus likely to reveal a wealth of unexpected correlated many-body phenomena. We hope that future theory and experiments will focus on studying this regime in detail. 

\textit{Note: Recently, Ref.~\cite{Kim2024_chiral_anyonSC} appeared which has a small overlap with our results. After the submission of our paper, Ref.~\cite{Divic2024_anyonSC_topcrit} appeared which studies a related mechanism for superconductivity in a Hofstadter-Hubbard model.}

\begin{acknowledgments}
We thank Zhihuan Dong, Tonghang Han, Zhaoyu Han, and Seth Musser for discussions. ZDS and TS are supported by NSF grant DMR-2206305, and partially through a Simons Investigator Award from the Simons Foundation.  
This work was also partly supported by the Simons Collaboration on Ultra-Quantum Matter, which is a grant from the Simons Foundation (Grant No. 651446, T.S.). 
\end{acknowledgments}

\onecolumngrid
\appendix

\section{Contrast with Laughlin's anyon superconductor: why is our story different?}\label{app:anyonSC}

In this appendix, we review the old work of anyon superconductivity (ASC) pioneered by Laughlin in Ref.~\onlinecite{Laughlin1988_anyonSC} and developed further in Refs.~\onlinecite{Fetter1989_anyonSC_RPA,Chen1989_anyonSC,Lee1989_anyonSC,halperin1989consequences,Wen1990_anyonSC}. The goal is to highlight the key conceptual differences between ASC and the two classes of superconductors that arise from doping an FQAH insulator. To that end we focus on the field theoretic analysis of Ref.~\onlinecite{Chen1989_anyonSC}. 

The starting point of Ref.~\onlinecite{Chen1989_anyonSC} is the following Lagrangian description of an anyon gas 
\begin{equation}
\label{chenL}
    L = L[f, a +A] + \frac{n}{4\pi} a da    \,, \quad L[f, a] = f^{\dagger} \left[(i\partial_{t} + a_t) + \frac{(i \nabla + \bs{a})^2}{2m} - \mu\right] f  \,,
\end{equation}
where $f$ is a fermion and $a$ is treated as a dynamical $U(1)$ gauge field. The Lagrangian is written in flat space-time. The $a_t$ equation of motion imposes a constraint
\begin{equation}
    \rho_f \equiv f^{\dagger} f = -\frac{n}{2\pi} \nabla \times \bs{a} \,. 
\end{equation}
This equation shows that fluctuations of $a$ attach $-1/n$ units of flux to the fermion $f$, turning it into an anyon with fractional statistics ${\theta = \pi (1 - \frac{1}{n})}$. Treating the fluctuating gauge field in a mean field approximation leads to the fermions seeing an average ``statistical" magnetic field with a filling factor $-n$. The fermions thus completely fill $|n|$ Landau levels. Integrating them out gives a low energy effective theory 
\begin{eqnarray}
    L_{eff} & = & -\frac{n}{4\pi} (a+A) d(a+A) + \frac{n}{4\pi} ada \nonumber \\
    & = & - \frac{n}{2\pi} Ada -\frac{n}{4\pi} AdA \nonumber 
\end{eqnarray}
Integrating out $a$, we get a charge-$n$ superconductor. 

To understand the relevance of the model to a microscopic system, let us ask what the local operators are in the theory defined by \eqref{chenL}. Operators such as $f^\dagger f$ and other gauge-invariant polynomials of $f, f^\dagger$ are local and are clearly bosonic ({\em i.e.} in relativistic parlance they have integer spin). More interesting are the monopole operators of $a$. Assuming standard flux quantization, the basic $2\pi$ monopole ${\cal M}_a$ is not by itself gauge invariant due to the Chern-Simons term. Rather, it carries gauge charge-$n$. A gauge-invariant monopole operator is obtained by binding $n$ fermions, schematically indicated as ${\cal M}_a f^n$. Due to the Chern-Simons term, the ${\cal M}_a$ field by itself has half-integer spin if $n$ is odd and integer spin if $n$ is even. Binding to $n$ fermions implies that the gauge-invariant monopole operator always has an integer spin. Note that it also has a physical electric charge $q_A = n$. Thus we conclude that all local operators in the theory are bosons and the smallest physical electric charge is $n$. 

The anyon superconductor model of Ref.~\onlinecite{Chen1989_anyonSC} should thus be viewed as a model of a microscopic system of physical charge-$n$ bosons in which the anyons have emerged at low energy. In particular, there is no electron operator in the low energy model. Consider microscopic models where the basic charged particle is an electron with physical charge $1$. Then, clearly all local operators with odd 
physical charge are fermionic while those with even physical charge are bosonic (known sometimes as the spin-charge relation~\cite{Seiberg2016_TQFTgappedbdry}). Thus, for odd $n$, the model in \eqref{chenL} can never emerge in such a microscopic electron system. In particular, the $n = 3$ case (which describes a fluid of anyons with $\frac{2\pi}{3}$ statistics) is not emergeable in an electron system, and hence the anyon superconductor found in that model should not be identified with the one we have described in this paper. 

We can view this conclusion from a more sophisticated perspective by placing the system on a general oriented space-time 3-manifold $M_3$ of the form $M_2 \times M_1$, and specifying whether $a, A$ are $U(1)$ gauge fields or $\text{spin}_{\mathbb{C}}$ connections. If we demand that the microscopic charge-$1$ particles are fermions, then we must take $A$ to be a $\text{spin}_{\mathbb{C}}$ connection. As $f$ couples to $a+A$, it follows that $a$ must be an ordinary $U(1)$ gauge field. Then for the Chern-Simons term to be properly gauge-invariant, $n$ must be even.  Thus we again conclude that for odd $n$ the theory cannot emerge from a microscopic electron system. For even $n$, it can emerge but the local operators that carry physical charge in the theory all have even charge, and hence are bosons. 

Is there an alternate coupling of the background $\text{spin}_{\mathbb{C}}$  connection in \eqref{chenL} that makes the theory equivalent to ours? The theory has a global $U(1)$ symmetry associated with conservation of the flux of $a$, and we can couple $A$ to the corresponding monopole current. The most general such theory, paying proper attention to quantization of Chern-Simons coefficients, is 
\begin{equation}\label{chen_modL}
\begin{aligned}
L & =  L[f,a] + \frac{n}{4\pi} ada + \frac{m}{2\pi} Ada + L[A,g] \,, \\
L[A,g] & = \frac{k}{4\pi} AdA + l \mathrm{CS}_g \,. 
\end{aligned}
\end{equation}
Here $n, m, k, l$ are integers which will be further constrained below. The gauge field $a$ should be viewed as a $\text{spin}_{\mathbb{C}}$  connection as it couples to $f$. Then the proper Chern-Simons term for $a$ is  
\begin{equation} 
 n\, \mathrm{CS}[a, g] = \frac{n}{4\pi} ada + 2n\, \mathrm{CS}_g \,.
\end{equation} 
Similarly as $A$ is a $\text{spin}_{\mathbb{C}}$ connection, the Chern-Simons term for it is 
\begin{equation}
 k\, \mathrm{CS}[A, g] = \frac{k}{4\pi} AdA + 2k\, \mathrm{CS}_g \,.
\end{equation}
Consider now the quantization condition for the coefficient of the $A da$ term. As $A$ and $a$ are both $\text{spin}_{\mathbb{C}}$ connections, $m$ must be an {\em even} integer. This is because, as already seen, the gauge-invariant monopole of $a$ must be a boson with integer spin. The $Ada$ term ensures that these monopoles carry physical charge $m$. But by the spin-charge relation, we must have $m$ even. We can obtain a stronger constraint as follows. 
Note that $a+2A$ and $a$ are both $\text{spin}_{\mathbb{C}}$ connections. Therefore a properly normalized term is 
\begin{equation} 
    p \left(\mathrm{CS}[a+ 2A, g] - \mathrm{CS}[a,g]\right) = \frac{2p}{2\pi} Ada + \frac{4p}{4\pi} AdA \,.
\end{equation} 
Subtracting $4p \mathrm{CS}[A,g]$ we get that 
\begin{equation} 
\frac{2p}{2\pi} Ada - 8 p \, \mathrm{CS}_g 
\end{equation} 
is well-defined. Setting $m = 2p$, we see that consistency of the Chern-Simons terms for $A,a$ requires that we get a contribution $2n + 2k - 8p$ to $l$. In addition, we can always add a term $16\, \hat{l}\, \mathrm{CS}_g$ to the Lagrangian with $\hat{l}$ an integer which encodes the response of an invertible bosonic topological order in 2+1 dimensions (the so-called $E_8$ states). Thus we finally get the constraints 
\begin{equation} 
\label{eq:CSconstraints} 
m = 2p \,, ~~~~~~l = 2n + 2k - 8p + 16\hat{l} \,. 
\end{equation} 
The Lagrangian in \eqref{chen_modL} with the constraints of \eqref{eq:CSconstraints} is the most general allowed coupling of background $A,g$ to the anyon action of \eqref{chenL}.

Now let us ask if the theory we actually have in the physical system of interest fits \eqref{chen_modL}. We start with \eqref{eq:dopedf3}, and shift the $U(1)$ gauge field $b$ to define a new $\text{spin}_{\mathbb{C}}$ connection $a = b - A$. Then \eqref{eq:dopedf3} becomes 
\begin{equation} 
L = L[\psi, a] + \frac{3}{4\pi} ada + \frac{2}{2\pi} Ada + \frac{2}{4\pi} AdA + 2CS[g] 
\end{equation} 
This has exactly the form of \eqref{chen_modL} with the identification $n = 3, k = 2, p = 1, \hat{l} = 0$. 

Thus we conclude that we can recast our theory in terms of field theories of the type considered in Ref.~\onlinecite{Chen1989_anyonSC} but with a peculiar coupling to the background gauge fields $A,g$. Indeed, if we just state that we have a finite density of anyons with $\frac{2\pi}{3}$ statistics, then a variety of couplings to $A,g$ are possible and will lead to distinct physical properties. It is thus more physical to start with a particular topological order that produces these anyons, and to then dope the system to induce a finite density as we have done.

 Similar comments apply to the hierarchical superconducting states in Ref.~\onlinecite{Lee1989_anyonSC}, where the anyons carry fractional statistics $\theta = \pi P/Q$ with $PQ$ even (this reduces to the case considered above for $P = n-1, Q = n$). With the coupling to the external gauge field $A$ specified in Ref.~\onlinecite{Lee1989_anyonSC},  the charge-$Q$ superconductor in the hierarchical construction cannot emerge from a microscopic electron system when $Q$ is odd, but can  emerge from a microscopic bosonic system. Therefore, we arrive at a robust charge/statistics relation for the usual anyon SC: it can only emerge from a microscopic system of interacting fermions when the anyons carry fractional statistics with \textit{even denominator}. However this charge/statistics relation  clearly does not apply to the doped FQAH states in our story.  We already discussed the chiral superconductor in Section~\ref{sec:scenario1_SC} at lattice filling $\nu = 2/3 - \delta$, which can be interpreted as doping a finite density $\delta$ of anyons into the $\nu = 2/3$ FQAH state. The doped anyon carries fractional statistics $\theta = 2\pi/3$ which has an odd denominator. Nevertheless, the effective Lagrangian describes a charge-2 superconductor. More generally, the charge-2 superconductors with non-abelian topological order and the charge-$|p|$ superconductors with abelian topological order that we constructed near a general Jain filling $\nu = p/(2p+1)$ are always even-charge superconductors that arise from doping anyons with odd-denominator fractional statistics.  

\section{Generalization to the principal Jain sequence \texorpdfstring{$\nu = p/(2p+1)$}{}}\label{app:generalization}

In this appendix, we generalize the construction in the main text to analyze the effects of doping on a general Jain state at lattice filling $\nu = p/(2p+1)$. As we will see, in the first scenario where the active band corresponds to $f_1$, lattice translation splits $f_1$ into $|2p+1|$ Fermi pockets labeled by $d_i$, each at filling $- \sgn(p)/2$ relative to the emergent flux it sees. Depending on microscopic interactions, $d_i$ either forms a Moore-Read topological order or a secondary CFL. For arbitrary $p$, putting $d_i$ in the Moore-Read state results in a charge-2 superconductor that spontaneously breaks the lattice translation symmetry. The superconductor carries a non-abelian topological order that generalizes the non-abelian topological order for $p = - 2$ described by \eqref{eq:MooreRead_SCfinal}. On the other hand, if $d_i$ forms the secondary CFL, gauge fluctuations drive a low temperature instability in which pairs $d_i, d_j$ with $i \neq j$ form exciton condensates. This condensate Higgses all the gauge fields and the true ground state is a translation-symmetry breaking Fermi liquid much like for $p = -2$. 

In the second scenario where the active band corresponds to $f_2/f_3$, the fate of the doped holes depends on the parity of $p$. For even $p$, the topological superconductor at $p = -2$ generalizes to a more exotic charge-$|p|$ superconductor coexisting with an intrinsic abelian topological order. For odd $p$, we instead obtain a generalized CFL that resembles the conventional CFL at even-denominator filling $-1/|p+1|$. In what follows, we will only sketch these results, since the detailed analysis largely parallels the case of $p = -2$. 

\subsection{Doping into the \texorpdfstring{$f_2/f_3$}{} band: an even/odd pattern}\label{subsec:general_p_scenario1}

We first consider the scenario where $f_3$ is the active parton, while $f_1, f_2$ continue to form Chern insulators with $C_1 = p$ and $C_2 = 1$. Since the anyon $\tilde f_3$ associated with $f_3$ carries charge $q = \nu e$ and self-statistics $\theta = \nu \pi$, it can be identified as the vison, which is nucleated by a $2\pi$ flux insertion. Doping into the $f_3$ band is therefore equivalent to doping a density $\delta$ of visons into the FQAH state at $\nu = 2/3$, which is natural when the vison has the lowest energy gap. 

At filling $\nu_1 = \nu_2 = p/(2p+1) - \delta$, in order for $f_1, f_2$ to form Chern insulators with $C_1 = p$ and $C_2 = 1$, the mean-field flux seen by $f_1, f_2$ must be reduced by $\delta/p, \delta$ respectively. This condition translates to the mean-field flux assignment
\begin{equation}
    \begin{aligned}
    \frac{1}{2\pi} \ev{\nabla \times \bs{a}} &= \frac{1}{2p+1} - \frac{\delta}{p} \,,\\
    \frac{1}{2\pi} \ev{\nabla \times \bs{b}} &= \frac{p+1}{2p+1} - \frac{p+1}{p} \delta \,.
    \end{aligned}
\end{equation}
In the $|2p+1|$-fold enlarged unit cell, the filling of $f_3$ is fixed at $|p| - |2p+1| \delta$, while the flux seen by $f_3$ is
\begin{equation}
    \Phi_{A-b} = |2p+1| \ev{\nabla \times (\bs{A} - \bs{b})} = 2 \pi \left[-|p+1| + |2p+1|\delta (p+1)/p\right] \,.
\end{equation}
By the Streda formula, part of the $f_3$ partons at total filling $|p| + |2p+1|\delta (p+1)/p$ will continue to form a Chern insulator with $C_3 = 1$, while the remaining $f_3$ partons at filling $- \delta (2p+1)^2/|p|$ see an effective magnetic flux $2\pi |2p+1|\delta (p+1)/p$. Due to the projective action of lattice translation, the $f_3$ parton splits into $|2p+1|$ degenerate species $\psi_i$, each at filling $- \delta (2p+1)/p$ and sees flux $2\pi |p+1| \delta (2p+1)/p$. This structure immediately suggests an even-odd pattern. 

For even $p$, $|p+1|$ is odd and the most natural state formed by each $\psi_i$ at filling $-1/|p+1|$ is the $U(1)_{-|p+1|}$ Laughlin state. In the deep IR limit, we therefore obtain an effective TQFT Lagrangian
\begin{equation}\label{eq:general_p_SC_TQFT}
    \begin{aligned}
    L &= \frac{p}{4\pi} a da + \frac{1}{4\pi} (b-a) d (b-a) + \frac{1}{4\pi} (A - b) d (A-b) + \sum_{i=1}^{|2p+1|} \frac{|p+1|}{4\pi} \alpha_i d \alpha_i + \frac{1}{2\pi} \alpha_i d (A-b) \,.
    \end{aligned}
\end{equation}
From here, a natural approach would be to mimic the discussion at $\nu = 2/3$ and integrate out the internal gauge fields $a, \alpha_i$ to get an effective theory for $b, A$. However, for $p \neq -2$, this turns out to be impossible. To see that, we consider the equations of motion for $a, b, \alpha_i$ together
\begin{equation}
    \begin{aligned}
        &\text{EOM for} \, a \,: (p+1) da = db \,, \\ 
        &\text{EOM for} \, b \,: 2 db - da - d A = \sum_{i=1}^{|2p+1|} d \alpha_i \,, \\
        &\text{EOM for} \, \alpha_i \,: |p+1| d\alpha_i = db - d A  \,.
    \end{aligned} 
\end{equation}
Crucially, for $p \neq -2$, the EOM for every internal gauge field involves an integer multiple $N$ of the field strength, where $|N| > 1$. Therefore, one cannot integrate out the gauge fields while maintaining flux quantization. 

Nevertheless, the equations of motion are powerful enough to determine that the TQFT in \eqref{eq:general_p_SC_TQFT} secretly describes a charge-$|p|$ superconductor. To see that, we first use the EOM for $a$ to fix $db = (p+1) da$. Substituting this into the EOM for $b, \alpha_i$, we find 
\begin{equation}
    (2p+1) da - A = \sum_{i=1}^{|2p+1|} d \alpha_i \,, \quad |p+1| d \alpha_i = (p+1) da - d A \,. 
\end{equation}
Next, we multiply the first equation by $|p+1|$ and sum over $i = 1, \ldots, |2p+1|$ in the second equation. Combining the results gives 
\begin{equation}
    \begin{aligned}
    \sum_{i=1}^{|2p+1|} |p+1| d \alpha_i &= (2p+1) |p+1| da - |p+1| dA \\
    &= |2p+1| (p+1) da - |2p+1| d A \,.
    \end{aligned}
\end{equation}
Finally, for $p \neq 0$, we have $\sgn(2p+1) = \sgn(p+1)$, which implies the simplified constraint
\begin{equation}
    (|2p+1| - |p+1|) d A = |p| d A = 0 \,.  
\end{equation}
Therefore, the collective fluctuations of internal gauge fields $\alpha_i, a, b$ Higgs the external $U(1)$ gauge field $A$ down to $\mathbb{Z}_{|p|}$, implying that the system is a charge-$|p|$ superconductor. When $p = -2$, this conclusion is consistent with Section~\ref{sec:scenario1_SC}. 

In terms of the microscopic electrons, this charge-$|p|$ superconductor is very exotic and beyond any mean-field description. To see that, let us consider the Landau-Ginzburg theory for a conventional charge-$|p|$ superconductor \textit{with no topological order} written in terms of a charge-$|p|$ order parameter field $\phi$ coupled to an external gauge field $A$ via the Lagrangian $L[\phi, |p| A]$. Under the particle-vortex duality where $\phi \rightarrow \phi_v$, the Lagrangian transforms to
\begin{equation}
    L[\phi, |p|A] + L_{\rm CS}[A] \quad \rightarrow \quad L[\phi_v, b] + \frac{|p|}{2\pi} b d A + L_{\rm CS}[A] \,. 
\end{equation}
where $L_{\rm CS}[A]$ can include Chern-Simons terms for $A$ generated by stacking invertible IQH states. In the superconducting phase, the vortex $\phi_v$ is massive and can be integrated out. Therefore, if we treat $A$ as a fluctuating gauge field, the TQFT for $b,A$ is simply a $\mathbb{Z}_{|p|}$ Dijkgraaf-Witten theory with ground state degeneracy $p^2$ on a spatial torus. 

To distinguish the superconductor in \eqref{eq:general_p_SC_TQFT} from the conventional charge-$|p|$ superconductor, it suffices to show that its torus ground state degeneracy differs from $p^2$. Since the TQFT in \eqref{eq:general_p_SC_TQFT} is a simple K-matrix theory, this computation is straightforward. We begin with a change of variables $b \rightarrow b + A$ under which the Lagrangian in \eqref{eq:general_p_SC_TQFT} transforms to
\begin{equation}
    L = \frac{p}{4\pi} a da + \frac{1}{4\pi} (b+A-a) d (b+A-a) + \frac{1}{4\pi} bdb + \sum_{i=1}^{|2p+1|} \frac{|p+1|}{4\pi} \alpha_i d \alpha_i - \frac{1}{2\pi} \alpha_i db \,.
\end{equation}
The K-matrix associated with this Lagrangian can be written in a block form
\begin{equation}
    \tilde K = \begin{pmatrix}
        \tilde K_{11} & \tilde K_{12} \\ \tilde K_{12}^T & \tilde K_{22}
    \end{pmatrix} \,, \quad \tilde K_{11} = \begin{pmatrix}
        p+1 & -1 & -1 \\ -1 & 2 & 1 \\ -1 & 1 & 1 
    \end{pmatrix} \,, \quad \tilde K_{12} = \begin{pmatrix}
        0 & \ldots & 0 \\ -1 & \ldots & -1 \\ 0 & \ldots & 0
    \end{pmatrix} \,, \quad \tilde K_{22} = |p+1| I_{|2p+1|} \,.
\end{equation}
This block structure allows us to compute the determinant analytically for all $p$:
\begin{equation}
    \begin{aligned}
        \det \tilde K &= \det \tilde K_{22} \cdot \det (\tilde K_{11} - \tilde K_{12} \tilde K_{22}^{-1} \tilde K_{12}^T) = |p+1|^{|2p+1| - 3} \det (|p+1| \tilde K_{11} - \tilde K_{12} \tilde K_{12}^T) \\
        &= |p+1|^{|2p+1|-3} \cdot \det \begin{pmatrix}
            (p+1) |p+1| & - |p+1| & - |p+1| \\ -|p+1| & 2|p+1| - |2p+1| & |p+1| \\ - |p+1| & |p+1| & |p+1|
        \end{pmatrix} \\
        &= |p+1|^{|2p+1| - 3} \cdot (p+1)^2 p^2 \cdot - \sgn(p+1) = - \sgn(p+1) |p+1|^{|2p+1| - 1} p^2 \,.
    \end{aligned}
\end{equation}
Note that the magnitude of the determinant, which controls the ground state degeneracy on a torus, can be written as a perfect square
\begin{equation}
    |\det \tilde K| = D^2 \,, \quad D = |p| \cdot |p+1|^{\frac{|2p+1|-1}{2}} \,.
\end{equation}
Since the torus ground state degeneracy $\text{GSD}(p) = D^2 > p^2$ for all $p \neq -2$, the topological order of the general charge-$|p|$ superconductor is distinct from $\mathbb{Z}_{|p|}$ Dijkgraaf-Witten theory. The existence of intrinsic topological order at general $p$ should not be surprising, since anyon condensation does not generally destroy the full topological order of the Jain state. 

In fact, we can prove that a general SC* state described by a $K$ matrix with a single zero-mode can always be factorized as a topologically trivial superconductor stacked with a gapped topological order. More precisely, we can demonstrate the following linear algebra fact: \newline 

\textit{Claim:}
    Let $K$ be an $N$ by $N$ matrix with entries in a principal ideal domain (PID). Suppose that the Smith normal form of $K$ is $UKV = \Lambda$, where $\Lambda$ is a diagonal matrix with a single zero diagonal entry (i.e. $K$ has a single zero mode). Without loss of generality, we can arrange the zero diagonal entry to be $\Lambda_{NN}$. Then we claim that the following block form holds
    \begin{equation}
        V^T K V = \begin{pmatrix}
            \tilde K_{(N-1) \times (N-1)} & \mathbf{0}_{(N-1) \times 1} \\ \mathbf{0}_{1 \times (N-1)} & 0 
        \end{pmatrix} \,.
    \end{equation} 
\begin{proof}
    Since $\Lambda$ is the Smith normal form, we know that $U, V$ are unimodular and invertible. Now let us consider the last column of $\Lambda$ which must be identically zero:
    \begin{equation}
        \Lambda_{jN} = (UKV)_{jN} = U_{jm} K_{mn} V_{nN} = 0 \,, \quad \forall j \,. 
    \end{equation}
    Since $U$ is invertible, this implies that $K_{mn} V_{nN} = 0$ for all $m$, meaning that $V_{nN}$ is precisely the zero mode of $K$. With this information, it is trivial to compute $V^T K V$
    \begin{equation}
        (V^T K V)_{ij} = V_{mi} K_{mn} V_{nj} = \begin{cases}
            V_{mN} (K_{mn} V_{nN}) = 0 & i = j = N \\
            (V_{mN} K_{mn}) V_{nj} = 0 & i = N, j \neq N \\
            V_{mi} (K_{mn} V_{nN}) = 0 & i \neq N, j = N \\
            V_{mi} K_{mn} V_{nj} & i, j \neq N  
        \end{cases} \,. 
    \end{equation}
    This computation agrees with the proposed block-diagonal form of $V^T K V$. 
\end{proof}
Now suppose we start with a superconductor described by the Chern-Simons Lagrangian
\begin{equation}
    L = - \frac{1}{4\pi} \alpha_I K_{IJ} d\alpha_J + \frac{1}{2\pi} q_I \alpha_I d A \,. 
\end{equation}
Given the Smith normal form $K = U^{-1} \Lambda V^{-1}$, we can define a new set of gauge fields $\beta_I = V^{-1}_{IJ} \alpha_J$ such that
\begin{equation}
    \begin{aligned}
    L &= - \frac{1}{4\pi} \alpha_I (U^{-1} \Lambda V^{-1})_{IJ} d \alpha_J + \frac{1}{2\pi} q_I V_{IJ} \beta_J d A  \\
    &= - \frac{1}{4\pi} \beta_I (V^T U^{-1} \Lambda)_{IJ} d \beta_J + \frac{1}{2\pi} q_I V_{IJ} \beta_J d A \,. 
    \end{aligned}
\end{equation}
Since $(V^T U^{-1} \Lambda)$ is block-diagonal, the system is a decoupled stack of a topologically trivial SC sector and a stacked topological order described by a rank $(N-1)$ K-matrix. In the $\beta_I$ basis, we can directly read off the charge of the superconductor 
\begin{equation}
    Q_{\rm SC} = q_I V_{IN} \,. 
\end{equation}
Similarly, we can extract the Hall response for the FQH sector:
\begin{equation}
    \sigma_{xy} = \sum_{i < N, j < N} (V^T q)_i (V^T U^{-1} \Lambda)^{-1}_{ij} (V^T q)_j = \sum_{i,j<N} \sum_{m, n \leq N} q_m V_{mi} \tilde K^{-1}_{ij} q_n V_{nj} \,.
\end{equation}
These formulae can be applied to each of the superconductors we found for $p \in 2 \mathbb{Z}$, and the stacked topological order can be easily identified.

In contrast, for odd $p$, each $\psi_i$ is at an even-denominator filling $-1/|p+1|$. It is then natural to attach $|p+1|$ fluxes to each $\psi_i$ and form a generalized CFL state. The resulting Lagrangian then becomes
\begin{equation}\label{eq:scenario1_oddp_CFL}
    \begin{aligned}
    L &= \frac{p}{4\pi} a da + \frac{1}{4\pi} (b-a) d (b-a) + \frac{1}{4\pi} (A-b) d (A-b) + \sum_{i=1}^{|2p+1|} L_{\rm CFL}[\psi_i, A-b] \,, 
    \end{aligned}
\end{equation}
where
\begin{equation}
    \begin{aligned}
    &L_{\rm CFL}[\psi_i, A-b] = L[\psi_i, A - b - \alpha_i] + \frac{|p+1|}{4\pi} \beta_i d \beta_i + \frac{1}{2\pi} \beta_i d \alpha_i \,. 
    \end{aligned}
\end{equation}
The resistivity in this CFL state is derived in Appendix~\ref{app:transport_oddp_CFL} and takes a somewhat complicated form
\begin{equation}
    \rho = \nu^{-1} (\sigma_1 + \sigma_{\psi})^{-1} - \frac{1-\nu}{\nu} (\sigma_1 + \sigma_{\psi})^{-1} (\sigma_{\psi}^T) \sigma_1^{-1} \,.
\end{equation}
Here, $\sigma_1$ is the conductivity of the $U(1)_1$ IQH state and $\sigma_{\psi}$ is the total parton conductivity for the $|2p+1|$ fermions $\psi_i$ given by
\begin{equation}
    \sigma_{\psi} = |2p+1| \left(- |p+1| \sigma_1^{-1} + \sigma_{\rm FL}^{-1} \right)^{-1} \,,
\end{equation}
where $\sigma_{\rm FL}$ is the conductivity of a Fermi liquid at filling $-\delta/|p|$ relative to the microscopic unit cell. In a system with weak disorder, we expect $\sigma_{\rm FL}$ to scale as $\tau_{\rm tr} \delta$ where $\tau_{\rm tr}$ is the transport scattering rate. In Appendix~\ref{app:transport_oddp_CFL}, we show that the physical Hall resistivity $\rho_{xy}$ approaches the Jain value when $\delta \rightarrow 0$, but gets enhanced by a factor of $\frac{1 + \nu}{1 - \nu}$ when $\tau_{\rm \tr} \rightarrow \infty$ (i.e. the ultraclean limit). This suggests the unusual trend where a decrease in the disorder strength leads to an increase in the Hall resistivity. We caution however that the interpolation between these limits can in general be complicated, as $\sigma_{\rm FL}$ itself can have a nontrivial Hall component. 

In the low temperature limit, the generalized CFL described by \eqref{eq:scenario1_oddp_CFL} could have additional pairing instabilities similar to the ones we encountered in Section~\ref{subsec:pairing_2ndCFL}. These instabilities can confine some of the internal gauge fields and destroy the non-Fermi liquid phenomenology associated with the generalized CFL. A detailed analysis of these instabilities is deferred to future work. 

\subsection{Doping into the \texorpdfstring{$f_1$}{} band: CDW metal and PDW superconductor}\label{subsec:general_p_scenario2}

We now revisit the second scenario at a general Jain filling $\nu = p/(2p+1)$, where $f_2, f_3$ continue to form Chern insulators with $C_2 = C_3 = 1$, while $f_1$ becomes itinerant. By the Streda formula, the mean-field fluxes are pinned to
\begin{equation}
    \begin{aligned}
    \frac{1}{2\pi} \ev{\nabla \times \bs{a}} &= \frac{1}{2p+1} + 2 \delta \,, \\ 
    \frac{1}{2\pi} \ev{\nabla \times \bs{b}} &= \frac{p+1}{2p+1} + \delta \,.
    \end{aligned}
\end{equation}
Upon integrating over $f_2, f_3$, we get an effective action
\begin{equation}
    L = L[f_1, a] + \frac{1}{4\pi} (b-a) d (b-a) + \frac{1}{4\pi} (A-b) d (A-b) \,.
\end{equation}
In the $|2p+1|$-fold enlarged unit cell, $f_1$ is at filling $\nu_1 = |p| -|2p+1| \delta$ while the flux of $a$ is $2\pi (\sgn(p) + 2 |2p+1| \delta)$. By the Streda formula, part of the $f_1$ partons with filling $|p| + 2p|2p+1| \delta$ continue to form a Chern insulator with $C_1 = p$, while the remaining $f_1$ partons with filling $- (2p+1) |2p+1| \delta$ see an effective net magnetic flux $4\pi |2p+1| \delta$. Due to the projective action of lattice translations, the $f_1$ band structure is $|2p+1|$-fold degenerate and the excess $f_1$ partons form $|2p+1|$ Fermi pockets labeled by $d_i$, each at filling $- \frac{1}{2} \sgn(p)$ relative to the effective magnetic flux $4\pi |2p+1| \delta$ it sees.


Like in Section~\ref{sec:scenario2_FL}, the natural state formed by $d_i$ at intermediate temperatures is a composite Fermi liquid at filling $-\sgn(p)/2$. To describe this state, we further fractionalize $d_i$ as $d_i = z_i \psi_i$ and introduce emergent gauge fields $\alpha_i$ such that $\psi_i$ couples to $\alpha_i$ and $z_i$ couples to $a - \alpha_i$. In terms of these additional fields, the effective Lagrangian takes the form
\begin{equation}
    \begin{aligned}
    L_{\rm eff} &= \sum_{i=1}^{|2p+1|} L[\psi_i, \alpha_i] + \frac{1}{2\pi} (a - \alpha_i) d \beta_i + \frac{2\, \sgn(p)}{4\pi} \beta_i d \beta_i +\frac{1}{4\pi} (b-a) d (b-a) + \frac{1}{4\pi} (A-b) d (A-b) \,. 
    \end{aligned}
\end{equation}
To study Fermi surface dynamics, we again neglect flux quantization and integrate over $a, b$ to simplify the effective Lagrangian 
\begin{equation}
    \begin{aligned}
        &L_{\rm eff} = \sum_{i=1}^{|2p+1|} L[\psi_i, \alpha_i] - \frac{1}{2\pi} \alpha_i d \beta_i + \frac{2 \, \sgn(p) }{4\pi} \beta_i d \beta_i + \frac{1}{8\pi} A d A \\
        &+ \alpha^*_{i,T} \left(\chi q^2 + \gamma \frac{|\omega|}{q}\right) \alpha_{i,T} - \frac{1}{4\pi} a_0 \nabla \times \left(\sum_{i=1}^{|2p+1|} \bs{\alpha_i} + \bs{A}\right) \,. 
    \end{aligned}
\end{equation}
The form of this Lagrangian is very similar to the effective Lagrangian in \eqref{eq:2ndCFL_Lag_simpler} for the $p=-2$ case. The $S_3$ symmetry in \eqref{eq:2ndCFL_Lag_simpler} gets enhanced to an $S_{|2p+1|}$ symmetry at general $p$. Repeating the analysis of pairing instabilities, one finds that all the intrapocket BCS interactions flow to finite fixed-point values, while all the interpocket BCS interactions flow to infinity. Since there is an odd number of pockets, the dominant pairing instability is the one in which all but one of the pockets go through interpocket pairing. However, when microscopic couplings conspire to generate a bare $V_{\rm BCS, intra}$ which is much stronger than $V_{\rm BCS, inter}$, intrapocket pairing occurs and each $d_i$ instead goes into a Moore-Read state. In what follows, we analyze these two cases separately. 

\subsubsection{Interpocket pairing: CDW metal + level-1 IQH state}

The structure of the pairing between $\psi_i, \psi_j$ depends on the sign of $p$. For $p < 0$, each pair $(d_i, d_j)$ can be regarded as a 1/2 + 1/2 quantum Hall bilayer and we assume that the partons $\psi_i, \psi_j$ prefer to pair in the $p_x+ ip_y$ channel. The resulting state is equivalent to an exciton condensate formed by $d_i, d_j$ which generates a level-1 Chern-Simons term for the gauge field $a$ to which they couple (see Appendix~\ref{app:CFpairing_review} for a derivation). On the other hand, for $p < 0$, $d_i$ and $d_j$ now form a $(-1/2) + (-1/2)$ quantum Hall bilayer and the preferred state is the time-reversal conjugate of the original exciton condensate. This means that the induced Chern-Simons term for $a$ has the opposite sign. 

What is the fate of the original Lagrangian after pairs of pockets form exciton condensates? Without loss of generality, let us choose $d_1$ as the unmatched pocket and have $(d_2, d_3), (d_4,d_5), \ldots$ form pairwise exciton condensates. After integrating out all the pockets except $d_1$, we obtain an effective Lagrangian
\begin{equation}
    \begin{aligned}
    L &= L_{\rm CFL, \sgn(p)}[d_1, a] - \sgn(p) \frac{|2p+1|-1}{2} \frac{1}{4\pi} a da \\
    &\hspace{0.5cm} + \frac{p}{4\pi} a da + \frac{1}{4\pi} (b-a) d (b-a) + \frac{1}{4\pi} (A-b) d (A-b) \\
    &= L_{\rm CFL, \sgn(p)}[d_1, a] + \frac{\sgn(p) + 1}{2} \frac{1}{4\pi} a da - \frac{1}{2\pi} b (da + d A) + \frac{2}{4\pi} b db + \frac{1}{4\pi} A d A \,,
    \end{aligned}
\end{equation}
where we introduced the notation
\begin{equation}
    L_{\rm CFL, \pm}[d, a] \rightarrow L[\psi, \alpha] + \frac{1}{2\pi} (a - \alpha) d \beta \pm \frac{2}{4\pi} \beta d \beta \,. 
\end{equation}
For all $p < 0$, $\sgn(p) + 1 = 0$ and the Lagrangian reduces to our result \eqref{eq:CFpairing_2/3_intermediate} for $p = -2$. Therefore, we obtain a Fermi liquid that spontaneously breaks the lattice translation symmetry. On the other hand, when $p > 0$, a level-1 Chern-Simons term for $a$ remains and the total Lagrangian is
\begin{equation}
    \begin{aligned}
    L &= L[\psi, \alpha] + \frac{1}{2\pi} (a-\alpha) d \beta + \frac{2}{4\pi} \beta d \beta + \frac{1}{4\pi} a da - \frac{1}{2\pi} b (da + d A) + \frac{2}{4\pi} bdb + \frac{1}{4\pi} A d A \,. 
    \end{aligned}
\end{equation}
Using the equation of motion for $a$, we can integrate out $a$ and simplify the Lagrangian to
\begin{equation}    
    \begin{aligned}
    L &= L[\psi,\alpha] - \frac{1}{2\pi} \beta d \alpha - \frac{1}{2\pi} b d A + \frac{1}{2\pi} b d \beta + \frac{1}{4\pi} A d A + \frac{1}{4\pi} \beta d \beta + \frac{1}{4\pi} b db \,. 
    \end{aligned}
\end{equation}
Now observe that $b, \beta$ together form the Halperin-111 state, which Higgses $A - \alpha$ and generates a Chern-Simons term for $A$ with level -1. The final Lagrangian is therefore
\begin{equation}
    L = L[\psi, A] + \frac{1}{4\pi} A d A - \frac{1}{4\pi} A d A = L[\psi, A] \,. 
\end{equation}
In contrast to \eqref{eq:2ndCFL_finalFL}, which is valid for $p < 0$, the answer for $p > 0$ is missing a Chern-Simons term for $A$ with level 1. Therefore, the Hall conductivity should be very small at low $\delta$. 

\subsubsection{Intrapocket pairing: PDW superconductor with non-abelian topological order}

We now turn to the second possibility in which secondary composite fermions $\psi_i$ undergo intrapocket pairing in the $p+ip$ channel. The Lagrangian for the $d_i$ sector therefore reduces to $|2p+1|$ copies of the Moore-Read state at filling $-\sgn(p)/2$. Gluing the TQFT for the Moore-Read state back to the remaining abelian Chern-Simons terms, we have the full Lagrangian
\begin{equation}
    \begin{aligned}
        L &= - \sgn(p) \sum_{i=1}^{|2p+1|} \bigg[-\frac{2}{4\pi} \Tr (\tilde c_i d \tilde c_i + \frac{2}{3} \tilde c_i^3) + \frac{2}{4\pi} (\Tr \tilde c_i) d (\Tr \tilde c_i) + \frac{1}{2\pi} (\Tr \tilde c_i) d \tilde b_i - \frac{1}{4\pi} \tilde b_i d \tilde b_i + \frac{1}{2\pi} a d \tilde b_i \bigg] \\
        &\hspace{1cm}+ \frac{p}{4\pi} a da + \frac{1}{4\pi} (b-a) d (b-a) + \frac{1}{4\pi} (A-b) d (A-b) \,.
    \end{aligned}
\end{equation}
Following the strategy in Section~\ref{subsubsec:pairing_intra}, we first integrate over $\tilde b_i$, setting $\tilde b_i = a + \Tr \tilde c_i$: 
\begin{equation}
    \begin{aligned}
        L &= - \sgn(p) \sum_{i=1}^{|2p+1|} \bigg[-\frac{2}{4\pi} \Tr (\tilde c_i d \tilde c_i + \frac{2}{3} \tilde c_i^3) + \frac{3}{4\pi} (\Tr \tilde c_i) d (\Tr \tilde c_i) + \frac{1}{2\pi} a d (\Tr \tilde c_i) \bigg]\\
        &\hspace{1cm} - \frac{p+1}{4\pi} a da + \frac{1}{4\pi} (b-a) d (b-a) + \frac{1}{4\pi} (A-b) d (A-b) \,.
    \end{aligned}
\end{equation}
From here, we can work out the equations of motion for the abelian factors $b, \Tr \tilde c_i, a$:
\begin{equation}\label{eq:MooreRead_23_EOM}
    \begin{aligned}
        2db - da - d A &= 0 \,, \\
        -4 \,\sgn(p) d \Tr \tilde c_i - 2\,\sgn(p) da &= 0 \,, \\
        -\sgn(p) \sum_i d \Tr \tilde c_i - p da - db &= 0 \,. 
    \end{aligned}
\end{equation}
Summing the second equation over $i$ and multiplying the third equation by 4, we obtain
\begin{equation}
    \begin{aligned}
    -4 \, \sgn(p) \sum_i d \Tr \tilde c_i - (4p+2)\, da &= 0 \,, \\ 
    -4 \, \sgn(p) \sum_i d \Tr \tilde c_i - 4p \, da - 4 \, db &= 0 \,. 
    \end{aligned}
\end{equation}
Subtracting these two equations gives $4b - 2 a = 0$. Substituting this relation into the first equation in \eqref{eq:MooreRead_23_EOM}, we immediately conclude that
\begin{equation}
    4b - 2 a = 2 dA = 0 \,. 
\end{equation}
This final equation implies the formation of a charge-2 superconductor. Since each Moore-Read state originates from the pairing of secondary composite fermions $\psi_i, \psi_j$ with finite average momentum, this superconducting state spontaneously breaks lattice translation symmetry and carries a more complicated non-abelian topological order. The anyon content for general $p$ is complicated, but can in principle be worked out using the methods in Appendix~\ref{app:nonabelian_TO}. 

\section{Topological order of the non-abelian superconductor}\label{app:nonabelian_TO}

In this appendix, we provide a more detailed understanding of the non-abelian topological order that accompanies the superconducting condensate in Section~\ref{subsubsec:pairing_intra}. The starting point of our analysis the effective Lagrangian
\begin{equation}
    L_{\rm eff} = L[d_i, a] - \frac{2}{4\pi} a da + \frac{1}{4\pi} (b-a) d (b-a) + \frac{1}{4\pi} (A-b) d (A-b) \,. 
\end{equation}
To obtain the superconductor, each $d_i$ goes into the Moore-Read state, which can be described by a non-abelian Chern-Simons theory $\left[U(2)_{2,-4} \times U(1)_{8}\right]/\mathbb{Z}_2$, where the $\mathbb{Z}_2$ quotient is implemented by gauging a discrete $\mathbb{Z}_2$ 1-form symmetry. The TQFT for $U(2)_{2,-4} \times U(1)_{8}$ can be written as
\begin{equation}
    L_{\rm MR}[c,b,a] = -\frac{2}{4\pi} L_{\rm CS}[c] + \frac{2}{4\pi} \Tr c d \Tr c - \frac{8}{4\pi} bd b - \frac{2}{2\pi} b da \,. 
\end{equation}
On the level of this TQFT, the $\mathbb{Z}_2$ quotient is implemented by declaring that the proper gauge fields are $\tilde c= c - b I$ and $\tilde b = 2 b$. Let us postpone the quotient and couple three copies of this TQFT to the FQAH state first:
\begin{equation}
    \begin{aligned}
    L_{\rm tot} &= \sum_i \left[-\frac{2}{4\pi} L_{\rm CS}[c_i] + \frac{2}{4\pi} \Tr c_i d \Tr c_i - \frac{8}{4\pi} b_id b_i - \frac{2}{2\pi} b_i da \right] \\
    &- \frac{2}{4\pi} a da + \frac{1}{4\pi} (b-a) d (b-a) + \frac{1}{4\pi} (A-b) d (A-b)   \,. 
    \end{aligned}
\end{equation}
Treating $A,b$ as internal gauge fields, we can integrate them out and simplify the Lagrangian to
\begin{equation}
    L_{\rm tot} = \sum_i \left[-\frac{2}{4\pi} L_{\rm CS}[c_i] + \frac{2}{4\pi} \Tr c_i d \Tr c_i - \frac{8}{4\pi} b_id b_i - \frac{2}{2\pi} b_i da \right]  - \frac{2}{4\pi} a da \,. 
\end{equation}
This Lagrangian describes two decoupled sectors. One sector contains three copies of the $U(2)_{2,-4}$ theory and the other sector is an abelian $K$-matrix theory with rank $4$. Each $U(2)_{2,-4}$ theory simply describes an Ising non-abelian topological order with three anyons $\{1, \sigma, \psi\}$. The $K$-matrix sector is more complicated. Via an explicit computation, we can show that
\begin{equation}
    K = \begin{pmatrix}
        -8 & 0 & 0 & 2 \\ 0 & -8 & 0 & 2 \\ 0 & 0 & -8 & 2 \\ 2 & 2 & 2 & -2
    \end{pmatrix} \,, \quad K^{-1} = \begin{pmatrix}
        - \frac{1}{4} & -\frac{1}{8} & -\frac{1}{8} & \frac{1}{2} \\ - \frac{1}{8} & - \frac{1}{4} & - \frac{1}{8} & \frac{1}{2} \\ - \frac{1}{8} & - \frac{1}{8} & - \frac{1}{4} & \frac{1}{2} \\ \frac{1}{2} & \frac{1}{2} & \frac{1}{2} & -2 
    \end{pmatrix} \,, \quad  D_K^2 = |\det K| = 256\,.
\end{equation}
Since the topological order is abelian, this implies that the $K$-matrix sector contains 256 distinct anyons. The physical topological order is obtained by taking the direct product of the $K$ matrix theory and three copies of the Ising topological order, before quotienting by $\mathbb{Z}_2^3$. Each of the Ising topological order has squared quantum dimension $4$, and the quotient by $\mathbb{Z}_2^3$ reduces the quantum dimension by $(2^3)^2 = 2^6$. Therefore, the squared quantum dimension after taking the $\mathbb{Z}_2^3$ quotient must be 
\begin{equation}\label{eq:MR_total_quantum_dimension}
    D^2_{\rm tot} = \frac{256 \cdot 4^3}{2^6} = 256 \,. 
\end{equation}
Our goal is to now enumerate all the physical anyons and reproduce the answer in \eqref{eq:MR_total_quantum_dimension}. We begin by labeling the 1024 potential anyons in the $K$-matrix sector by a 4-dimensional vector $(n_i, n)$ where $n_i \in \{0,\ldots, 7\}$ and $n \in \{0,1\}$. It turns out that this list has a 4-fold redundancy. To see that, we compute the self and mutual statistics between all pairs of anyons:
\begin{equation}
    \begin{aligned}
    \theta_{(n_i, n)} &= \pi (n_i, n)  (n_i, n)^T + n \pi \\
    &=- \frac{\pi}{4} \left[8 n^2 + n_1^2 + n_2^2 + n_3^3 - 4 n (n_1+n_2+n_3) + n_1n_2+n_2n_3+n_1n_3\right] + n \pi \,, \\
    \theta_{(n_i,n), (m_i,m)} &= 2\pi (n_i, n) K^{-1} (m_i, m)^T \\
    &= - \frac{\pi}{4} \bigg[-4n (m_1+m_2+m_3 + 4m) + n_1 (2m_1 + m_2 + m_3 - 4m) \\
    &\hspace{1cm} + n_2 (m_1 + 2m_2 + m_3 - 4m) + n_3 (m_1+m_2+2m_3-4m) \bigg] \,.
    \end{aligned}
\end{equation}
If $n=0, n_1 = n_2 = n_3 = 2$, then $\theta_{(n_i,n)} = \theta_{(n_i,n),(m_i,m)} = 0 \mod 2\pi$ for all choices of $m_i, m$. Therefore, $(2,2,2,0)$ is a trivial boson and tuples $(n_i, n)$ that differ by $(2,2,2,0)$ should be identified. Since $(2,2,2,0)$ has order 4, we conclude that the anyon count gets reduced by a factor of 4. The total number of anyons is therefore $D_K^2 = 8^3 \cdot 2/4 = 256$, which agrees with the $K$-matrix determinant. The complete list of anyons in the $K$-matrix sector can therefore be represented as $v_1^{n_1} v_2^{n_2} v_3^{n_3} v^n$ with $n_i \in \{0,\ldots, 7\}, n \in \{0,1\}$ and subject to the equivalence relation $v_1^2 v_2^2 v_3^2 = \mathbf{1}$. 

Now let us combine the $K$-matrix sector with three copies of $U(2)_{2,-4}$, whose anyons are labeled by $\times_{i=1}^3 \{1_i, \psi_i, \sigma_i\}$. The $\mathbb{Z}_2^3$ quotient corresponds to condensing the anyons $\psi_1 v_1^4, \psi_2 v_2^4, \psi_3 v_3^4$. This condensation identifies pairs of anyons related by $\psi_1 v_1^4, \psi_2 v_2^4, \psi_3 v_3^4$. We are therefore left with a reduced anyon set
\begin{equation}
    \mathcal{A} = \times_i \{1, v_i, v_i^2, v_i^3, \sigma_i, \sigma_i v_i, \sigma_i v_i^2, \sigma_i v_i^3, \sigma_1 v_1^3, \psi_i, \psi_i v_i, \psi_i v_i^2, \psi_i v_i^3\} \times \{n=0, 1\} \,. 
\end{equation}
Next, we need to remove anyons that have nontrivial braiding with the condensing anyons and get confined by the condensation. In the $K$-matrix sector, we have the mutual statistics
\begin{equation}
    \theta_{v_1^4, v_1^{m_1} v_2^{m_2} v_3^{m_3} v^m} = - \pi (2m_1 + m_2 + m_3 - 4m) \,.
\end{equation}
In the non-abelian sector, $\psi_i$ and $\sigma_i$ have a mutual $-1$ braiding phase. The confinement pattern follows from these two pieces of information. If an anyon doesn't contain $\sigma_i$, it is confined by the condensation of $\psi_i v_i^4$ if $\sum_{j \neq i} m_j$ is odd; if an anyon contains $\sigma_i$, it is confined by the condensation of $\psi_i v_i^4$ if $\sum_{j \neq i} m_j$ is even. Therefore, in order for an anyon to remain deconfined, it must satisfy the following conditions
\begin{equation}
    \text{Allowed anyons} = \begin{cases}
        m_1 + m_2, m_1+m_3,m_2+m_3 \text{ even} & \text{no } \sigma_i \\
        m_2+m_3 \text{ odd}, m_1 + m_2, m_1+m_3 \text{ even} & \sigma_1 \\ 
        m_1+m_3 \text{ odd}, m_1 + m_2, m_2+m_3 \text{ even} & \sigma_2 \\ 
        m_1+m_2 \text{ odd}, m_1 + m_3, m_2+m_3 \text{ even} & \sigma_3 \\ 
        m_1+m_3, m_2+m_3 \text{ odd}, m_1 + m_2 \text{ even} & \sigma_1\sigma_2 \\ 
        m_1+m_2, m_2+m_3 \text{ odd}, m_1 + m_3 \text{ even} & \sigma_1\sigma_3 \\ 
        m_1+m_2, m_1+m_3 \text{ odd}, m_1 + m_2 \text{ even} & \sigma_2\sigma_3 \\ 
        m_1+m_3, m_2+m_3, m_1+m_2 \text{ odd} & \sigma_1\sigma_2\sigma_3 \\ 
    \end{cases} \,.
\end{equation}
For an anyon that contains an odd number of $\sigma_i$'s, the constraints on $m_i$ are impossible to satisfy. Therefore, only anyons with an even number of $\sigma_i$'s survive. The list of deconfined anyons therefore simplifies to
\begin{equation}
    \text{Allowed anyons} = \begin{cases}
        m_i \text{ all even or all odd } & \text{no } \sigma_i \\
        m_1+m_3, m_2+m_3 \text{ odd } & \sigma_1\sigma_2 \\ 
        m_1+m_2, m_2+m_3 \text{ odd } & \sigma_1\sigma_3 \\ 
        m_1+m_2, m_1+m_3 \text{ odd } & \sigma_2\sigma_3 
    \end{cases} \,.
\end{equation}
The first case allows $8$ choices in the Ising sector and $16 \times 2$ choices in the abelian sector, giving 256 abelian anyons in total. The second/third/fourth case allows 2 choices in the Ising sector and $16 \times 2$ choices in the abelian sector, giving 64 non-abelian anyons. 

Finally, we have to account for the possibility of splitting, which occurs when the fusion product of two anyons contains the vacuum with multiplicity $N > 1$. It turns out that anyon splitting does not occur in this theory. To see that, it is instructive to consider a representative non-abelian anyon in the $\sigma_1 \sigma_2$ class, $\sigma_1 \sigma_2 v_1 v_2$. By explicit computation, we see that
\begin{equation}
    (\sigma_1 \sigma_2 v_1 v_2) \times (\sigma_1 \sigma_2 v_1^3 v_2^3) = (1 + \psi_1) (1 + \psi_2) v_1^4 v_2^4 = (1 + \psi_1 + \psi_2 + \psi_1 \psi_2) v_1^4 v_2^4 \equiv 1 + \psi_1 \psi_2 + \psi_1 + \psi_2 \,. 
\end{equation}
Since the vacuum appears with multiplicity 1, splitting does not occur for this pair of anyons. Moreover, since four different fusion products appear on the RHS, $\sigma_1 \sigma_2 v_1 v_2$ is a non-abelian anyon with squared quantum dimension 4. A similar calculation shows that this squared dimension is in fact shared by all the non-abelian anyons in this theory. Therefore, the quotient by $\mathbb{Z}_2^3$ does not split any of the non-abelian anyons. After accounting for the 4-fold redundancy in the $K$-matrix sector (i.e. $v_1^2 v_2^2 v_3^2$ is a trivial boson), we arrive at the total quantum dimension
\begin{equation}
    D^2_{\rm tot} = \frac{256 + 3 \cdot (64 \cdot 4)}{4} = 256 \,.
\end{equation}
It is gratifying to see that the explicit enumeration of anyons in this theory matches the quantum dimension formula \eqref{eq:MR_total_quantum_dimension} derived from general arguments. 

\section{Brief review of exciton condensation in quantum Hall bilayers}\label{app:CFpairing_review}

In this appendix, we provide a brief review of exciton condensation in the 1/2 + 1/2 quantum Hall bilayer to justify some of the arguments in Section~\ref{subsec:pairing_2ndCFL} and Appendix~\ref{subsec:general_p_scenario2}. Consider two layers of electrons $d_i$ subject to a strong magnetic field such that both layers are at half-filling of the lowest Landau level. In the absence of interlayer interactions, each of the two layers forms a CFL. To describe the CFL, we fractionalize the electrons as $d_i = z_i \psi_i$ and put $z_i$ in a $U(1)_{2}$ state. The total Lagrangian then takes the form
\begin{equation}
    L = \sum_{i=1}^2 \left[L[\psi_i, a_i] - \frac{2}{4\pi} b_i db_i + \frac{1}{2\pi} b_i d (A_i - a_i)\right] \,,
\end{equation}
where $A_1, A_2$ are external electromagnetic fields that can be applied to the two layers separately. A renormalization group analysis in~\cite{Sodemann2017_CFLbilayer} shows that there is a tendency towards pairing between the composite fermions $\psi_1, \psi_2$. However, the RG analysis does not fix the pairing channel. When the interlayer distance $d$ is comparable or smaller than the magnetic length $l_B$, numerical evidence from~\cite{Milovanovic2007_CFpairing,Moller2008_CFpairing,Moller2009_CFpairing,Milovanovic2015_CFpairing} suggests that the preferred pairing channel is $p+ip$. We will assume that the same channel persists at arbitrary $d/l_B$.

What are the universal properties of the paired state? As shown in~\cite{Sodemann2017_CFLbilayer}, the pairing between $\psi_1, \psi_2$ in the $p+ip$ channel generates the TQFT
\begin{equation}
    \frac{1}{4\pi}(\beta_1 - \beta_2) d (\beta_1 - \beta_2) + \frac{1}{2\pi} a_1 d \beta_1 + \frac{1}{2\pi} a_2 d \beta_2 \,.
\end{equation}
If we glue this TQFT onto the $U(1)_{2}$ state formed by $z_i$, we obtain
\begin{equation}
    L = \frac{1}{4\pi}(\beta_1 - \beta_2) d (\beta_1 - \beta_2) + \sum_i \left[\frac{1}{2\pi} a_i d \beta_i - \frac{2}{4\pi} b_i db_i + \frac{1}{2\pi} b_i d (A_i - a_i) \right] \,. 
\end{equation}
The equations of motion for $a_i$ enforce the constraint $b_i = \beta_i$. After integrating out $a_i, b_i$, we land on the Halperin-111 state:
\begin{equation}
    L = - \frac{1}{4\pi} (\beta_1 + \beta_2) d (\beta_1 + \beta_2) + \frac{1}{2\pi} \beta_1 d A_1 + \frac{1}{2\pi} \beta_2 d A_2 \,. 
\end{equation}
Finally, integrating over $\beta_1$ enforces the constraint $\beta_1 + \beta_2 = A_1$ and simplifies the Lagrangian to
\begin{equation}
    L = - \frac{1}{4\pi} A_1 d A_1 + \frac{1}{2\pi} (A_1 - \beta_2) d A_1 + \frac{1}{2\pi} \beta_2 d A_2 = \frac{1}{4\pi} A_1 d A_1 + \frac{1}{2\pi} \beta_2 (A_2 - A_1) \,. 
\end{equation}
In this final Lagrangian, $\beta_2$ acts as a Lagrange multiplier that Higgses $A_2 - A_1$. This signals the condensation of an interlayer exciton. The remaining level = 1 Chern-Simons term describes the quantized Hall conductance of the exciton condensate in response to a total electromagnetic field. 

The discussion so far assumes that $d_1, d_2$ are physical electrons. To connect with Section~\ref{subsec:scenario2_CFLformation}, we need to treat $d_1, d_2$ as composite fermions and identify $A_1, A_2$ with a single fluctuating gauge field $a$. When $p < 0$, $d_1, d_2$ are at half-filling relative to the flux they see and the above discussion for 1/2 + 1/2 bilayer applies. The exciton condensate generates a level = 1 Chern-Simons term for $a$, which is the answer quoted in the main text. When $p > 0$, each of the $d_i$ fermions is instead at $-1/2$ filling relative to the flux it sees. Such a system is related to the 1/2 + 1/2 quantum Hall bilayer by time-reversal and generates a Chern-Simons term for $a$ with level ${= -1}$ instead.

\section{Transport properties of the odd-p CFL state}\label{app:transport_oddp_CFL}

In this appendix, we derive the transport properties associated with the generalized CFL state in \eqref{eq:scenario1_oddp_CFL}. We start with the generalized CFL Lagrangian
\begin{equation}
    \begin{aligned}
    L &= \frac{p}{4\pi} a da + \frac{1}{4\pi} (b-a) d (b-a) + \frac{1}{4\pi} (A-b) d (A-b) + \sum_{i=1}^{|2p+1|} L_{\rm CFL}[\psi_i, A-b] \,, 
    \end{aligned}
\end{equation}
where
\begin{equation}
    \begin{aligned}
    &L_{\rm CFL}[\psi_i, A-b] = L[f_i, A - b - \alpha_i] + \frac{|p+1|}{4\pi} \beta_i d \beta_i + \frac{1}{2\pi} \beta_i d \alpha_i \,. 
    \end{aligned}
\end{equation}
To compute the conductivity, we neglect flux quantization and integrate out the gauge field $a$ which only appears in Chern-Simons terms. Using the equations of motion $(p+1) da = db$ for $a$, the remaining Lagrangian takes the form 
\begin{equation}
    L = \frac{1}{4\pi (1-\nu)} b db - \frac{1}{2\pi} A db + \frac{1}{4\pi} A d A + \sum_{i=1}^{|2p+1|} L_{\rm CFL}[\psi_i, A - b] \,.
\end{equation}
Due to the $S_{|2p+1|}$ permutation symmetry of the different CFL species, the expectation value of the total CFL current $\ev{J_{\psi}} = \sum_i \ev{J_{\psi_i}}$ is equal to $|2p+1| \ev{J_{\psi_i}}$. Let $\Pi_{\psi} = |2p+1| \Pi_{\psi_i}$ be the total response function of the CFL and let $\Pi_1$ be the response function of a level-1 IQH state. Then we can integrate out all the fermions $\psi_i$ to generate an effective Lagrangian involving only $b, A$:
\begin{equation}
    \begin{aligned}
    L &= \frac{1}{2} b^* (1-\nu)^{-1} \Pi_1 b - b^* \Pi_1 A + \frac{1}{2} A^* \Pi_1 A + \frac{1}{2} (A-b)^* \Pi_{\psi} (A-b) \\
    &= \frac{1}{2} b^* \left[(1-\nu)^{-1} \Pi_1 + \Pi_{\psi}\right] b - b^* (\Pi_1 + \Pi_{\psi}) A  + \frac{1}{2} A^* (\Pi_1 + \Pi_{\psi}) A \,. 
    \end{aligned}
\end{equation}
In the above expression, we have passed to momentum space so that $b, b^{\dagger}$ should be interpreted as $b_{\nu}(q), b^*_{\mu}(q)$ and $\Pi_1$ should be interpreted as $\Pi_1^{\mu\nu}(q)$. The equations of motion allow us to relate $b$ and $A$
\begin{equation}
    b =  \left[(1-\nu)^{-1} \Pi_1 + \Pi_{\psi}\right]^{-1} (\Pi_1 + \Pi_{\psi}) A \,. 
\end{equation}
Plugging this relation back into the Lagrangian, we find
\begin{equation}
    \begin{aligned}
    L[A] &= - \frac{1}{2} A^T (\Pi_1 + \Pi_{\psi})^T \left[(1-\nu)^{-1} \Pi_1 + \Pi_{\psi}\right]^{T, -1} (\Pi_1+\Pi_{\psi}) A + \frac{1}{2} A^T (\Pi_1 + \Pi_{\psi}) A \\
    &= \frac{\nu}{2} A^T \Pi_1^T \left[\Pi_1^T + (1-\nu) \Pi_{\psi}^T\right]^{-1} (\Pi_1 + \Pi_{\psi}) A \,. 
    \end{aligned}
\end{equation}
Now differentiating with respect to $A$ twice and then taking the matrix inverse, we obtain the physical resistivity
\begin{equation}
    \rho_{\rm phy} = \nu^{-1} (\sigma_1 + \sigma_{\psi})^{-1} \left[\sigma_1^T + (1-\nu) \sigma_{\psi}^T\right] (\sigma_1^T)^{-1} = \nu^{-1} (\sigma_1 + \sigma_{\psi})^{-1} - \frac{1-\nu}{\nu} (\sigma_1 + \sigma_{\psi})^{-1} (\sigma_{\psi}^T) \sigma_1^{-1} \,,
\end{equation}
where in the last line, we used the fact that $\sigma_1^T = - \sigma_1$ for the level-1 IQH conductivity. Finally, let us recall that the composite fermion conductivity at filling $-1/|p+1|$ is given by the standard Ioffe-Larkin rule
\begin{equation}
    \sigma_{\psi_i} = \frac{1}{|2p+1|} \sigma_{\psi} = \left(\sigma_{-|p+1|}^{-1} + \sigma_{\rm FL}^{-1}\right)^{-1} \,,
\end{equation}
where $\sigma_{\rm FL}$ is the Fermi liquid conductivity associated with each $f_i$ and $\sigma_{-|p+1|}$ is the conductivity of the $U(1)_{-|p+1|}$ Laughlin state which can be related to the IQH conductivity as
\begin{equation}
    \sigma_{-|p+1|} = - \frac{1}{|p+1|} \sigma_1 \,. 
\end{equation}
Putting everything together, we recover the formula quoted in the main text:
\begin{equation}
    \sigma_{\psi} = |2p+1| \left(- |p+1| \sigma_1^{-1} + \sigma_{\rm FL}^{-1} \right)^{-1}
\end{equation}
In the presence of momentum relaxation, $\sigma_{\rm FL}$ scales as $\frac{\tau_{\rm tr} \delta}{|p| m_{\rm tr}}$ where $m_{\rm tr}$ is the transport effective mass at the top of the active band. If we take the clean limit where $\tau_{\rm tr} \rightarrow \infty$, then $\sigma_{\rm FL}$ diverges and $\sigma_{\psi} \approx - (1-\nu)^{-1} \sigma_1$. This implies the resistivity
\begin{equation}
    \rho_{\rm phy}(\tau_{\rm tr} \rightarrow \infty) \approx (1-\nu)^{-1} \sigma_1^{-1} - \frac{1-\nu}{\nu} (1-\nu)^{-1} \sigma_1^{-1} (1-\nu)^{-1} \sigma_1 \sigma_1^{-1} = \nu^{-1} \frac{1 + \nu}{1 - \nu} \sigma_1^{-1} \,. 
\end{equation}
For $0 < \nu < 1$, the Hall resistivity is enhanced by a positive factor $\frac{1 + \nu}{1 - \nu}$ in the ultra-clean limit. If we instead keep $\tau_{\rm tr}$ finite and take the low density limit $\delta \ll 1$, then $\sigma_{\psi} \approx 0$ and we recover the resistivity for the Jain state at filling $\nu$:
\begin{equation}
    \rho_{\rm phy} (\delta \rightarrow 0) \approx \nu^{-1} \sigma_1^{-1} \,.
\end{equation}

\bibliography{dope_FQAH}

\end{document}

%% file: ACPreamble.tex
\usepackage{graphicx}
\usepackage{dcolumn}
\usepackage{bm}
\usepackage{multirow}

\usepackage[normalem]{ulem}

\usepackage{amsmath}
\usepackage{amsthm}
\usepackage{amstext}
\usepackage{amssymb}
\usepackage{mathrsfs}
\usepackage{amsfonts}
\usepackage{amsbsy} 

\usepackage{braket}

\usepackage[all,matrix,cmtip]{xy}
\usepackage{mathtools}

\usepackage{color}
\definecolor{red}{rgb}{1,0,0}
\definecolor{blue}{rgb}{0,0,1}
\definecolor{dblue}{rgb}{0,0,0.4}
\definecolor{green}{rgb}{0,1,0}
\definecolor{black}{rgb}{0,0,0}
\definecolor{white}{rgb}{1,1,1}
\definecolor{pastelblue}{RGB}{20,93,160}

\definecolor{brn}{rgb}{.8,.4,.0}
\definecolor{redo}{rgb}{1,.5,.0}
\definecolor{ddgrn}{rgb}{0,0.4,0}
\definecolor{dgrn}{rgb}{0,0.55,0}
\definecolor{dbl}{rgb}{0,0,0.5}

\usepackage[colorlinks,citecolor=pastelblue,linkcolor=pastelblue,urlcolor=pastelblue]{hyperref}

\newcommand{\sgn}{{\rm sgn}}

\newcommand{\bpm}{\begin{pmatrix}}
	\newcommand{\epm}{\end{pmatrix}}
\newcommand{\bmm}{\begin{matrix}}
	\newcommand{\emm}{\end{matrix}}
\newcommand{\bvm}{\begin{vmatrix}}
	\newcommand{\evm}{\end{vmatrix}}

\usepackage{euscript}

\makeatletter
\newsavebox{\@brx}
\newcommand{\llangle}[1][]{\savebox{\@brx}{\(\m@th{#1\langle}\)}%
	\mathopen{\copy\@brx\kern-0.5\wd\@brx\usebox{\@brx}}}
\newcommand{\rrangle}[1][]{\savebox{\@brx}{\(\m@th{#1\rangle}\)}%
	\mathclose{\copy\@brx\kern-0.5\wd\@brx\usebox{\@brx}}}
\makeatother

%% file: draft_prx_accepted.bbl
\begin{thebibliography}{75}%
\makeatletter
\providecommand \@ifxundefined [1]{%
 \@ifx{#1\undefined}
}%
\providecommand \@ifnum [1]{%
 \ifnum #1\expandafter \@firstoftwo
 \else \expandafter \@secondoftwo
 \fi
}%
\providecommand \@ifx [1]{%
 \ifx #1\expandafter \@firstoftwo
 \else \expandafter \@secondoftwo
 \fi
}%
\providecommand \natexlab [1]{#1}%
\providecommand \enquote  [1]{``#1''}%
\providecommand \bibnamefont  [1]{#1}%
\providecommand \bibfnamefont [1]{#1}%
\providecommand \citenamefont [1]{#1}%
\providecommand \href@noop [0]{\@secondoftwo}%
\providecommand \href [0]{\begingroup \@sanitize@url \@href}%
\providecommand \@href[1]{\@@startlink{#1}\@@href}%
\providecommand \@@href[1]{\endgroup#1\@@endlink}%
\providecommand \@sanitize@url [0]{\catcode `\\12\catcode `\$12\catcode
  `\&12\catcode `\#12\catcode `\^12\catcode `\_12\catcode `\%12\relax}%
\providecommand \@@startlink[1]{}%
\providecommand \@@endlink[0]{}%
\providecommand \url  [0]{\begingroup\@sanitize@url \@url }%
\providecommand \@url [1]{\endgroup\@href {#1}{\urlprefix }}%
\providecommand \urlprefix  [0]{URL }%
\providecommand \Eprint [0]{\href }%
\providecommand \doibase [0]{https://doi.org/}%
\providecommand \selectlanguage [0]{\@gobble}%
\providecommand \bibinfo  [0]{\@secondoftwo}%
\providecommand \bibfield  [0]{\@secondoftwo}%
\providecommand \translation [1]{[#1]}%
\providecommand \BibitemOpen [0]{}%
\providecommand \bibitemStop [0]{}%
\providecommand \bibitemNoStop [0]{.\EOS\space}%
\providecommand \EOS [0]{\spacefactor3000\relax}%
\providecommand \BibitemShut  [1]{\csname bibitem#1\endcsname}%
\let\auto@bib@innerbib\@empty
\bibitem [{\citenamefont {Laughlin}(1983)}]{Laughlin1983_FQHtheory}%
  \BibitemOpen
  \bibfield  {author} {\bibinfo {author} {\bibfnamefont {R.~B.}\ \bibnamefont
  {Laughlin}},\ }\bibfield  {title} {\bibinfo {title} {Anomalous quantum hall
  effect: An incompressible quantum fluid with fractionally charged
  excitations},\ }\href {https://doi.org/10.1103/PhysRevLett.50.1395}
  {\bibfield  {journal} {\bibinfo  {journal} {Phys. Rev. Lett.}\ }\textbf
  {\bibinfo {volume} {50}},\ \bibinfo {pages} {1395} (\bibinfo {year}
  {1983})}\BibitemShut {NoStop}%
\bibitem [{\citenamefont {Jain}(1989)}]{Jain1989_CFframework}%
  \BibitemOpen
  \bibfield  {author} {\bibinfo {author} {\bibfnamefont {J.~K.}\ \bibnamefont
  {Jain}},\ }\bibfield  {title} {\bibinfo {title} {Composite-fermion approach
  for the fractional quantum hall effect},\ }\href
  {https://doi.org/10.1103/PhysRevLett.63.199} {\bibfield  {journal} {\bibinfo
  {journal} {Phys. Rev. Lett.}\ }\textbf {\bibinfo {volume} {63}},\ \bibinfo
  {pages} {199} (\bibinfo {year} {1989})}\BibitemShut {NoStop}%
\bibitem [{\citenamefont {Stormer}\ \emph {et~al.}(1999)\citenamefont
  {Stormer}, \citenamefont {Tsui},\ and\ \citenamefont
  {Gossard}}]{Stormer1999_FQHreview}%
  \BibitemOpen
  \bibfield  {author} {\bibinfo {author} {\bibfnamefont {H.~L.}\ \bibnamefont
  {Stormer}}, \bibinfo {author} {\bibfnamefont {D.~C.}\ \bibnamefont {Tsui}},\
  and\ \bibinfo {author} {\bibfnamefont {A.~C.}\ \bibnamefont {Gossard}},\
  }\bibfield  {title} {\bibinfo {title} {The fractional quantum hall effect},\
  }\href {https://doi.org/10.1103/RevModPhys.71.S298} {\bibfield  {journal}
  {\bibinfo  {journal} {Rev. Mod. Phys.}\ }\textbf {\bibinfo {volume} {71}},\
  \bibinfo {pages} {S298} (\bibinfo {year} {1999})}\BibitemShut {NoStop}%
\bibitem [{\citenamefont {Halperin}\ \emph {et~al.}(1993)\citenamefont
  {Halperin}, \citenamefont {Lee},\ and\ \citenamefont
  {Read}}]{Halperin1993_HLRtheory}%
  \BibitemOpen
  \bibfield  {author} {\bibinfo {author} {\bibfnamefont {B.~I.}\ \bibnamefont
  {Halperin}}, \bibinfo {author} {\bibfnamefont {P.~A.}\ \bibnamefont {Lee}},\
  and\ \bibinfo {author} {\bibfnamefont {N.}~\bibnamefont {Read}},\ }\bibfield
  {title} {\bibinfo {title} {Theory of the half-filled landau level},\ }\href
  {https://doi.org/10.1103/PhysRevB.47.7312} {\bibfield  {journal} {\bibinfo
  {journal} {Phys. Rev. B}\ }\textbf {\bibinfo {volume} {47}},\ \bibinfo
  {pages} {7312} (\bibinfo {year} {1993})}\BibitemShut {NoStop}%
\bibitem [{\citenamefont {Thouless}\ \emph {et~al.}(1982)\citenamefont
  {Thouless}, \citenamefont {Kohmoto}, \citenamefont {Nightingale},\ and\
  \citenamefont {den Nijs}}]{Thouless1982_TKNN}%
  \BibitemOpen
  \bibfield  {author} {\bibinfo {author} {\bibfnamefont {D.~J.}\ \bibnamefont
  {Thouless}}, \bibinfo {author} {\bibfnamefont {M.}~\bibnamefont {Kohmoto}},
  \bibinfo {author} {\bibfnamefont {M.~P.}\ \bibnamefont {Nightingale}},\ and\
  \bibinfo {author} {\bibfnamefont {M.}~\bibnamefont {den Nijs}},\ }\bibfield
  {title} {\bibinfo {title} {Quantized hall conductance in a two-dimensional
  periodic potential},\ }\href {https://doi.org/10.1103/PhysRevLett.49.405}
  {\bibfield  {journal} {\bibinfo  {journal} {Phys. Rev. Lett.}\ }\textbf
  {\bibinfo {volume} {49}},\ \bibinfo {pages} {405} (\bibinfo {year}
  {1982})}\BibitemShut {NoStop}%
\bibitem [{\citenamefont {Haldane}(1988)}]{Haldane1988_QAH}%
  \BibitemOpen
  \bibfield  {author} {\bibinfo {author} {\bibfnamefont {F.~D.~M.}\
  \bibnamefont {Haldane}},\ }\bibfield  {title} {\bibinfo {title} {Model for a
  quantum hall effect without landau levels: Condensed-matter realization of
  the "parity anomaly"},\ }\href {https://doi.org/10.1103/PhysRevLett.61.2015}
  {\bibfield  {journal} {\bibinfo  {journal} {Phys. Rev. Lett.}\ }\textbf
  {\bibinfo {volume} {61}},\ \bibinfo {pages} {2015} (\bibinfo {year}
  {1988})}\BibitemShut {NoStop}%
\bibitem [{\citenamefont {{Neupert}}\ \emph {et~al.}(2011)\citenamefont
  {{Neupert}}, \citenamefont {{Santos}}, \citenamefont {{Chamon}},\ and\
  \citenamefont {{Mudry}}}]{Neupert2010_FQAH}%
  \BibitemOpen
  \bibfield  {author} {\bibinfo {author} {\bibfnamefont {T.}~\bibnamefont
  {{Neupert}}}, \bibinfo {author} {\bibfnamefont {L.}~\bibnamefont {{Santos}}},
  \bibinfo {author} {\bibfnamefont {C.}~\bibnamefont {{Chamon}}},\ and\
  \bibinfo {author} {\bibfnamefont {C.}~\bibnamefont {{Mudry}}},\ }\bibfield
  {title} {\bibinfo {title} {{Fractional Quantum Hall States at Zero Magnetic
  Field}},\ }\href {https://doi.org/10.1103/PhysRevLett.106.236804} {\bibfield
  {journal} {\bibinfo  {journal} {\prl}\ }\textbf {\bibinfo {volume} {106}},\
  \bibinfo {eid} {236804} (\bibinfo {year} {2011})},\ \Eprint
  {https://arxiv.org/abs/1012.4723} {arXiv:1012.4723 [cond-mat.str-el]}
  \BibitemShut {NoStop}%
\bibitem [{\citenamefont {Sun}\ \emph {et~al.}(2011)\citenamefont {Sun},
  \citenamefont {Gu}, \citenamefont {Katsura},\ and\ \citenamefont
  {Das~Sarma}}]{Sun2011_FCI}%
  \BibitemOpen
  \bibfield  {author} {\bibinfo {author} {\bibfnamefont {K.}~\bibnamefont
  {Sun}}, \bibinfo {author} {\bibfnamefont {Z.}~\bibnamefont {Gu}}, \bibinfo
  {author} {\bibfnamefont {H.}~\bibnamefont {Katsura}},\ and\ \bibinfo {author}
  {\bibfnamefont {S.}~\bibnamefont {Das~Sarma}},\ }\bibfield  {title} {\bibinfo
  {title} {Nearly flatbands with nontrivial topology},\ }\href
  {https://doi.org/10.1103/PhysRevLett.106.236803} {\bibfield  {journal}
  {\bibinfo  {journal} {Phys. Rev. Lett.}\ }\textbf {\bibinfo {volume} {106}},\
  \bibinfo {pages} {236803} (\bibinfo {year} {2011})}\BibitemShut {NoStop}%
\bibitem [{\citenamefont {{Sheng}}\ \emph {et~al.}(2011)\citenamefont
  {{Sheng}}, \citenamefont {{Gu}}, \citenamefont {{Sun}},\ and\ \citenamefont
  {{Sheng}}}]{Sheng2011_FCI}%
  \BibitemOpen
  \bibfield  {author} {\bibinfo {author} {\bibfnamefont {D.~N.}\ \bibnamefont
  {{Sheng}}}, \bibinfo {author} {\bibfnamefont {Z.-C.}\ \bibnamefont {{Gu}}},
  \bibinfo {author} {\bibfnamefont {K.}~\bibnamefont {{Sun}}},\ and\ \bibinfo
  {author} {\bibfnamefont {L.}~\bibnamefont {{Sheng}}},\ }\bibfield  {title}
  {\bibinfo {title} {{Fractional quantum Hall effect in the absence of Landau
  levels}},\ }\href {https://doi.org/10.1038/ncomms1380} {\bibfield  {journal}
  {\bibinfo  {journal} {Nature Communications}\ }\textbf {\bibinfo {volume}
  {2}},\ \bibinfo {eid} {389} (\bibinfo {year} {2011})},\ \Eprint
  {https://arxiv.org/abs/1102.2658} {arXiv:1102.2658 [cond-mat.str-el]}
  \BibitemShut {NoStop}%
\bibitem [{\citenamefont {{Tang}}\ \emph {et~al.}(2011)\citenamefont {{Tang}},
  \citenamefont {{Mei}},\ and\ \citenamefont {{Wen}}}]{Tang2011_FQAH}%
  \BibitemOpen
  \bibfield  {author} {\bibinfo {author} {\bibfnamefont {E.}~\bibnamefont
  {{Tang}}}, \bibinfo {author} {\bibfnamefont {J.-W.}\ \bibnamefont {{Mei}}},\
  and\ \bibinfo {author} {\bibfnamefont {X.-G.}\ \bibnamefont {{Wen}}},\
  }\bibfield  {title} {\bibinfo {title} {{High-Temperature Fractional Quantum
  Hall States}},\ }\href {https://doi.org/10.1103/PhysRevLett.106.236802}
  {\bibfield  {journal} {\bibinfo  {journal} {\prl}\ }\textbf {\bibinfo
  {volume} {106}},\ \bibinfo {eid} {236802} (\bibinfo {year} {2011})},\ \Eprint
  {https://arxiv.org/abs/1012.2930} {arXiv:1012.2930 [cond-mat.str-el]}
  \BibitemShut {NoStop}%
\bibitem [{\citenamefont {{Wang}}\ \emph {et~al.}(2011)\citenamefont {{Wang}},
  \citenamefont {{Gu}}, \citenamefont {{Gong}},\ and\ \citenamefont
  {{Sheng}}}]{Wang2011_FQAH}%
  \BibitemOpen
  \bibfield  {author} {\bibinfo {author} {\bibfnamefont {Y.-F.}\ \bibnamefont
  {{Wang}}}, \bibinfo {author} {\bibfnamefont {Z.-C.}\ \bibnamefont {{Gu}}},
  \bibinfo {author} {\bibfnamefont {C.-D.}\ \bibnamefont {{Gong}}},\ and\
  \bibinfo {author} {\bibfnamefont {D.~N.}\ \bibnamefont {{Sheng}}},\
  }\bibfield  {title} {\bibinfo {title} {{Fractional Quantum Hall Effect of
  Hard-Core Bosons in Topological Flat Bands}},\ }\href
  {https://doi.org/10.1103/PhysRevLett.107.146803} {\bibfield  {journal}
  {\bibinfo  {journal} {\prl}\ }\textbf {\bibinfo {volume} {107}},\ \bibinfo
  {eid} {146803} (\bibinfo {year} {2011})},\ \Eprint
  {https://arxiv.org/abs/1103.1686} {arXiv:1103.1686 [cond-mat.str-el]}
  \BibitemShut {NoStop}%
\bibitem [{\citenamefont {{Regnault}}\ and\ \citenamefont
  {{Bernevig}}(2011)}]{Regnault2011_FCI}%
  \BibitemOpen
  \bibfield  {author} {\bibinfo {author} {\bibfnamefont {N.}~\bibnamefont
  {{Regnault}}}\ and\ \bibinfo {author} {\bibfnamefont {B.~A.}\ \bibnamefont
  {{Bernevig}}},\ }\bibfield  {title} {\bibinfo {title} {{Fractional Chern
  Insulator}},\ }\href {https://doi.org/10.1103/PhysRevX.1.021014} {\bibfield
  {journal} {\bibinfo  {journal} {Physical Review X}\ }\textbf {\bibinfo
  {volume} {1}},\ \bibinfo {eid} {021014} (\bibinfo {year} {2011})},\ \Eprint
  {https://arxiv.org/abs/1105.4867} {arXiv:1105.4867 [cond-mat.str-el]}
  \BibitemShut {NoStop}%
\bibitem [{\citenamefont {{Bergholtz}}\ and\ \citenamefont
  {{Liu}}(2013)}]{Bergholtz2013_FCIreview}%
  \BibitemOpen
  \bibfield  {author} {\bibinfo {author} {\bibfnamefont {E.~J.}\ \bibnamefont
  {{Bergholtz}}}\ and\ \bibinfo {author} {\bibfnamefont {Z.}~\bibnamefont
  {{Liu}}},\ }\bibfield  {title} {\bibinfo {title} {{Topological Flat Band
  Models and Fractional Chern Insulators}},\ }\href
  {https://doi.org/10.1142/S021797921330017X} {\bibfield  {journal} {\bibinfo
  {journal} {International Journal of Modern Physics B}\ }\textbf {\bibinfo
  {volume} {27}},\ \bibinfo {eid} {1330017} (\bibinfo {year} {2013})},\ \Eprint
  {https://arxiv.org/abs/1308.0343} {arXiv:1308.0343 [cond-mat.str-el]}
  \BibitemShut {NoStop}%
\bibitem [{\citenamefont {{Parameswaran}}\ \emph {et~al.}(2013)\citenamefont
  {{Parameswaran}}, \citenamefont {{Roy}},\ and\ \citenamefont
  {{Sondhi}}}]{Parameswaran2013_FCIreview}%
  \BibitemOpen
  \bibfield  {author} {\bibinfo {author} {\bibfnamefont {S.~A.}\ \bibnamefont
  {{Parameswaran}}}, \bibinfo {author} {\bibfnamefont {R.}~\bibnamefont
  {{Roy}}},\ and\ \bibinfo {author} {\bibfnamefont {S.~L.}\ \bibnamefont
  {{Sondhi}}},\ }\bibfield  {title} {\bibinfo {title} {{Fractional quantum Hall
  physics in topological flat bands}},\ }\href
  {https://doi.org/10.1016/j.crhy.2013.04.003} {\bibfield  {journal} {\bibinfo
  {journal} {Comptes Rendus Physique}\ }\textbf {\bibinfo {volume} {14}},\
  \bibinfo {pages} {816} (\bibinfo {year} {2013})},\ \Eprint
  {https://arxiv.org/abs/1302.6606} {arXiv:1302.6606 [cond-mat.str-el]}
  \BibitemShut {NoStop}%
\bibitem [{\citenamefont {{Zhang}}\ \emph {et~al.}(2019)\citenamefont
  {{Zhang}}, \citenamefont {{Mao}}, \citenamefont {{Cao}}, \citenamefont
  {{Jarillo-Herrero}},\ and\ \citenamefont
  {{Senthil}}}]{Zhang2018_nearly_flat}%
  \BibitemOpen
  \bibfield  {author} {\bibinfo {author} {\bibfnamefont {Y.-H.}\ \bibnamefont
  {{Zhang}}}, \bibinfo {author} {\bibfnamefont {D.}~\bibnamefont {{Mao}}},
  \bibinfo {author} {\bibfnamefont {Y.}~\bibnamefont {{Cao}}}, \bibinfo
  {author} {\bibfnamefont {P.}~\bibnamefont {{Jarillo-Herrero}}},\ and\
  \bibinfo {author} {\bibfnamefont {T.}~\bibnamefont {{Senthil}}},\ }\bibfield
  {title} {\bibinfo {title} {{Nearly flat Chern bands in moir{\'e}
  superlattices}},\ }\href {https://doi.org/10.1103/PhysRevB.99.075127}
  {\bibfield  {journal} {\bibinfo  {journal} {\prb}\ }\textbf {\bibinfo
  {volume} {99}},\ \bibinfo {eid} {075127} (\bibinfo {year} {2019})},\ \Eprint
  {https://arxiv.org/abs/1805.08232} {arXiv:1805.08232 [cond-mat.str-el]}
  \BibitemShut {NoStop}%
\bibitem [{\citenamefont {{Ledwith}}\ \emph {et~al.}(2020)\citenamefont
  {{Ledwith}}, \citenamefont {{Tarnopolsky}}, \citenamefont {{Khalaf}},\ and\
  \citenamefont {{Vishwanath}}}]{Ledwith2019_FCITBG}%
  \BibitemOpen
  \bibfield  {author} {\bibinfo {author} {\bibfnamefont {P.~J.}\ \bibnamefont
  {{Ledwith}}}, \bibinfo {author} {\bibfnamefont {G.}~\bibnamefont
  {{Tarnopolsky}}}, \bibinfo {author} {\bibfnamefont {E.}~\bibnamefont
  {{Khalaf}}},\ and\ \bibinfo {author} {\bibfnamefont {A.}~\bibnamefont
  {{Vishwanath}}},\ }\bibfield  {title} {\bibinfo {title} {{Fractional Chern
  insulator states in twisted bilayer graphene: An analytical approach}},\
  }\href {https://doi.org/10.1103/PhysRevResearch.2.023237} {\bibfield
  {journal} {\bibinfo  {journal} {Physical Review Research}\ }\textbf {\bibinfo
  {volume} {2}},\ \bibinfo {eid} {023237} (\bibinfo {year} {2020})},\ \Eprint
  {https://arxiv.org/abs/1912.09634} {arXiv:1912.09634 [cond-mat.str-el]}
  \BibitemShut {NoStop}%
\bibitem [{\citenamefont {{Repellin}}\ and\ \citenamefont
  {{Senthil}}(2020)}]{Repellin2019_TBGChern}%
  \BibitemOpen
  \bibfield  {author} {\bibinfo {author} {\bibfnamefont {C.}~\bibnamefont
  {{Repellin}}}\ and\ \bibinfo {author} {\bibfnamefont {T.}~\bibnamefont
  {{Senthil}}},\ }\bibfield  {title} {\bibinfo {title} {{Chern bands of twisted
  bilayer graphene: Fractional Chern insulators and spin phase transition}},\
  }\href {https://doi.org/10.1103/PhysRevResearch.2.023238} {\bibfield
  {journal} {\bibinfo  {journal} {Physical Review Research}\ }\textbf {\bibinfo
  {volume} {2}},\ \bibinfo {eid} {023238} (\bibinfo {year} {2020})},\ \Eprint
  {https://arxiv.org/abs/1912.11469} {arXiv:1912.11469 [cond-mat.str-el]}
  \BibitemShut {NoStop}%
\bibitem [{\citenamefont {{Wilhelm}}\ \emph {et~al.}(2021)\citenamefont
  {{Wilhelm}}, \citenamefont {{Lang}},\ and\ \citenamefont
  {{L{\"a}uchli}}}]{Wilhelm2020_TBGFCI}%
  \BibitemOpen
  \bibfield  {author} {\bibinfo {author} {\bibfnamefont {P.}~\bibnamefont
  {{Wilhelm}}}, \bibinfo {author} {\bibfnamefont {T.~C.}\ \bibnamefont
  {{Lang}}},\ and\ \bibinfo {author} {\bibfnamefont {A.~M.}\ \bibnamefont
  {{L{\"a}uchli}}},\ }\bibfield  {title} {\bibinfo {title} {{Interplay of
  fractional Chern insulator and charge density wave phases in twisted bilayer
  graphene}},\ }\href {https://doi.org/10.1103/PhysRevB.103.125406} {\bibfield
  {journal} {\bibinfo  {journal} {\prb}\ }\textbf {\bibinfo {volume} {103}},\
  \bibinfo {eid} {125406} (\bibinfo {year} {2021})},\ \Eprint
  {https://arxiv.org/abs/2012.09829} {arXiv:2012.09829 [cond-mat.str-el]}
  \BibitemShut {NoStop}%
\bibitem [{\citenamefont {{Abouelkomsan}}\ \emph {et~al.}(2020)\citenamefont
  {{Abouelkomsan}}, \citenamefont {{Liu}},\ and\ \citenamefont
  {{Bergholtz}}}]{Abouelkomsan2019_MoireFCI}%
  \BibitemOpen
  \bibfield  {author} {\bibinfo {author} {\bibfnamefont {A.}~\bibnamefont
  {{Abouelkomsan}}}, \bibinfo {author} {\bibfnamefont {Z.}~\bibnamefont
  {{Liu}}},\ and\ \bibinfo {author} {\bibfnamefont {E.~J.}\ \bibnamefont
  {{Bergholtz}}},\ }\bibfield  {title} {\bibinfo {title} {{Particle-Hole
  Duality, Emergent Fermi Liquids, and Fractional Chern Insulators in Moir{\'e}
  Flatbands}},\ }\href {https://doi.org/10.1103/PhysRevLett.124.106803}
  {\bibfield  {journal} {\bibinfo  {journal} {\prl}\ }\textbf {\bibinfo
  {volume} {124}},\ \bibinfo {eid} {106803} (\bibinfo {year} {2020})},\ \Eprint
  {https://arxiv.org/abs/1912.04907} {arXiv:1912.04907 [cond-mat.mes-hall]}
  \BibitemShut {NoStop}%
\bibitem [{\citenamefont {{Wu}}\ \emph {et~al.}(2019)\citenamefont {{Wu}},
  \citenamefont {{Lovorn}}, \citenamefont {{Tutuc}}, \citenamefont {{Martin}},\
  and\ \citenamefont {{MacDonald}}}]{Wu2018_TMD_topo}%
  \BibitemOpen
  \bibfield  {author} {\bibinfo {author} {\bibfnamefont {F.}~\bibnamefont
  {{Wu}}}, \bibinfo {author} {\bibfnamefont {T.}~\bibnamefont {{Lovorn}}},
  \bibinfo {author} {\bibfnamefont {E.}~\bibnamefont {{Tutuc}}}, \bibinfo
  {author} {\bibfnamefont {I.}~\bibnamefont {{Martin}}},\ and\ \bibinfo
  {author} {\bibfnamefont {A.~H.}\ \bibnamefont {{MacDonald}}},\ }\bibfield
  {title} {\bibinfo {title} {{Topological Insulators in Twisted Transition
  Metal Dichalcogenide Homobilayers}},\ }\href
  {https://doi.org/10.1103/PhysRevLett.122.086402} {\bibfield  {journal}
  {\bibinfo  {journal} {\prl}\ }\textbf {\bibinfo {volume} {122}},\ \bibinfo
  {eid} {086402} (\bibinfo {year} {2019})},\ \Eprint
  {https://arxiv.org/abs/1807.03311} {arXiv:1807.03311 [cond-mat.mes-hall]}
  \BibitemShut {NoStop}%
\bibitem [{\citenamefont {{Yu}}\ \emph {et~al.}(2019)\citenamefont {{Yu}},
  \citenamefont {{Chen}},\ and\ \citenamefont {{Yao}}}]{Yu2019_TMD_topo}%
  \BibitemOpen
  \bibfield  {author} {\bibinfo {author} {\bibfnamefont {H.}~\bibnamefont
  {{Yu}}}, \bibinfo {author} {\bibfnamefont {M.}~\bibnamefont {{Chen}}},\ and\
  \bibinfo {author} {\bibfnamefont {W.}~\bibnamefont {{Yao}}},\ }\bibfield
  {title} {\bibinfo {title} {{Giant magnetic field from moir{\'e} induced Berry
  phase in homobilayer semiconductors}},\ }\href
  {https://doi.org/10.48550/arXiv.1906.05499} {\bibfield  {journal} {\bibinfo
  {journal} {arXiv e-prints}\ ,\ \bibinfo {eid} {arXiv:1906.05499}} (\bibinfo
  {year} {2019})},\ \Eprint {https://arxiv.org/abs/1906.05499}
  {arXiv:1906.05499 [cond-mat.mes-hall]} \BibitemShut {NoStop}%
\bibitem [{\citenamefont {{Devakul}}\ \emph {et~al.}(2021)\citenamefont
  {{Devakul}}, \citenamefont {{Cr{\'e}pel}}, \citenamefont {{Zhang}},\ and\
  \citenamefont {{Fu}}}]{Devakul2021_TMD_topo}%
  \BibitemOpen
  \bibfield  {author} {\bibinfo {author} {\bibfnamefont {T.}~\bibnamefont
  {{Devakul}}}, \bibinfo {author} {\bibfnamefont {V.}~\bibnamefont
  {{Cr{\'e}pel}}}, \bibinfo {author} {\bibfnamefont {Y.}~\bibnamefont
  {{Zhang}}},\ and\ \bibinfo {author} {\bibfnamefont {L.}~\bibnamefont
  {{Fu}}},\ }\bibfield  {title} {\bibinfo {title} {{Magic in twisted transition
  metal dichalcogenide bilayers}},\ }\href
  {https://doi.org/10.1038/s41467-021-27042-9} {\bibfield  {journal} {\bibinfo
  {journal} {Nature Communications}\ }\textbf {\bibinfo {volume} {12}},\
  \bibinfo {eid} {6730} (\bibinfo {year} {2021})},\ \Eprint
  {https://arxiv.org/abs/2106.11954} {arXiv:2106.11954 [cond-mat.mes-hall]}
  \BibitemShut {NoStop}%
\bibitem [{\citenamefont {{Bultinck}}\ \emph {et~al.}(2020)\citenamefont
  {{Bultinck}}, \citenamefont {{Chatterjee}},\ and\ \citenamefont
  {{Zaletel}}}]{Bultinck2020_QAHFM_TBG}%
  \BibitemOpen
  \bibfield  {author} {\bibinfo {author} {\bibfnamefont {N.}~\bibnamefont
  {{Bultinck}}}, \bibinfo {author} {\bibfnamefont {S.}~\bibnamefont
  {{Chatterjee}}},\ and\ \bibinfo {author} {\bibfnamefont {M.~P.}\ \bibnamefont
  {{Zaletel}}},\ }\bibfield  {title} {\bibinfo {title} {{Mechanism for
  Anomalous Hall Ferromagnetism in Twisted Bilayer Graphene}},\ }\href
  {https://doi.org/10.1103/PhysRevLett.124.166601} {\bibfield  {journal}
  {\bibinfo  {journal} {\prl}\ }\textbf {\bibinfo {volume} {124}},\ \bibinfo
  {eid} {166601} (\bibinfo {year} {2020})},\ \Eprint
  {https://arxiv.org/abs/1901.08110} {arXiv:1901.08110 [cond-mat.str-el]}
  \BibitemShut {NoStop}%
\bibitem [{\citenamefont {Zhang}\ \emph {et~al.}(2019)\citenamefont {Zhang},
  \citenamefont {Mao},\ and\ \citenamefont {Senthil}}]{zhang2019twisted}%
  \BibitemOpen
  \bibfield  {author} {\bibinfo {author} {\bibfnamefont {Y.-H.}\ \bibnamefont
  {Zhang}}, \bibinfo {author} {\bibfnamefont {D.}~\bibnamefont {Mao}},\ and\
  \bibinfo {author} {\bibfnamefont {T.}~\bibnamefont {Senthil}},\ }\bibfield
  {title} {\bibinfo {title} {Twisted bilayer graphene aligned with hexagonal
  boron nitride: Anomalous hall effect and a lattice model},\ }\href@noop {}
  {\bibfield  {journal} {\bibinfo  {journal} {Physical Review Research}\
  }\textbf {\bibinfo {volume} {1}},\ \bibinfo {pages} {033126} (\bibinfo {year}
  {2019})}\BibitemShut {NoStop}%
\bibitem [{\citenamefont {{Repellin}}\ \emph {et~al.}(2020)\citenamefont
  {{Repellin}}, \citenamefont {{Dong}}, \citenamefont {{Zhang}},\ and\
  \citenamefont {{Senthil}}}]{Repellin2020_moireFM_theory}%
  \BibitemOpen
  \bibfield  {author} {\bibinfo {author} {\bibfnamefont {C.}~\bibnamefont
  {{Repellin}}}, \bibinfo {author} {\bibfnamefont {Z.}~\bibnamefont {{Dong}}},
  \bibinfo {author} {\bibfnamefont {Y.-H.}\ \bibnamefont {{Zhang}}},\ and\
  \bibinfo {author} {\bibfnamefont {T.}~\bibnamefont {{Senthil}}},\ }\bibfield
  {title} {\bibinfo {title} {{Ferromagnetism in Narrow Bands of Moir{\'e}
  Superlattices}},\ }\href {https://doi.org/10.1103/PhysRevLett.124.187601}
  {\bibfield  {journal} {\bibinfo  {journal} {\prl}\ }\textbf {\bibinfo
  {volume} {124}},\ \bibinfo {eid} {187601} (\bibinfo {year} {2020})},\ \Eprint
  {https://arxiv.org/abs/1907.11723} {arXiv:1907.11723 [cond-mat.str-el]}
  \BibitemShut {NoStop}%
\bibitem [{\citenamefont {Liu}\ and\ \citenamefont
  {Dai}(2021)}]{liu2021orbital}%
  \BibitemOpen
  \bibfield  {author} {\bibinfo {author} {\bibfnamefont {J.}~\bibnamefont
  {Liu}}\ and\ \bibinfo {author} {\bibfnamefont {X.}~\bibnamefont {Dai}},\
  }\bibfield  {title} {\bibinfo {title} {Orbital magnetic states in moir{\'e}
  graphene systems},\ }\href@noop {} {\bibfield  {journal} {\bibinfo  {journal}
  {Nature Reviews Physics}\ }\textbf {\bibinfo {volume} {3}},\ \bibinfo {pages}
  {367} (\bibinfo {year} {2021})}\BibitemShut {NoStop}%
\bibitem [{\citenamefont {{Sharpe}}\ \emph {et~al.}(2019)\citenamefont
  {{Sharpe}}, \citenamefont {{Fox}}, \citenamefont {{Barnard}}, \citenamefont
  {{Finney}}, \citenamefont {{Watanabe}}, \citenamefont {{Taniguchi}},
  \citenamefont {{Kastner}},\ and\ \citenamefont
  {{Goldhaber-Gordon}}}]{Sharpe2019_TBG_FM}%
  \BibitemOpen
  \bibfield  {author} {\bibinfo {author} {\bibfnamefont {A.~L.}\ \bibnamefont
  {{Sharpe}}}, \bibinfo {author} {\bibfnamefont {E.~J.}\ \bibnamefont {{Fox}}},
  \bibinfo {author} {\bibfnamefont {A.~W.}\ \bibnamefont {{Barnard}}}, \bibinfo
  {author} {\bibfnamefont {J.}~\bibnamefont {{Finney}}}, \bibinfo {author}
  {\bibfnamefont {K.}~\bibnamefont {{Watanabe}}}, \bibinfo {author}
  {\bibfnamefont {T.}~\bibnamefont {{Taniguchi}}}, \bibinfo {author}
  {\bibfnamefont {M.~A.}\ \bibnamefont {{Kastner}}},\ and\ \bibinfo {author}
  {\bibfnamefont {D.}~\bibnamefont {{Goldhaber-Gordon}}},\ }\bibfield  {title}
  {\bibinfo {title} {{Emergent ferromagnetism near three-quarters filling in
  twisted bilayer graphene}},\ }\href {https://doi.org/10.1126/science.aaw3780}
  {\bibfield  {journal} {\bibinfo  {journal} {Science}\ }\textbf {\bibinfo
  {volume} {365}},\ \bibinfo {pages} {605} (\bibinfo {year} {2019})},\ \Eprint
  {https://arxiv.org/abs/1901.03520} {arXiv:1901.03520 [cond-mat.mes-hall]}
  \BibitemShut {NoStop}%
\bibitem [{\citenamefont {{Anderson}}\ \emph {et~al.}(2023)\citenamefont
  {{Anderson}}, \citenamefont {{Fan}}, \citenamefont {{Cai}}, \citenamefont
  {{Holtzmann}}, \citenamefont {{Taniguchi}}, \citenamefont {{Watanabe}},
  \citenamefont {{Xiao}}, \citenamefont {{Yao}},\ and\ \citenamefont
  {{Xu}}}]{Anderson2023_TMD_FM}%
  \BibitemOpen
  \bibfield  {author} {\bibinfo {author} {\bibfnamefont {E.}~\bibnamefont
  {{Anderson}}}, \bibinfo {author} {\bibfnamefont {F.-R.}\ \bibnamefont
  {{Fan}}}, \bibinfo {author} {\bibfnamefont {J.}~\bibnamefont {{Cai}}},
  \bibinfo {author} {\bibfnamefont {W.}~\bibnamefont {{Holtzmann}}}, \bibinfo
  {author} {\bibfnamefont {T.}~\bibnamefont {{Taniguchi}}}, \bibinfo {author}
  {\bibfnamefont {K.}~\bibnamefont {{Watanabe}}}, \bibinfo {author}
  {\bibfnamefont {D.}~\bibnamefont {{Xiao}}}, \bibinfo {author} {\bibfnamefont
  {W.}~\bibnamefont {{Yao}}},\ and\ \bibinfo {author} {\bibfnamefont
  {X.}~\bibnamefont {{Xu}}},\ }\bibfield  {title} {\bibinfo {title}
  {{Programming correlated magnetic states with gate-controlled moir{\'e}
  geometry}},\ }\href {https://doi.org/10.1126/science.adg4268} {\bibfield
  {journal} {\bibinfo  {journal} {Science}\ }\textbf {\bibinfo {volume}
  {381}},\ \bibinfo {pages} {325} (\bibinfo {year} {2023})},\ \Eprint
  {https://arxiv.org/abs/2303.17038} {arXiv:2303.17038 [cond-mat.mes-hall]}
  \BibitemShut {NoStop}%
\bibitem [{\citenamefont {{Cai}}\ \emph {et~al.}(2023)\citenamefont {{Cai}},
  \citenamefont {{Anderson}}, \citenamefont {{Wang}}, \citenamefont {{Zhang}},
  \citenamefont {{Liu}}, \citenamefont {{Holtzmann}}, \citenamefont {{Zhang}},
  \citenamefont {{Fan}}, \citenamefont {{Taniguchi}}, \citenamefont
  {{Watanabe}}, \citenamefont {{Ran}}, \citenamefont {{Cao}}, \citenamefont
  {{Fu}}, \citenamefont {{Xiao}}, \citenamefont {{Yao}},\ and\ \citenamefont
  {{Xu}}}]{Cai2023_FQAHTMD}%
  \BibitemOpen
  \bibfield  {author} {\bibinfo {author} {\bibfnamefont {J.}~\bibnamefont
  {{Cai}}}, \bibinfo {author} {\bibfnamefont {E.}~\bibnamefont {{Anderson}}},
  \bibinfo {author} {\bibfnamefont {C.}~\bibnamefont {{Wang}}}, \bibinfo
  {author} {\bibfnamefont {X.}~\bibnamefont {{Zhang}}}, \bibinfo {author}
  {\bibfnamefont {X.}~\bibnamefont {{Liu}}}, \bibinfo {author} {\bibfnamefont
  {W.}~\bibnamefont {{Holtzmann}}}, \bibinfo {author} {\bibfnamefont
  {Y.}~\bibnamefont {{Zhang}}}, \bibinfo {author} {\bibfnamefont
  {F.}~\bibnamefont {{Fan}}}, \bibinfo {author} {\bibfnamefont
  {T.}~\bibnamefont {{Taniguchi}}}, \bibinfo {author} {\bibfnamefont
  {K.}~\bibnamefont {{Watanabe}}}, \bibinfo {author} {\bibfnamefont
  {Y.}~\bibnamefont {{Ran}}}, \bibinfo {author} {\bibfnamefont
  {T.}~\bibnamefont {{Cao}}}, \bibinfo {author} {\bibfnamefont
  {L.}~\bibnamefont {{Fu}}}, \bibinfo {author} {\bibfnamefont {D.}~\bibnamefont
  {{Xiao}}}, \bibinfo {author} {\bibfnamefont {W.}~\bibnamefont {{Yao}}},\ and\
  \bibinfo {author} {\bibfnamefont {X.}~\bibnamefont {{Xu}}},\ }\bibfield
  {title} {\bibinfo {title} {{Signatures of fractional quantum anomalous Hall
  states in twisted MoTe$_{2}$}},\ }\href
  {https://doi.org/10.1038/s41586-023-06289-w} {\bibfield  {journal} {\bibinfo
  {journal} {\nat}\ }\textbf {\bibinfo {volume} {622}},\ \bibinfo {pages} {63}
  (\bibinfo {year} {2023})},\ \Eprint {https://arxiv.org/abs/2304.08470}
  {arXiv:2304.08470 [cond-mat.mes-hall]} \BibitemShut {NoStop}%
\bibitem [{\citenamefont {{Park}}\ \emph {et~al.}(2023)\citenamefont {{Park}},
  \citenamefont {{Cai}}, \citenamefont {{Anderson}}, \citenamefont {{Zhang}},
  \citenamefont {{Zhu}}, \citenamefont {{Liu}}, \citenamefont {{Wang}},
  \citenamefont {{Holtzmann}}, \citenamefont {{Hu}}, \citenamefont {{Liu}},
  \citenamefont {{Taniguchi}}, \citenamefont {{Watanabe}}, \citenamefont
  {{Chu}}, \citenamefont {{Cao}}, \citenamefont {{Fu}}, \citenamefont {{Yao}},
  \citenamefont {{Chang}}, \citenamefont {{Cobden}}, \citenamefont {{Xiao}},\
  and\ \citenamefont {{Xu}}}]{Park2023_FQAHTMD}%
  \BibitemOpen
  \bibfield  {author} {\bibinfo {author} {\bibfnamefont {H.}~\bibnamefont
  {{Park}}}, \bibinfo {author} {\bibfnamefont {J.}~\bibnamefont {{Cai}}},
  \bibinfo {author} {\bibfnamefont {E.}~\bibnamefont {{Anderson}}}, \bibinfo
  {author} {\bibfnamefont {Y.}~\bibnamefont {{Zhang}}}, \bibinfo {author}
  {\bibfnamefont {J.}~\bibnamefont {{Zhu}}}, \bibinfo {author} {\bibfnamefont
  {X.}~\bibnamefont {{Liu}}}, \bibinfo {author} {\bibfnamefont
  {C.}~\bibnamefont {{Wang}}}, \bibinfo {author} {\bibfnamefont
  {W.}~\bibnamefont {{Holtzmann}}}, \bibinfo {author} {\bibfnamefont
  {C.}~\bibnamefont {{Hu}}}, \bibinfo {author} {\bibfnamefont {Z.}~\bibnamefont
  {{Liu}}}, \bibinfo {author} {\bibfnamefont {T.}~\bibnamefont {{Taniguchi}}},
  \bibinfo {author} {\bibfnamefont {K.}~\bibnamefont {{Watanabe}}}, \bibinfo
  {author} {\bibfnamefont {J.-H.}\ \bibnamefont {{Chu}}}, \bibinfo {author}
  {\bibfnamefont {T.}~\bibnamefont {{Cao}}}, \bibinfo {author} {\bibfnamefont
  {L.}~\bibnamefont {{Fu}}}, \bibinfo {author} {\bibfnamefont {W.}~\bibnamefont
  {{Yao}}}, \bibinfo {author} {\bibfnamefont {C.-Z.}\ \bibnamefont {{Chang}}},
  \bibinfo {author} {\bibfnamefont {D.}~\bibnamefont {{Cobden}}}, \bibinfo
  {author} {\bibfnamefont {D.}~\bibnamefont {{Xiao}}},\ and\ \bibinfo {author}
  {\bibfnamefont {X.}~\bibnamefont {{Xu}}},\ }\bibfield  {title} {\bibinfo
  {title} {{Observation of fractionally quantized anomalous Hall effect}},\
  }\href {https://doi.org/10.1038/s41586-023-06536-0} {\bibfield  {journal}
  {\bibinfo  {journal} {\nat}\ }\textbf {\bibinfo {volume} {622}},\ \bibinfo
  {pages} {74} (\bibinfo {year} {2023})},\ \Eprint
  {https://arxiv.org/abs/2308.02657} {arXiv:2308.02657 [cond-mat.mes-hall]}
  \BibitemShut {NoStop}%
\bibitem [{\citenamefont {{Xu}}\ \emph {et~al.}(2023)\citenamefont {{Xu}},
  \citenamefont {{Sun}}, \citenamefont {{Jia}}, \citenamefont {{Liu}},
  \citenamefont {{Xu}}, \citenamefont {{Li}}, \citenamefont {{Gu}},
  \citenamefont {{Watanabe}}, \citenamefont {{Taniguchi}}, \citenamefont
  {{Tong}}, \citenamefont {{Jia}}, \citenamefont {{Shi}}, \citenamefont
  {{Jiang}}, \citenamefont {{Zhang}}, \citenamefont {{Liu}},\ and\
  \citenamefont {{Li}}}]{Xu2023_FQAHTMD}%
  \BibitemOpen
  \bibfield  {author} {\bibinfo {author} {\bibfnamefont {F.}~\bibnamefont
  {{Xu}}}, \bibinfo {author} {\bibfnamefont {Z.}~\bibnamefont {{Sun}}},
  \bibinfo {author} {\bibfnamefont {T.}~\bibnamefont {{Jia}}}, \bibinfo
  {author} {\bibfnamefont {C.}~\bibnamefont {{Liu}}}, \bibinfo {author}
  {\bibfnamefont {C.}~\bibnamefont {{Xu}}}, \bibinfo {author} {\bibfnamefont
  {C.}~\bibnamefont {{Li}}}, \bibinfo {author} {\bibfnamefont {Y.}~\bibnamefont
  {{Gu}}}, \bibinfo {author} {\bibfnamefont {K.}~\bibnamefont {{Watanabe}}},
  \bibinfo {author} {\bibfnamefont {T.}~\bibnamefont {{Taniguchi}}}, \bibinfo
  {author} {\bibfnamefont {B.}~\bibnamefont {{Tong}}}, \bibinfo {author}
  {\bibfnamefont {J.}~\bibnamefont {{Jia}}}, \bibinfo {author} {\bibfnamefont
  {Z.}~\bibnamefont {{Shi}}}, \bibinfo {author} {\bibfnamefont
  {S.}~\bibnamefont {{Jiang}}}, \bibinfo {author} {\bibfnamefont
  {Y.}~\bibnamefont {{Zhang}}}, \bibinfo {author} {\bibfnamefont
  {X.}~\bibnamefont {{Liu}}},\ and\ \bibinfo {author} {\bibfnamefont
  {T.}~\bibnamefont {{Li}}},\ }\bibfield  {title} {\bibinfo {title}
  {{Observation of Integer and Fractional Quantum Anomalous Hall Effects in
  Twisted Bilayer MoTe$_{2}$}},\ }\href
  {https://doi.org/10.1103/PhysRevX.13.031037} {\bibfield  {journal} {\bibinfo
  {journal} {Physical Review X}\ }\textbf {\bibinfo {volume} {13}},\ \bibinfo
  {eid} {031037} (\bibinfo {year} {2023})},\ \Eprint
  {https://arxiv.org/abs/2308.06177} {arXiv:2308.06177 [cond-mat.mes-hall]}
  \BibitemShut {NoStop}%
\bibitem [{\citenamefont {{Zeng}}\ \emph {et~al.}(2023)\citenamefont {{Zeng}},
  \citenamefont {{Xia}}, \citenamefont {{Kang}}, \citenamefont {{Zhu}},
  \citenamefont {{Kn{\"u}ppel}}, \citenamefont {{Vaswani}}, \citenamefont
  {{Watanabe}}, \citenamefont {{Taniguchi}}, \citenamefont {{Mak}},\ and\
  \citenamefont {{Shan}}}]{Zeng2023_FQAHTMD}%
  \BibitemOpen
  \bibfield  {author} {\bibinfo {author} {\bibfnamefont {Y.}~\bibnamefont
  {{Zeng}}}, \bibinfo {author} {\bibfnamefont {Z.}~\bibnamefont {{Xia}}},
  \bibinfo {author} {\bibfnamefont {K.}~\bibnamefont {{Kang}}}, \bibinfo
  {author} {\bibfnamefont {J.}~\bibnamefont {{Zhu}}}, \bibinfo {author}
  {\bibfnamefont {P.}~\bibnamefont {{Kn{\"u}ppel}}}, \bibinfo {author}
  {\bibfnamefont {C.}~\bibnamefont {{Vaswani}}}, \bibinfo {author}
  {\bibfnamefont {K.}~\bibnamefont {{Watanabe}}}, \bibinfo {author}
  {\bibfnamefont {T.}~\bibnamefont {{Taniguchi}}}, \bibinfo {author}
  {\bibfnamefont {K.~F.}\ \bibnamefont {{Mak}}},\ and\ \bibinfo {author}
  {\bibfnamefont {J.}~\bibnamefont {{Shan}}},\ }\bibfield  {title} {\bibinfo
  {title} {{Integer and fractional Chern insulators in twisted bilayer
  MoTe2}},\ }\href {https://doi.org/10.48550/arXiv.2305.00973} {\bibfield
  {journal} {\bibinfo  {journal} {arXiv e-prints}\ ,\ \bibinfo {eid}
  {arXiv:2305.00973}} (\bibinfo {year} {2023})},\ \Eprint
  {https://arxiv.org/abs/2305.00973} {arXiv:2305.00973 [cond-mat.mes-hall]}
  \BibitemShut {NoStop}%
\bibitem [{\citenamefont {{Lu}}\ \emph {et~al.}(2024)\citenamefont {{Lu}},
  \citenamefont {{Han}}, \citenamefont {{Yao}}, \citenamefont {{Reddy}},
  \citenamefont {{Yang}}, \citenamefont {{Seo}}, \citenamefont {{Watanabe}},
  \citenamefont {{Taniguchi}}, \citenamefont {{Fu}},\ and\ \citenamefont
  {{Ju}}}]{Lu2023_FQAHPenta}%
  \BibitemOpen
  \bibfield  {author} {\bibinfo {author} {\bibfnamefont {Z.}~\bibnamefont
  {{Lu}}}, \bibinfo {author} {\bibfnamefont {T.}~\bibnamefont {{Han}}},
  \bibinfo {author} {\bibfnamefont {Y.}~\bibnamefont {{Yao}}}, \bibinfo
  {author} {\bibfnamefont {A.~P.}\ \bibnamefont {{Reddy}}}, \bibinfo {author}
  {\bibfnamefont {J.}~\bibnamefont {{Yang}}}, \bibinfo {author} {\bibfnamefont
  {J.}~\bibnamefont {{Seo}}}, \bibinfo {author} {\bibfnamefont
  {K.}~\bibnamefont {{Watanabe}}}, \bibinfo {author} {\bibfnamefont
  {T.}~\bibnamefont {{Taniguchi}}}, \bibinfo {author} {\bibfnamefont
  {L.}~\bibnamefont {{Fu}}},\ and\ \bibinfo {author} {\bibfnamefont
  {L.}~\bibnamefont {{Ju}}},\ }\bibfield  {title} {\bibinfo {title}
  {{Fractional quantum anomalous Hall effect in multilayer graphene}},\ }\href
  {https://doi.org/10.1038/s41586-023-07010-7} {\bibfield  {journal} {\bibinfo
  {journal} {\nat}\ }\textbf {\bibinfo {volume} {626}},\ \bibinfo {pages} {759}
  (\bibinfo {year} {2024})},\ \Eprint {https://arxiv.org/abs/2309.17436}
  {arXiv:2309.17436 [cond-mat.mes-hall]} \BibitemShut {NoStop}%
\bibitem [{\citenamefont {Lee}\ \emph {et~al.}(2006)\citenamefont {Lee},
  \citenamefont {Nagaosa},\ and\ \citenamefont {Wen}}]{Lee2004_dopedMott}%
  \BibitemOpen
  \bibfield  {author} {\bibinfo {author} {\bibfnamefont {P.~A.}\ \bibnamefont
  {Lee}}, \bibinfo {author} {\bibfnamefont {N.}~\bibnamefont {Nagaosa}},\ and\
  \bibinfo {author} {\bibfnamefont {X.-G.}\ \bibnamefont {Wen}},\ }\bibfield
  {title} {\bibinfo {title} {Doping a mott insulator: Physics of
  high-temperature superconductivity},\ }\href
  {https://doi.org/10.1103/RevModPhys.78.17} {\bibfield  {journal} {\bibinfo
  {journal} {Rev. Mod. Phys.}\ }\textbf {\bibinfo {volume} {78}},\ \bibinfo
  {pages} {17} (\bibinfo {year} {2006})}\BibitemShut {NoStop}%
\bibitem [{\citenamefont {{Cheng}}\ \emph {et~al.}(2016)\citenamefont
  {{Cheng}}, \citenamefont {{Zaletel}}, \citenamefont {{Barkeshli}},
  \citenamefont {{Vishwanath}},\ and\ \citenamefont
  {{Bonderson}}}]{Cheng2016_SET_translation}%
  \BibitemOpen
  \bibfield  {author} {\bibinfo {author} {\bibfnamefont {M.}~\bibnamefont
  {{Cheng}}}, \bibinfo {author} {\bibfnamefont {M.}~\bibnamefont {{Zaletel}}},
  \bibinfo {author} {\bibfnamefont {M.}~\bibnamefont {{Barkeshli}}}, \bibinfo
  {author} {\bibfnamefont {A.}~\bibnamefont {{Vishwanath}}},\ and\ \bibinfo
  {author} {\bibfnamefont {P.}~\bibnamefont {{Bonderson}}},\ }\bibfield
  {title} {\bibinfo {title} {{Translational Symmetry and Microscopic
  Constraints on Symmetry-Enriched Topological Phases: A View from the
  Surface}},\ }\href {https://doi.org/10.1103/PhysRevX.6.041068} {\bibfield
  {journal} {\bibinfo  {journal} {Physical Review X}\ }\textbf {\bibinfo
  {volume} {6}},\ \bibinfo {eid} {041068} (\bibinfo {year} {2016})},\ \Eprint
  {https://arxiv.org/abs/1511.02263} {arXiv:1511.02263 [cond-mat.str-el]}
  \BibitemShut {NoStop}%
\bibitem [{\citenamefont {{Bultinck}}\ and\ \citenamefont
  {{Cheng}}(2018)}]{Bultinck2018_LSM_zerofield}%
  \BibitemOpen
  \bibfield  {author} {\bibinfo {author} {\bibfnamefont {N.}~\bibnamefont
  {{Bultinck}}}\ and\ \bibinfo {author} {\bibfnamefont {M.}~\bibnamefont
  {{Cheng}}},\ }\bibfield  {title} {\bibinfo {title} {{Filling constraints on
  fermionic topological order in zero magnetic field}},\ }\href
  {https://doi.org/10.1103/PhysRevB.98.161119} {\bibfield  {journal} {\bibinfo
  {journal} {\prb}\ }\textbf {\bibinfo {volume} {98}},\ \bibinfo {eid} {161119}
  (\bibinfo {year} {2018})},\ \Eprint {https://arxiv.org/abs/1808.00324}
  {arXiv:1808.00324 [cond-mat.str-el]} \BibitemShut {NoStop}%
\bibitem [{\citenamefont {Laughlin}(1988)}]{Laughlin1988_anyonSC}%
  \BibitemOpen
  \bibfield  {author} {\bibinfo {author} {\bibfnamefont {R.~B.}\ \bibnamefont
  {Laughlin}},\ }\bibfield  {title} {\bibinfo {title} {Superconducting ground
  state of noninteracting particles obeying fractional statistics},\ }\href
  {https://doi.org/10.1103/PhysRevLett.60.2677} {\bibfield  {journal} {\bibinfo
   {journal} {Phys. Rev. Lett.}\ }\textbf {\bibinfo {volume} {60}},\ \bibinfo
  {pages} {2677} (\bibinfo {year} {1988})}\BibitemShut {NoStop}%
\bibitem [{\citenamefont {Fetter}\ \emph {et~al.}(1989)\citenamefont {Fetter},
  \citenamefont {Hanna},\ and\ \citenamefont
  {Laughlin}}]{Fetter1989_anyonSC_RPA}%
  \BibitemOpen
  \bibfield  {author} {\bibinfo {author} {\bibfnamefont {A.~L.}\ \bibnamefont
  {Fetter}}, \bibinfo {author} {\bibfnamefont {C.~B.}\ \bibnamefont {Hanna}},\
  and\ \bibinfo {author} {\bibfnamefont {R.~B.}\ \bibnamefont {Laughlin}},\
  }\bibfield  {title} {\bibinfo {title} {Random-phase approximation in the
  fractional-statistics gas},\ }\href
  {https://doi.org/10.1103/PhysRevB.39.9679} {\bibfield  {journal} {\bibinfo
  {journal} {Phys. Rev. B}\ }\textbf {\bibinfo {volume} {39}},\ \bibinfo
  {pages} {9679} (\bibinfo {year} {1989})}\BibitemShut {NoStop}%
\bibitem [{\citenamefont {Lee}\ and\ \citenamefont
  {Fisher}(1989)}]{Lee1989_anyonSC}%
  \BibitemOpen
  \bibfield  {author} {\bibinfo {author} {\bibfnamefont {D.-H.}\ \bibnamefont
  {Lee}}\ and\ \bibinfo {author} {\bibfnamefont {M.~P.~A.}\ \bibnamefont
  {Fisher}},\ }\bibfield  {title} {\bibinfo {title} {Anyon superconductivity
  and the fractional quantum hall effect},\ }\href
  {https://doi.org/10.1103/PhysRevLett.63.903} {\bibfield  {journal} {\bibinfo
  {journal} {Phys. Rev. Lett.}\ }\textbf {\bibinfo {volume} {63}},\ \bibinfo
  {pages} {903} (\bibinfo {year} {1989})}\BibitemShut {NoStop}%
\bibitem [{\citenamefont {{Chen}}\ \emph {et~al.}(1989)\citenamefont {{Chen}},
  \citenamefont {{Wilczek}}, \citenamefont {{Witten}},\ and\ \citenamefont
  {{Halperin}}}]{Chen1989_anyonSC}%
  \BibitemOpen
  \bibfield  {author} {\bibinfo {author} {\bibfnamefont {Y.-H.}\ \bibnamefont
  {{Chen}}}, \bibinfo {author} {\bibfnamefont {F.}~\bibnamefont {{Wilczek}}},
  \bibinfo {author} {\bibfnamefont {E.}~\bibnamefont {{Witten}}},\ and\
  \bibinfo {author} {\bibfnamefont {B.~I.}\ \bibnamefont {{Halperin}}},\
  }\bibfield  {title} {\bibinfo {title} {{On Anyon Superconductivity}},\ }\href
  {https://doi.org/10.1142/S0217751X89001631} {\bibfield  {journal} {\bibinfo
  {journal} {International Journal of Modern Physics A}\ }\textbf {\bibinfo
  {volume} {4}},\ \bibinfo {pages} {3983} (\bibinfo {year} {1989})}\BibitemShut
  {NoStop}%
\bibitem [{\citenamefont {Halperin}\ \emph {et~al.}(1989)\citenamefont
  {Halperin}, \citenamefont {March-Russell},\ and\ \citenamefont
  {Wilczek}}]{halperin1989consequences}%
  \BibitemOpen
  \bibfield  {author} {\bibinfo {author} {\bibfnamefont {B.~I.}\ \bibnamefont
  {Halperin}}, \bibinfo {author} {\bibfnamefont {J.}~\bibnamefont
  {March-Russell}},\ and\ \bibinfo {author} {\bibfnamefont {F.}~\bibnamefont
  {Wilczek}},\ }\bibfield  {title} {\bibinfo {title} {Consequences of
  time-reversal-symmetry violation in models of high-t c superconductors},\
  }\href@noop {} {\bibfield  {journal} {\bibinfo  {journal} {Physical Review
  B}\ }\textbf {\bibinfo {volume} {40}},\ \bibinfo {pages} {8726} (\bibinfo
  {year} {1989})}\BibitemShut {NoStop}%
\bibitem [{\citenamefont {Wen}\ and\ \citenamefont
  {Zee}(1990)}]{Wen1990_anyonSC}%
  \BibitemOpen
  \bibfield  {author} {\bibinfo {author} {\bibfnamefont {X.~G.}\ \bibnamefont
  {Wen}}\ and\ \bibinfo {author} {\bibfnamefont {A.}~\bibnamefont {Zee}},\
  }\bibfield  {title} {\bibinfo {title} {Compressibility and superfluidity in
  the fractional-statistics liquid},\ }\href
  {https://doi.org/10.1103/PhysRevB.41.240} {\bibfield  {journal} {\bibinfo
  {journal} {Phys. Rev. B}\ }\textbf {\bibinfo {volume} {41}},\ \bibinfo
  {pages} {240} (\bibinfo {year} {1990})}\BibitemShut {NoStop}%
\bibitem [{\citenamefont {Senthil}\ and\ \citenamefont
  {Fisher}(2000)}]{Senthil2000_SC*}%
  \BibitemOpen
  \bibfield  {author} {\bibinfo {author} {\bibfnamefont {T.}~\bibnamefont
  {Senthil}}\ and\ \bibinfo {author} {\bibfnamefont {M.~P.~A.}\ \bibnamefont
  {Fisher}},\ }\bibfield  {title} {\bibinfo {title} {${Z}_{2}$ gauge theory of
  electron fractionalization in strongly correlated systems},\ }\href
  {https://doi.org/10.1103/PhysRevB.62.7850} {\bibfield  {journal} {\bibinfo
  {journal} {Phys. Rev. B}\ }\textbf {\bibinfo {volume} {62}},\ \bibinfo
  {pages} {7850} (\bibinfo {year} {2000})}\BibitemShut {NoStop}%
\bibitem [{\citenamefont {{Tang}}\ and\ \citenamefont
  {{Wen}}(2013)}]{Tang2013_anyon_hop}%
  \BibitemOpen
  \bibfield  {author} {\bibinfo {author} {\bibfnamefont {E.}~\bibnamefont
  {{Tang}}}\ and\ \bibinfo {author} {\bibfnamefont {X.-G.}\ \bibnamefont
  {{Wen}}},\ }\bibfield  {title} {\bibinfo {title} {{Superconductivity with
  intrinsic topological order induced by pure Coulomb interaction and
  time-reversal symmetry breaking}},\ }\href
  {https://doi.org/10.1103/PhysRevB.88.195117} {\bibfield  {journal} {\bibinfo
  {journal} {\prb}\ }\textbf {\bibinfo {volume} {88}},\ \bibinfo {eid} {195117}
  (\bibinfo {year} {2013})},\ \Eprint {https://arxiv.org/abs/1306.1528}
  {arXiv:1306.1528 [cond-mat.str-el]} \BibitemShut {NoStop}%
\bibitem [{\citenamefont {{Han}}\ \emph {et~al.}(2024)\citenamefont {{Han}},
  \citenamefont {{Lu}}, \citenamefont {{Yao}}, \citenamefont {{Shi}},
  \citenamefont {{Yang}}, \citenamefont {{Seo}}, \citenamefont {{Ye}},
  \citenamefont {{Wu}}, \citenamefont {{Zhou}}, \citenamefont {{Liu}},
  \citenamefont {{Shi}}, \citenamefont {{Hua}}, \citenamefont {{Watanabe}},
  \citenamefont {{Taniguchi}}, \citenamefont {{Xiong}}, \citenamefont {{Fu}},\
  and\ \citenamefont {{Ju}}}]{Han2024_chiralSC_penta}%
  \BibitemOpen
  \bibfield  {author} {\bibinfo {author} {\bibfnamefont {T.}~\bibnamefont
  {{Han}}}, \bibinfo {author} {\bibfnamefont {Z.}~\bibnamefont {{Lu}}},
  \bibinfo {author} {\bibfnamefont {Y.}~\bibnamefont {{Yao}}}, \bibinfo
  {author} {\bibfnamefont {L.}~\bibnamefont {{Shi}}}, \bibinfo {author}
  {\bibfnamefont {J.}~\bibnamefont {{Yang}}}, \bibinfo {author} {\bibfnamefont
  {J.}~\bibnamefont {{Seo}}}, \bibinfo {author} {\bibfnamefont
  {S.}~\bibnamefont {{Ye}}}, \bibinfo {author} {\bibfnamefont {Z.}~\bibnamefont
  {{Wu}}}, \bibinfo {author} {\bibfnamefont {M.}~\bibnamefont {{Zhou}}},
  \bibinfo {author} {\bibfnamefont {H.}~\bibnamefont {{Liu}}}, \bibinfo
  {author} {\bibfnamefont {G.}~\bibnamefont {{Shi}}}, \bibinfo {author}
  {\bibfnamefont {Z.}~\bibnamefont {{Hua}}}, \bibinfo {author} {\bibfnamefont
  {K.}~\bibnamefont {{Watanabe}}}, \bibinfo {author} {\bibfnamefont
  {T.}~\bibnamefont {{Taniguchi}}}, \bibinfo {author} {\bibfnamefont
  {P.}~\bibnamefont {{Xiong}}}, \bibinfo {author} {\bibfnamefont
  {L.}~\bibnamefont {{Fu}}},\ and\ \bibinfo {author} {\bibfnamefont
  {L.}~\bibnamefont {{Ju}}},\ }\bibfield  {title} {\bibinfo {title}
  {{Signatures of Chiral Superconductivity in Rhombohedral Graphene}},\ }\href
  {https://doi.org/10.48550/arXiv.2408.15233} {\bibfield  {journal} {\bibinfo
  {journal} {arXiv e-prints}\ ,\ \bibinfo {eid} {arXiv:2408.15233}} (\bibinfo
  {year} {2024})},\ \Eprint {https://arxiv.org/abs/2408.15233}
  {arXiv:2408.15233 [cond-mat.mes-hall]} \BibitemShut {NoStop}%
\bibitem [{\citenamefont {{Spanton}}\ \emph {et~al.}(2018)\citenamefont
  {{Spanton}}, \citenamefont {{Zibrov}}, \citenamefont {{Zhou}}, \citenamefont
  {{Taniguchi}}, \citenamefont {{Watanabe}}, \citenamefont {{Zaletel}},\ and\
  \citenamefont {{Young}}}]{Spanton2017_FCI}%
  \BibitemOpen
  \bibfield  {author} {\bibinfo {author} {\bibfnamefont {E.~M.}\ \bibnamefont
  {{Spanton}}}, \bibinfo {author} {\bibfnamefont {A.~A.}\ \bibnamefont
  {{Zibrov}}}, \bibinfo {author} {\bibfnamefont {H.}~\bibnamefont {{Zhou}}},
  \bibinfo {author} {\bibfnamefont {T.}~\bibnamefont {{Taniguchi}}}, \bibinfo
  {author} {\bibfnamefont {K.}~\bibnamefont {{Watanabe}}}, \bibinfo {author}
  {\bibfnamefont {M.~P.}\ \bibnamefont {{Zaletel}}},\ and\ \bibinfo {author}
  {\bibfnamefont {A.~F.}\ \bibnamefont {{Young}}},\ }\bibfield  {title}
  {\bibinfo {title} {{Observation of fractional Chern insulators in a van der
  Waals heterostructure}},\ }\href {https://doi.org/10.1126/science.aan8458}
  {\bibfield  {journal} {\bibinfo  {journal} {Science}\ }\textbf {\bibinfo
  {volume} {360}},\ \bibinfo {pages} {62} (\bibinfo {year} {2018})},\ \Eprint
  {https://arxiv.org/abs/1706.06116} {arXiv:1706.06116 [cond-mat.str-el]}
  \BibitemShut {NoStop}%
\bibitem [{\citenamefont {{Xie}}\ \emph {et~al.}(2021)\citenamefont {{Xie}},
  \citenamefont {{Pierce}}, \citenamefont {{Park}}, \citenamefont {{Parker}},
  \citenamefont {{Khalaf}}, \citenamefont {{Ledwith}}, \citenamefont {{Cao}},
  \citenamefont {{Lee}}, \citenamefont {{Chen}}, \citenamefont {{Forrester}},
  \citenamefont {{Watanabe}}, \citenamefont {{Taniguchi}}, \citenamefont
  {{Vishwanath}}, \citenamefont {{Jarillo-Herrero}},\ and\ \citenamefont
  {{Yacoby}}}]{Xie2021_FCI}%
  \BibitemOpen
  \bibfield  {author} {\bibinfo {author} {\bibfnamefont {Y.}~\bibnamefont
  {{Xie}}}, \bibinfo {author} {\bibfnamefont {A.~T.}\ \bibnamefont {{Pierce}}},
  \bibinfo {author} {\bibfnamefont {J.~M.}\ \bibnamefont {{Park}}}, \bibinfo
  {author} {\bibfnamefont {D.~E.}\ \bibnamefont {{Parker}}}, \bibinfo {author}
  {\bibfnamefont {E.}~\bibnamefont {{Khalaf}}}, \bibinfo {author}
  {\bibfnamefont {P.}~\bibnamefont {{Ledwith}}}, \bibinfo {author}
  {\bibfnamefont {Y.}~\bibnamefont {{Cao}}}, \bibinfo {author} {\bibfnamefont
  {S.~H.}\ \bibnamefont {{Lee}}}, \bibinfo {author} {\bibfnamefont
  {S.}~\bibnamefont {{Chen}}}, \bibinfo {author} {\bibfnamefont {P.~R.}\
  \bibnamefont {{Forrester}}}, \bibinfo {author} {\bibfnamefont
  {K.}~\bibnamefont {{Watanabe}}}, \bibinfo {author} {\bibfnamefont
  {T.}~\bibnamefont {{Taniguchi}}}, \bibinfo {author} {\bibfnamefont
  {A.}~\bibnamefont {{Vishwanath}}}, \bibinfo {author} {\bibfnamefont
  {P.}~\bibnamefont {{Jarillo-Herrero}}},\ and\ \bibinfo {author}
  {\bibfnamefont {A.}~\bibnamefont {{Yacoby}}},\ }\bibfield  {title} {\bibinfo
  {title} {{Fractional Chern insulators in magic-angle twisted bilayer
  graphene}},\ }\href {https://doi.org/10.1038/s41586-021-04002-3} {\bibfield
  {journal} {\bibinfo  {journal} {\nat}\ }\textbf {\bibinfo {volume} {600}},\
  \bibinfo {pages} {439} (\bibinfo {year} {2021})},\ \Eprint
  {https://arxiv.org/abs/2107.10854} {arXiv:2107.10854 [cond-mat.mes-hall]}
  \BibitemShut {NoStop}%
\bibitem [{\citenamefont {{Aronson}}\ \emph {et~al.}(2024)\citenamefont
  {{Aronson}}, \citenamefont {{Han}}, \citenamefont {{Lu}}, \citenamefont
  {{Yao}}, \citenamefont {{Watanabe}}, \citenamefont {{Taniguchi}},
  \citenamefont {{Ju}},\ and\ \citenamefont {{Ashoori}}}]{Aronson2024_FCI}%
  \BibitemOpen
  \bibfield  {author} {\bibinfo {author} {\bibfnamefont {S.~H.}\ \bibnamefont
  {{Aronson}}}, \bibinfo {author} {\bibfnamefont {T.}~\bibnamefont {{Han}}},
  \bibinfo {author} {\bibfnamefont {Z.}~\bibnamefont {{Lu}}}, \bibinfo {author}
  {\bibfnamefont {Y.}~\bibnamefont {{Yao}}}, \bibinfo {author} {\bibfnamefont
  {K.}~\bibnamefont {{Watanabe}}}, \bibinfo {author} {\bibfnamefont
  {T.}~\bibnamefont {{Taniguchi}}}, \bibinfo {author} {\bibfnamefont
  {L.}~\bibnamefont {{Ju}}},\ and\ \bibinfo {author} {\bibfnamefont {R.~C.}\
  \bibnamefont {{Ashoori}}},\ }\bibfield  {title} {\bibinfo {title}
  {{Displacement field-controlled fractional Chern insulators and charge
  density waves in a graphene/hBN moir{\'e} superlattice}},\ }\href
  {https://doi.org/10.48550/arXiv.2408.11220} {\bibfield  {journal} {\bibinfo
  {journal} {arXiv e-prints}\ ,\ \bibinfo {eid} {arXiv:2408.11220}} (\bibinfo
  {year} {2024})},\ \Eprint {https://arxiv.org/abs/2408.11220}
  {arXiv:2408.11220 [cond-mat.mes-hall]} \BibitemShut {NoStop}%
\bibitem [{\citenamefont {Jalabert}\ and\ \citenamefont
  {Sachdev}(1991)}]{jalabert1991spontaneous}%
  \BibitemOpen
  \bibfield  {author} {\bibinfo {author} {\bibfnamefont {R.~A.}\ \bibnamefont
  {Jalabert}}\ and\ \bibinfo {author} {\bibfnamefont {S.}~\bibnamefont
  {Sachdev}},\ }\bibfield  {title} {\bibinfo {title} {Spontaneous alignment of
  frustrated bonds in an anisotropic, three-dimensional ising model},\
  }\href@noop {} {\bibfield  {journal} {\bibinfo  {journal} {Physical Review
  B}\ }\textbf {\bibinfo {volume} {44}},\ \bibinfo {pages} {686} (\bibinfo
  {year} {1991})}\BibitemShut {NoStop}%
\bibitem [{\citenamefont {{Senthil}}\ and\ \citenamefont
  {{Fisher}}(2000)}]{senthil2000z}%
  \BibitemOpen
  \bibfield  {author} {\bibinfo {author} {\bibfnamefont {T.}~\bibnamefont
  {{Senthil}}}\ and\ \bibinfo {author} {\bibfnamefont {M.~P.~A.}\ \bibnamefont
  {{Fisher}}},\ }\bibfield  {title} {\bibinfo {title} {{Z$_{2}$ gauge theory of
  electron fractionalization in strongly correlated systems}},\ }\href
  {https://doi.org/10.1103/PhysRevB.62.7850} {\bibfield  {journal} {\bibinfo
  {journal} {\prb}\ }\textbf {\bibinfo {volume} {62}},\ \bibinfo {pages} {7850}
  (\bibinfo {year} {2000})},\ \Eprint {https://arxiv.org/abs/cond-mat/9910224}
  {arXiv:cond-mat/9910224 [cond-mat.str-el]} \BibitemShut {NoStop}%
\bibitem [{\citenamefont {Lu}\ \emph {et~al.}(2020)\citenamefont {Lu},
  \citenamefont {Ran},\ and\ \citenamefont {Oshikawa}}]{Lu2017_filling_Hall}%
  \BibitemOpen
  \bibfield  {author} {\bibinfo {author} {\bibfnamefont {Y.-M.}\ \bibnamefont
  {Lu}}, \bibinfo {author} {\bibfnamefont {Y.}~\bibnamefont {Ran}},\ and\
  \bibinfo {author} {\bibfnamefont {M.}~\bibnamefont {Oshikawa}},\ }\bibfield
  {title} {\bibinfo {title} {Filling-enforced constraint on the quantized hall
  conductivity on a periodic lattice},\ }\href
  {https://doi.org/https://doi.org/10.1016/j.aop.2019.168060} {\bibfield
  {journal} {\bibinfo  {journal} {Annals of Physics}\ }\textbf {\bibinfo
  {volume} {413}},\ \bibinfo {pages} {168060} (\bibinfo {year}
  {2020})}\BibitemShut {NoStop}%
\bibitem [{\citenamefont {{Song}}\ \emph {et~al.}(2024)\citenamefont {{Song}},
  \citenamefont {{Zhang}},\ and\ \citenamefont
  {{Senthil}}}]{Song2023_QPT_FQAH}%
  \BibitemOpen
  \bibfield  {author} {\bibinfo {author} {\bibfnamefont {X.-Y.}\ \bibnamefont
  {{Song}}}, \bibinfo {author} {\bibfnamefont {Y.-H.}\ \bibnamefont
  {{Zhang}}},\ and\ \bibinfo {author} {\bibfnamefont {T.}~\bibnamefont
  {{Senthil}}},\ }\bibfield  {title} {\bibinfo {title} {{Phase transitions out
  of quantum Hall states in moir{\'e} materials}},\ }\href
  {https://doi.org/10.1103/PhysRevB.109.085143} {\bibfield  {journal} {\bibinfo
   {journal} {\prb}\ }\textbf {\bibinfo {volume} {109}},\ \bibinfo {eid}
  {085143} (\bibinfo {year} {2024})},\ \Eprint
  {https://arxiv.org/abs/2308.10903} {arXiv:2308.10903 [cond-mat.str-el]}
  \BibitemShut {NoStop}%
\bibitem [{\citenamefont {Senthil}\ \emph {et~al.}(2019)\citenamefont
  {Senthil}, \citenamefont {Son}, \citenamefont {Wang},\ and\ \citenamefont
  {Xu}}]{senthil2019duality}%
  \BibitemOpen
  \bibfield  {author} {\bibinfo {author} {\bibfnamefont {T.}~\bibnamefont
  {Senthil}}, \bibinfo {author} {\bibfnamefont {D.~T.}\ \bibnamefont {Son}},
  \bibinfo {author} {\bibfnamefont {C.}~\bibnamefont {Wang}},\ and\ \bibinfo
  {author} {\bibfnamefont {C.}~\bibnamefont {Xu}},\ }\bibfield  {title}
  {\bibinfo {title} {Duality between (2+ 1) d quantum critical points},\
  }\href@noop {} {\bibfield  {journal} {\bibinfo  {journal} {Physics Reports}\
  }\textbf {\bibinfo {volume} {827}},\ \bibinfo {pages} {1} (\bibinfo {year}
  {2019})}\BibitemShut {NoStop}%
\bibitem [{\citenamefont {{Metlitski}}(2015)}]{metlitski2015s}%
  \BibitemOpen
  \bibfield  {author} {\bibinfo {author} {\bibfnamefont {M.~A.}\ \bibnamefont
  {{Metlitski}}},\ }\bibfield  {title} {\bibinfo {title} {{$S$-duality of
  $u(1)$ gauge theory with $\theta =\pi$ on non-orientable manifolds:
  Applications to topological insulators and superconductors}},\ }\href
  {https://doi.org/10.48550/arXiv.1510.05663} {\bibfield  {journal} {\bibinfo
  {journal} {arXiv e-prints}\ ,\ \bibinfo {eid} {arXiv:1510.05663}} (\bibinfo
  {year} {2015})},\ \Eprint {https://arxiv.org/abs/1510.05663}
  {arXiv:1510.05663 [hep-th]} \BibitemShut {NoStop}%
\bibitem [{\citenamefont {{Seiberg}}\ and\ \citenamefont
  {{Witten}}(2016)}]{Seiberg2016_TQFTgappedbdry}%
  \BibitemOpen
  \bibfield  {author} {\bibinfo {author} {\bibfnamefont {N.}~\bibnamefont
  {{Seiberg}}}\ and\ \bibinfo {author} {\bibfnamefont {E.}~\bibnamefont
  {{Witten}}},\ }\bibfield  {title} {\bibinfo {title} {{Gapped boundary phases
  of topological insulators via weak coupling}},\ }\href
  {https://doi.org/10.1093/ptep/ptw083} {\bibfield  {journal} {\bibinfo
  {journal} {Progress of Theoretical and Experimental Physics}\ }\textbf
  {\bibinfo {volume} {2016}},\ \bibinfo {eid} {12C101} (\bibinfo {year}
  {2016})},\ \Eprint {https://arxiv.org/abs/1602.04251} {arXiv:1602.04251
  [cond-mat.str-el]} \BibitemShut {NoStop}%
\bibitem [{\citenamefont {{Seiberg}}\ \emph {et~al.}(2016)\citenamefont
  {{Seiberg}}, \citenamefont {{Senthil}}, \citenamefont {{Wang}},\ and\
  \citenamefont {{Witten}}}]{seiberg2016duality}%
  \BibitemOpen
  \bibfield  {author} {\bibinfo {author} {\bibfnamefont {N.}~\bibnamefont
  {{Seiberg}}}, \bibinfo {author} {\bibfnamefont {T.}~\bibnamefont
  {{Senthil}}}, \bibinfo {author} {\bibfnamefont {C.}~\bibnamefont {{Wang}}},\
  and\ \bibinfo {author} {\bibfnamefont {E.}~\bibnamefont {{Witten}}},\
  }\bibfield  {title} {\bibinfo {title} {{A duality web in 2 + 1 dimensions and
  condensed matter physics}},\ }\href
  {https://doi.org/10.1016/j.aop.2016.08.007} {\bibfield  {journal} {\bibinfo
  {journal} {Annals of Physics}\ }\textbf {\bibinfo {volume} {374}},\ \bibinfo
  {pages} {395} (\bibinfo {year} {2016})},\ \Eprint
  {https://arxiv.org/abs/1606.01989} {arXiv:1606.01989 [hep-th]} \BibitemShut
  {NoStop}%
\bibitem [{\citenamefont {{Moore}}\ and\ \citenamefont
  {{Read}}(1991)}]{Moore1991_nonabelian}%
  \BibitemOpen
  \bibfield  {author} {\bibinfo {author} {\bibfnamefont {G.}~\bibnamefont
  {{Moore}}}\ and\ \bibinfo {author} {\bibfnamefont {N.}~\bibnamefont
  {{Read}}},\ }\bibfield  {title} {\bibinfo {title} {{Nonabelions in the
  fractional quantum hall effect}},\ }\href
  {https://doi.org/10.1016/0550-3213(91)90407-O} {\bibfield  {journal}
  {\bibinfo  {journal} {Nuclear Physics B}\ }\textbf {\bibinfo {volume}
  {360}},\ \bibinfo {pages} {362} (\bibinfo {year} {1991})}\BibitemShut
  {NoStop}%
\bibitem [{\citenamefont {{Read}}\ and\ \citenamefont
  {{Green}}(2000)}]{Read1999_pair}%
  \BibitemOpen
  \bibfield  {author} {\bibinfo {author} {\bibfnamefont {N.}~\bibnamefont
  {{Read}}}\ and\ \bibinfo {author} {\bibfnamefont {D.}~\bibnamefont
  {{Green}}},\ }\bibfield  {title} {\bibinfo {title} {{Paired states of
  fermions in two dimensions with breaking of parity and time-reversal
  symmetries and the fractional quantum Hall effect}},\ }\href
  {https://doi.org/10.1103/PhysRevB.61.10267} {\bibfield  {journal} {\bibinfo
  {journal} {\prb}\ }\textbf {\bibinfo {volume} {61}},\ \bibinfo {pages}
  {10267} (\bibinfo {year} {2000})},\ \Eprint
  {https://arxiv.org/abs/cond-mat/9906453} {arXiv:cond-mat/9906453
  [cond-mat.mes-hall]} \BibitemShut {NoStop}%
\bibitem [{\citenamefont {{L{\"o}hneysen}}\ \emph {et~al.}(2007)\citenamefont
  {{L{\"o}hneysen}}, \citenamefont {{Rosch}}, \citenamefont {{Vojta}},\ and\
  \citenamefont {{W{\"o}lfle}}}]{Lohneysen2006_HertzMillis_review}%
  \BibitemOpen
  \bibfield  {author} {\bibinfo {author} {\bibfnamefont {H.~V.}\ \bibnamefont
  {{L{\"o}hneysen}}}, \bibinfo {author} {\bibfnamefont {A.}~\bibnamefont
  {{Rosch}}}, \bibinfo {author} {\bibfnamefont {M.}~\bibnamefont {{Vojta}}},\
  and\ \bibinfo {author} {\bibfnamefont {P.}~\bibnamefont {{W{\"o}lfle}}},\
  }\bibfield  {title} {\bibinfo {title} {{Fermi-liquid instabilities at
  magnetic quantum phase transitions}},\ }\href
  {https://doi.org/10.1103/RevModPhys.79.1015} {\bibfield  {journal} {\bibinfo
  {journal} {Reviews of Modern Physics}\ }\textbf {\bibinfo {volume} {79}},\
  \bibinfo {pages} {1015} (\bibinfo {year} {2007})},\ \Eprint
  {https://arxiv.org/abs/cond-mat/0606317} {arXiv:cond-mat/0606317
  [cond-mat.str-el]} \BibitemShut {NoStop}%
\bibitem [{\citenamefont {{Mross}}\ \emph {et~al.}(2010)\citenamefont
  {{Mross}}, \citenamefont {{McGreevy}}, \citenamefont {{Liu}},\ and\
  \citenamefont {{Senthil}}}]{Mross2010_NFL}%
  \BibitemOpen
  \bibfield  {author} {\bibinfo {author} {\bibfnamefont {D.~F.}\ \bibnamefont
  {{Mross}}}, \bibinfo {author} {\bibfnamefont {J.}~\bibnamefont {{McGreevy}}},
  \bibinfo {author} {\bibfnamefont {H.}~\bibnamefont {{Liu}}},\ and\ \bibinfo
  {author} {\bibfnamefont {T.}~\bibnamefont {{Senthil}}},\ }\bibfield  {title}
  {\bibinfo {title} {{Controlled expansion for certain non-Fermi-liquid
  metals}},\ }\href {https://doi.org/10.1103/PhysRevB.82.045121} {\bibfield
  {journal} {\bibinfo  {journal} {\prb}\ }\textbf {\bibinfo {volume} {82}},\
  \bibinfo {eid} {045121} (\bibinfo {year} {2010})},\ \Eprint
  {https://arxiv.org/abs/1003.0894} {arXiv:1003.0894 [cond-mat.str-el]}
  \BibitemShut {NoStop}%
\bibitem [{\citenamefont {Metlitski}\ \emph {et~al.}(2015)\citenamefont
  {Metlitski}, \citenamefont {Mross}, \citenamefont {Sachdev},\ and\
  \citenamefont {Senthil}}]{Metlitski2014_PairingNFL}%
  \BibitemOpen
  \bibfield  {author} {\bibinfo {author} {\bibfnamefont {M.~A.}\ \bibnamefont
  {Metlitski}}, \bibinfo {author} {\bibfnamefont {D.~F.}\ \bibnamefont
  {Mross}}, \bibinfo {author} {\bibfnamefont {S.}~\bibnamefont {Sachdev}},\
  and\ \bibinfo {author} {\bibfnamefont {T.}~\bibnamefont {Senthil}},\
  }\bibfield  {title} {\bibinfo {title} {Cooper pairing in non-fermi liquids},\
  }\href {https://doi.org/10.1103/PhysRevB.91.115111} {\bibfield  {journal}
  {\bibinfo  {journal} {Phys. Rev. B}\ }\textbf {\bibinfo {volume} {91}},\
  \bibinfo {pages} {115111} (\bibinfo {year} {2015})}\BibitemShut {NoStop}%
\bibitem [{\citenamefont {Shi}\ \emph {et~al.}(2022)\citenamefont {Shi},
  \citenamefont {Goldman}, \citenamefont {Else},\ and\ \citenamefont
  {Senthil}}]{Shi2022_gifts}%
  \BibitemOpen
  \bibfield  {author} {\bibinfo {author} {\bibfnamefont {Z.~D.}\ \bibnamefont
  {Shi}}, \bibinfo {author} {\bibfnamefont {H.}~\bibnamefont {Goldman}},
  \bibinfo {author} {\bibfnamefont {D.~V.}\ \bibnamefont {Else}},\ and\
  \bibinfo {author} {\bibfnamefont {T.}~\bibnamefont {Senthil}},\ }\bibfield
  {title} {\bibinfo {title} {{Gifts from anomalies: Exact results for Landau
  phase transitions in metals}},\ }\href
  {https://doi.org/10.21468/SciPostPhys.13.5.102} {\bibfield  {journal}
  {\bibinfo  {journal} {SciPost Phys.}\ }\textbf {\bibinfo {volume} {13}},\
  \bibinfo {pages} {102} (\bibinfo {year} {2022})},\ \Eprint
  {https://arxiv.org/abs/2204.07585} {2204.07585} \BibitemShut {NoStop}%
\bibitem [{\citenamefont {{Guo}}\ \emph {et~al.}(2022)\citenamefont {{Guo}},
  \citenamefont {{Patel}}, \citenamefont {{Esterlis}},\ and\ \citenamefont
  {{Sachdev}}}]{Guo2022_clean_NFLtransport}%
  \BibitemOpen
  \bibfield  {author} {\bibinfo {author} {\bibfnamefont {H.}~\bibnamefont
  {{Guo}}}, \bibinfo {author} {\bibfnamefont {A.~A.}\ \bibnamefont {{Patel}}},
  \bibinfo {author} {\bibfnamefont {I.}~\bibnamefont {{Esterlis}}},\ and\
  \bibinfo {author} {\bibfnamefont {S.}~\bibnamefont {{Sachdev}}},\ }\bibfield
  {title} {\bibinfo {title} {{Large-N theory of critical Fermi surfaces. II.
  Conductivity}},\ }\href {https://doi.org/10.1103/PhysRevB.106.115151}
  {\bibfield  {journal} {\bibinfo  {journal} {\prb}\ }\textbf {\bibinfo
  {volume} {106}},\ \bibinfo {eid} {115151} (\bibinfo {year} {2022})},\ \Eprint
  {https://arxiv.org/abs/2207.08841} {arXiv:2207.08841 [cond-mat.str-el]}
  \BibitemShut {NoStop}%
\bibitem [{\citenamefont {{Shi}}(2024)}]{Shi2024_transport_exp}%
  \BibitemOpen
  \bibfield  {author} {\bibinfo {author} {\bibfnamefont {Z.~D.}\ \bibnamefont
  {{Shi}}},\ }\bibfield  {title} {\bibinfo {title} {{Controlled expansion for
  transport in a class of non-Fermi liquids}},\ }\href
  {https://doi.org/10.1103/PhysRevB.109.195110} {\bibfield  {journal} {\bibinfo
   {journal} {\prb}\ }\textbf {\bibinfo {volume} {109}},\ \bibinfo {eid}
  {195110} (\bibinfo {year} {2024})},\ \Eprint
  {https://arxiv.org/abs/2311.12922} {arXiv:2311.12922 [cond-mat.str-el]}
  \BibitemShut {NoStop}%
\bibitem [{\citenamefont {{Haldane}}(2004)}]{Haldane2004_FSBerry}%
  \BibitemOpen
  \bibfield  {author} {\bibinfo {author} {\bibfnamefont {F.~D.}\ \bibnamefont
  {{Haldane}}},\ }\bibfield  {title} {\bibinfo {title} {{Berry Curvature on the
  Fermi Surface: Anomalous Hall Effect as a Topological Fermi-Liquid
  Property}},\ }\href {https://doi.org/10.1103/PhysRevLett.93.206602}
  {\bibfield  {journal} {\bibinfo  {journal} {\prl}\ }\textbf {\bibinfo
  {volume} {93}},\ \bibinfo {eid} {206602} (\bibinfo {year} {2004})},\ \Eprint
  {https://arxiv.org/abs/cond-mat/0408417} {arXiv:cond-mat/0408417
  [cond-mat.mes-hall]} \BibitemShut {NoStop}%
\bibitem [{\citenamefont {{Sodemann}}\ \emph {et~al.}(2017)\citenamefont
  {{Sodemann}}, \citenamefont {{Kimchi}}, \citenamefont {{Wang}},\ and\
  \citenamefont {{Senthil}}}]{Sodemann2017_CFLbilayer}%
  \BibitemOpen
  \bibfield  {author} {\bibinfo {author} {\bibfnamefont {I.}~\bibnamefont
  {{Sodemann}}}, \bibinfo {author} {\bibfnamefont {I.}~\bibnamefont
  {{Kimchi}}}, \bibinfo {author} {\bibfnamefont {C.}~\bibnamefont {{Wang}}},\
  and\ \bibinfo {author} {\bibfnamefont {T.}~\bibnamefont {{Senthil}}},\
  }\bibfield  {title} {\bibinfo {title} {{Composite fermion duality for
  half-filled multicomponent Landau levels}},\ }\href
  {https://doi.org/10.1103/PhysRevB.95.085135} {\bibfield  {journal} {\bibinfo
  {journal} {\prb}\ }\textbf {\bibinfo {volume} {95}},\ \bibinfo {eid} {085135}
  (\bibinfo {year} {2017})},\ \Eprint {https://arxiv.org/abs/1609.08616}
  {arXiv:1609.08616 [cond-mat.str-el]} \BibitemShut {NoStop}%
\bibitem [{\citenamefont {Patri}\ \emph {et~al.}(2024)\citenamefont {Patri},
  \citenamefont {Dong},\ and\ \citenamefont {Senthil}}]{patri2024extended}%
  \BibitemOpen
  \bibfield  {author} {\bibinfo {author} {\bibfnamefont {A.~S.}\ \bibnamefont
  {Patri}}, \bibinfo {author} {\bibfnamefont {Z.}~\bibnamefont {Dong}},\ and\
  \bibinfo {author} {\bibfnamefont {T.}~\bibnamefont {Senthil}},\ }\bibfield
  {title} {\bibinfo {title} {Extended quantum anomalous hall effect in
  moir$\backslash$'e structures: phase transitions and transport},\ }\href@noop
  {} {\bibfield  {journal} {\bibinfo  {journal} {arXiv preprint
  arXiv:2408.11818}\ } (\bibinfo {year} {2024})}\BibitemShut {NoStop}%
\bibitem [{\citenamefont {Senthil}\ \emph {et~al.}(2003)\citenamefont
  {Senthil}, \citenamefont {Sachdev},\ and\ \citenamefont
  {Vojta}}]{senthil2003fractionalized}%
  \BibitemOpen
  \bibfield  {author} {\bibinfo {author} {\bibfnamefont {T.}~\bibnamefont
  {Senthil}}, \bibinfo {author} {\bibfnamefont {S.}~\bibnamefont {Sachdev}},\
  and\ \bibinfo {author} {\bibfnamefont {M.}~\bibnamefont {Vojta}},\ }\bibfield
   {title} {\bibinfo {title} {Fractionalized fermi liquids},\ }\href@noop {}
  {\bibfield  {journal} {\bibinfo  {journal} {Physical review letters}\
  }\textbf {\bibinfo {volume} {90}},\ \bibinfo {pages} {216403} (\bibinfo
  {year} {2003})}\BibitemShut {NoStop}%
\bibitem [{\citenamefont {Senthil}\ \emph {et~al.}(2004)\citenamefont
  {Senthil}, \citenamefont {Vojta},\ and\ \citenamefont
  {Sachdev}}]{senthil2004weak}%
  \BibitemOpen
  \bibfield  {author} {\bibinfo {author} {\bibfnamefont {T.}~\bibnamefont
  {Senthil}}, \bibinfo {author} {\bibfnamefont {M.}~\bibnamefont {Vojta}},\
  and\ \bibinfo {author} {\bibfnamefont {S.}~\bibnamefont {Sachdev}},\
  }\bibfield  {title} {\bibinfo {title} {Weak magnetism and non-fermi liquids
  near heavy-fermion critical points},\ }\href@noop {} {\bibfield  {journal}
  {\bibinfo  {journal} {Physical Review B}\ }\textbf {\bibinfo {volume} {69}},\
  \bibinfo {pages} {035111} (\bibinfo {year} {2004})}\BibitemShut {NoStop}%
\bibitem [{\citenamefont {Kim}\ \emph {et~al.}(2024)\citenamefont {Kim},
  \citenamefont {Timmel}, \citenamefont {Ju},\ and\ \citenamefont
  {Wen}}]{Kim2024_chiral_anyonSC}%
  \BibitemOpen
  \bibfield  {author} {\bibinfo {author} {\bibfnamefont {M.}~\bibnamefont
  {Kim}}, \bibinfo {author} {\bibfnamefont {A.}~\bibnamefont {Timmel}},
  \bibinfo {author} {\bibfnamefont {L.}~\bibnamefont {Ju}},\ and\ \bibinfo
  {author} {\bibfnamefont {X.-G.}\ \bibnamefont {Wen}},\ }\href
  {https://arxiv.org/abs/2409.18067} {\bibinfo {title} {Topological chiral
  superconductivity}} (\bibinfo {year} {2024}),\ \Eprint
  {https://arxiv.org/abs/2409.18067} {arXiv:2409.18067 [cond-mat.str-el]}
  \BibitemShut {NoStop}%
\bibitem [{\citenamefont {{Divic}}\ \emph {et~al.}(2024)\citenamefont
  {{Divic}}, \citenamefont {{Cr{\'e}pel}}, \citenamefont {{Soejima}},
  \citenamefont {{Song}}, \citenamefont {{Millis}}, \citenamefont {{Zaletel}},\
  and\ \citenamefont {{Vishwanath}}}]{Divic2024_anyonSC_topcrit}%
  \BibitemOpen
  \bibfield  {author} {\bibinfo {author} {\bibfnamefont {S.}~\bibnamefont
  {{Divic}}}, \bibinfo {author} {\bibfnamefont {V.}~\bibnamefont
  {{Cr{\'e}pel}}}, \bibinfo {author} {\bibfnamefont {T.}~\bibnamefont
  {{Soejima}}}, \bibinfo {author} {\bibfnamefont {X.-Y.}\ \bibnamefont
  {{Song}}}, \bibinfo {author} {\bibfnamefont {A.}~\bibnamefont {{Millis}}},
  \bibinfo {author} {\bibfnamefont {M.~P.}\ \bibnamefont {{Zaletel}}},\ and\
  \bibinfo {author} {\bibfnamefont {A.}~\bibnamefont {{Vishwanath}}},\
  }\bibfield  {title} {\bibinfo {title} {{Anyon Superconductivity from
  Topological Criticality in a Hofstadter-Hubbard Model}},\ }\href
  {https://doi.org/10.48550/arXiv.2410.18175} {\bibfield  {journal} {\bibinfo
  {journal} {arXiv e-prints}\ ,\ \bibinfo {eid} {arXiv:2410.18175}} (\bibinfo
  {year} {2024})},\ \Eprint {https://arxiv.org/abs/2410.18175}
  {arXiv:2410.18175 [cond-mat.str-el]} \BibitemShut {NoStop}%
\bibitem [{\citenamefont {{Milovanovi{\'c}}}\ and\ \citenamefont
  {{Papi{\'c}}}(2009)}]{Milovanovic2007_CFpairing}%
  \BibitemOpen
  \bibfield  {author} {\bibinfo {author} {\bibfnamefont {M.~V.}\ \bibnamefont
  {{Milovanovi{\'c}}}}\ and\ \bibinfo {author} {\bibfnamefont {Z.}~\bibnamefont
  {{Papi{\'c}}}},\ }\bibfield  {title} {\bibinfo {title} {{Nonperturbative
  approach to the quantum Hall bilayer}},\ }\href
  {https://doi.org/10.1103/PhysRevB.79.115319} {\bibfield  {journal} {\bibinfo
  {journal} {\prb}\ }\textbf {\bibinfo {volume} {79}},\ \bibinfo {eid} {115319}
  (\bibinfo {year} {2009})},\ \Eprint {https://arxiv.org/abs/0710.0478}
  {arXiv:0710.0478 [cond-mat.mes-hall]} \BibitemShut {NoStop}%
\bibitem [{\citenamefont {{M{\"o}ller}}\ \emph {et~al.}(2008)\citenamefont
  {{M{\"o}ller}}, \citenamefont {{Simon}},\ and\ \citenamefont
  {{Rezayi}}}]{Moller2008_CFpairing}%
  \BibitemOpen
  \bibfield  {author} {\bibinfo {author} {\bibfnamefont {G.}~\bibnamefont
  {{M{\"o}ller}}}, \bibinfo {author} {\bibfnamefont {S.~H.}\ \bibnamefont
  {{Simon}}},\ and\ \bibinfo {author} {\bibfnamefont {E.~H.}\ \bibnamefont
  {{Rezayi}}},\ }\bibfield  {title} {\bibinfo {title} {{Paired Composite
  Fermion Phase of Quantum Hall Bilayers at
  {\ensuremath{\nu}}=(1)/(2)+(1)/(2)}},\ }\href
  {https://doi.org/10.1103/PhysRevLett.101.176803} {\bibfield  {journal}
  {\bibinfo  {journal} {\prl}\ }\textbf {\bibinfo {volume} {101}},\ \bibinfo
  {eid} {176803} (\bibinfo {year} {2008})},\ \Eprint
  {https://arxiv.org/abs/0804.1286} {arXiv:0804.1286 [cond-mat.mes-hall]}
  \BibitemShut {NoStop}%
\bibitem [{\citenamefont {{M{\"o}ller}}\ \emph {et~al.}(2009)\citenamefont
  {{M{\"o}ller}}, \citenamefont {{Simon}},\ and\ \citenamefont
  {{Rezayi}}}]{Moller2009_CFpairing}%
  \BibitemOpen
  \bibfield  {author} {\bibinfo {author} {\bibfnamefont {G.}~\bibnamefont
  {{M{\"o}ller}}}, \bibinfo {author} {\bibfnamefont {S.~H.}\ \bibnamefont
  {{Simon}}},\ and\ \bibinfo {author} {\bibfnamefont {E.~H.}\ \bibnamefont
  {{Rezayi}}},\ }\bibfield  {title} {\bibinfo {title} {{Trial wave functions
  for {\ensuremath{\nu}}=(1)/(2)+(1)/(2) quantum Hall bilayers}},\ }\href
  {https://doi.org/10.1103/PhysRevB.79.125106} {\bibfield  {journal} {\bibinfo
  {journal} {\prb}\ }\textbf {\bibinfo {volume} {79}},\ \bibinfo {eid} {125106}
  (\bibinfo {year} {2009})},\ \Eprint {https://arxiv.org/abs/0811.4116}
  {arXiv:0811.4116 [cond-mat.mes-hall]} \BibitemShut {NoStop}%
\bibitem [{\citenamefont {{Milovanovi{\'c}}}\ \emph {et~al.}(2015)\citenamefont
  {{Milovanovi{\'c}}}, \citenamefont {{Dobardzi{\'c}}},\ and\ \citenamefont
  {{Papi{\'c}}}}]{Milovanovic2015_CFpairing}%
  \BibitemOpen
  \bibfield  {author} {\bibinfo {author} {\bibfnamefont {M.~V.}\ \bibnamefont
  {{Milovanovi{\'c}}}}, \bibinfo {author} {\bibfnamefont {E.}~\bibnamefont
  {{Dobardzi{\'c}}}},\ and\ \bibinfo {author} {\bibfnamefont {Z.}~\bibnamefont
  {{Papi{\'c}}}},\ }\bibfield  {title} {\bibinfo {title} {{Meron deconfinement
  in the quantum Hall bilayer at intermediate distances}},\ }\href
  {https://doi.org/10.1103/PhysRevB.92.195311} {\bibfield  {journal} {\bibinfo
  {journal} {\prb}\ }\textbf {\bibinfo {volume} {92}},\ \bibinfo {eid} {195311}
  (\bibinfo {year} {2015})},\ \Eprint {https://arxiv.org/abs/1509.01921}
  {arXiv:1509.01921 [cond-mat.str-el]} \BibitemShut {NoStop}%
\end{thebibliography}%
